\documentclass[aps,prx,nofootinbib,twocolumn,amsmath,amssymb,superscriptaddress,notitlepage,longbibliography]{revtex4-1}

\usepackage{latexsym}
\usepackage{graphicx}
\usepackage{times,psfrag}
\usepackage{enumerate}
\usepackage{amsmath}
\usepackage{dsfont}
\usepackage{dcolumn}
\usepackage{bm,bbm}
\usepackage{hyperref}
\usepackage{color}
\usepackage{latexsym,amsmath,amssymb,bm,euscript}
\usepackage{dsfont}
\usepackage{textcomp}
\usepackage{tabularx}
\usepackage{setspace}
\usepackage{sidecap}
\usepackage{placeins}
\usepackage{threeparttable}
\usepackage{multirow}
\usepackage{physics}
\usepackage{tikz-feynman}
\usepackage{pifont}%
\usepackage[T1]{fontenc}
\usepackage[utf8]{inputenc}

\let \oldbm \bm
\renewcommand{\vec}[1]{\oldbm{#1}}

\def\bk{{\vec k}}
\def\bb{{\vec b}}
\def\ba{{\vec a}}

\def\bq{{\vec q}}
\def\bt{{\vec t}}

\def\bR{{\vec R}}

\def\bp{{\bf p}}

\def\bn{{\vec n}}
\def\bm{{\vec m}}

\def\br{{\vec r}}
\def\bq{{\vec q}}
\def\bt{{\vec t}}

\def\sgn{\mathop{\mathrm{sgn}}}
\def\tr{\mathop{\mathrm{tr}}}
\def\Z{\mathds{Z}}
\def\T{\mathcal{T}}
\def\Q{\mathcal{Q}}
\def\O{\mathcal{O}}

\def\I{\mathcal{I}}
\def\G{\mathcal{G}}

\def\H{\mathcal{H}}
\def\K{\mathcal{K}}

\def\inv{^{-1}}
\def\half{{1\over2}}

\def\W{\mathcal{W}}
\renewcommand{\ket}[1]{\left| #1 \right>} 
\renewcommand{\bra}[1]{\left< #1 \right|}

\newcommand{\beq}{\begin{equation}}
\newcommand{\eeq}{\end{equation}}
\newcommand{\beqarray}{\begin{eqnarray}}
\newcommand{\eeqarray}{\end{eqnarray}}

\allowdisplaybreaks

\begin{document}

\author{Eslam Khalaf}
\affiliation{Department of Physics, Harvard University, Cambridge, MA 02138}
\author{Wladimir A. Benalcazar}
\affiliation{Department of Physics, The Pennsylvania State University, University Park, PA 16802, USA}
\author{Taylor L. Hughes}
\affiliation{Department of Physics and Institute for Condensed Matter Theory, University of Illinois at Urbana-Champaign, IL 61801, USA}
\author{Raquel Queiroz}
\affiliation{Department of Condensed Matter Physics,
Weizmann Institute of Science,
Rehovot 7610001, Israel}

\title{Boundary-obstructed topological phases}
\begin{abstract}
Symmetry protected topological (SPT) phases are gapped phases of matter that cannot be deformed to a trivial phase without breaking the symmetry or closing the bulk gap. Here, we introduce a new notion of a topological obstruction that is not captured by bulk energy gap closings in periodic boundary conditions. More specifically, given a symmetric boundary termination we say two bulk Hamiltonians belong to distinct boundary obstructed topological phases (BOTPs) if they can be deformed to each other on a system with periodic boundaries, but cannot be deformed to each other in the open system without closing the gap at at least one high symmetry surface. BOTPs are not topological phases of matter in the standard sense since they are adiabatically deformable to each other on a torus but, similar to SPTs, they are associated with boundary signatures in open systems such as surface states or fractional corner charges. In contrast to SPTs, these boundary signatures are not anomalous and can be removed by symmetrically adding lower dimensional SPTs on the boundary, but they are stable as long as the spectral gap at high-symmetry edges/surfaces remains open. We show that the double-mirror quadrupole model of [Science, 357(6346), 2018] is a prototypical example of such phases, and present a detailed analysis of several aspects of boundary obstructions in this model. In addition, we introduce several three-dimensional models having boundary obstructions, which are characterized either by surface states or fractional corner charges. Furthermore, we provide a complete characterization of boundary obstructed phases in terms of symmetry representations. Namely, two distinct BOTP phases correspond to equivalent band representations in the periodic system which become inequivalent upon restricting the symmetry group to that of the open system. This is used to shown that for a given open boundary, there is only one class of BOTPs which corresponds to a local representation of the symmetry of the open system and thus can be designated as the trivial phase. All other BOTP classes do not correspond to local representation of the open system and as a result necessarily exhibit a filling anomaly or gapless surface states. 

\end{abstract}

\date{\today}
\maketitle

\tableofcontents

\section{Introduction}

A symmetry protected topological (SPT) phase is a gapped phase of matter that cannot be adiabatically deformed to a trivial phase without breaking the symmetry or closing the bulk energy gap \cite{Kane05a, Kane05b, Bernevig06, Qi08, Moore09, Hasan10, Qi11, Molenkamp13}. For free fermion systems with internal symmetries, a complete understanding of SPTs was achieved in the pioneering work of Refs.~\cite{Kitaev09, Schnyder09,Qi08,Ryu10}. In any given symmetry class and dimension ($d$), an SPT hosts anomalous gapless $(d-1)$-dimensional surface states whose existence is tied to the non-trivial bulk topology. This connection between bulk and boundary topological properties is known as a bulk-boundary correspondence.

For spatial or crystalline symmetries, the relationship between bulk and boundary topological signatures can be more subtle \cite{Alexandradinata, Trifunovic18}. On one hand, there are topological crystalline phases, such as mirror Chern insulators \cite{Dziawa12, Tanaka12}, that exhibit a conventional bulk-boundary correspondence signaled by the appearance of 2D gapless surface states on any mirror invariant surface plane. On the other hand, less traditional types of surface states, known as higher-order surface states \cite{Benalcazar14,Benalcazar17, Benalcazar17b, Langbehn17, Schindler17, Song17, Fang17, Khalaf17, Geier18, Khalaf18, Trifunovic18}, which have a lower dimensionality, e.g., states confined to corners or hinges of the sample, are also possible in systems with crystalline symmetry. Even in the absence of any mid-gap surface (or higher-order boundary) states, additional signatures such as fractional charge at corners \cite{Benalcazar18,peterson2020} or defects \cite{Benalcazar14, TeoHughes17, Liu18,Li19} can be used to distinguish different SPTs. Furthermore, there are some bulk SPTs protected by crystalline symmetries that are not associated with surface states or boundary fractional charges at all. For instance, two atomic insulators corresponding to filling the same Wyckoff positions, but with orbitals that transform \emph{differently} under site symmetries, cannot be smoothly deformed to each other, but they cannot be distinguished by surface states or corner charges either \cite{Po17, bradlyn2017topological, Liu18, AhnJung}. Thus, for SPTs protected by crystalline symmetries, bulk topological distinctions do not always imply boundary signatures in the form of gapless states or fractional charges at the boundary.

The purpose of this work is to investigate the reverse question. Rather than asking whether the distinction between two bulk SPTs can be captured by a boundary signature, we ask whether two systems which exhibit different boundary signatures, e.g.,  low-energy modes or fractional charges, necessarily correspond to topologically distinct systems in the bulk. There is an obvious counter-example to this statement corresponding to the case where non-trivial SPTs are placed on the boundary of a trivial bulk. For example, one can imagine gluing 2D layers on the surface of a 3D trivial insulator. If these layers are gapless in their two-dimensional bulk, then the 3D system will have surface states. If  the layers are 2D topological insulators instead, then the system will exhibit gapless hinge modes. However, these boundary signatures are not associated with a bulk property. {These phases have been dubbed} ``extrinsic higher-order" phases {and their surface states can be removed or ``peeled off'' by means of symmetric boundary manipulations}\cite{Geier18, Trifunovic18}. {However, this definition does not distinguish the cases where the boundary modes are completely decoupled from the bulk, e.g., a completely trivial bulk is attached to a non-trivial boundary, from those where the boundary modes are associated with the bulk in some way such that they can be created or destroyed by tuning bulk parameters. }

{ One of the first examples} of {the latter phenomenon} is provided by the quadrupole model of Ref.~\onlinecite{Benalcazar17}.  This is a 2D model with gapped bulk and edges that hosts quantized fractional charge at the corners. Two symmetry variants of the model were considered: one with fourfold rotation symmetry $C_{4z}$ which satisfies $(C_{4z})^4=-1,$ and the other with two anticommuting mirror symmetries. In the former case, the quadrupole phase is a topological phase in the standard sense: it is an obstructed atomic limit \cite{bradlyn2017topological}, { which denotes an atomic insulator whose charge centers are displaced relative to the underlying positive atoms, and which is }  separated from the trivial atomic limit by a bulk gap-closing phase transition. However, in the latter case, the quadrupole phase is not a topological phase, in the standard sense, since the value of the quadrupole moment can be changed without closing the bulk gap (in periodic boundary conditions). 
The topology in this model is captured by a more subtle distinction contained in the spectrum of the Wilson loop operators (Wannier spectrum), or in the entanglement spectrum\cite{Schindler17,dubinkin2020entanglement}, rather than the bulk energy spectrum. This was argued to imply that the two phases of the model are separated by an edge (rather than a bulk) phase transition when considered with open boundary conditions \cite{Benalcazar17, Benalcazar17b}. Indeed, since both the gaps in the bulk and on the edge protect the quantized corner charge, one can imagine  the corner charge delocalizing along the edges and changing values if the edge gap closes.

Despite the number of subsequent works that have studied several aspects of the quadrupole model and its generalizations, the subtle topological distinction captured by the double-mirror quadrupole insulator (DMQI) has ramifications that have been mostly overlooked. In particular, several questions regarding the nature of topological distinctions that are not captured by a bulk phase transition remain unanswered. These include: (i) how can one, in general, define a topological distinction that does not involve a gap closing phase transition in the bulk? (ii) are there other examples of models in two or three dimensions that exhibit similar phenomenology? (iii) under what conditions does an obstruction in connecting two Wannier spectra indicate a gap-closing transition at the boundary?, and (iv) how is this related to boundary signatures, e.g., fractional corner charge, or boundary states, in a general setting?

In this work, we answer these questions by introducing the concept of a boundary obstructed topological phase (BOTP) that captures distinctions between Hamiltonians that can be adiabatically connected with periodic boundary conditions, but cannot be connected for   symmetric surface terminations with open boundary conditions.  We provide a general definition of such distinctions, and show how they can be understood in terms of the Wannier spectra as well as real-space symmetry representations. We show that the presence of topologically robust boundary modes in BOTPs depends on the lattice termination, and, inversely, that {for any} BOTP there is a choice of boundary characterized by topologically robust modes protected by a boundary gap closing.

In order to make the presentation as clear as possible, we will first start with detailed explanations of different aspects of the boundary obstruction concept that are realized in the DMQI. Then we will move on to discuss the general aspects of boundary obstructions. After reviewing the DMQI model in Sec.~\ref{sec:Review}, we provide a real space framework for its boundary obstruction in Sec.~\ref{sec:2DReal}, followed by a detailed analysis of the Wannier spectra, and how they relate to the physical edge spectra of particular surface terminations in Sec.~\ref{sec:edge_Wannier}. Afterwards, we establish how the boundary obstruction in the DMQI model is related to the existence of fractional corner charge via a detailed symmetry analysis of the model with open boundaries in Sec.~\ref{sec:filling_anomaly}. Finally, in Sec. \ref{sec:symmeigsDMQI} we show how the notion of boundary obstructions in the DMQI can be understood in terms of restricting the bulk (periodic) symmetry representations to representations of the point group of the open system.

After the detailed study of the DMQI model, we introduce the general definition of boundary obstructed phases, and discuss their stability  from a homotopic point of view in Sec.~\ref{sec:GenDef}.  In Sec.~\ref{sec:bandreps} we present a definition of BOTPs from the point of view of the symmetry representations in the open system, and show how to diagnose them through the symmetry representations of the Wannier spectrum. 

Afterwards, we discuss other possible 2D models with boundary obstructions in Sec.~\ref{sec:2DProof} and show that, apart from the DMQI (and variants of it), all such models require specific, complicated surface terminations to probe the boundary obstructions and the associated corner charges. In Sec. \ref{sec:3D} we introduce a general recipe to generate 3D models with various types of boundary obstructions from a 2D building block. Such obstructions can be associated with fractional corner charges, fractional  hinge charge density (Sec.~\ref{sec:3DFA}), or boundary states (Sec.~\ref{sec:3DSS}) depending on the choice of the 2D building block. For example, in the latter case we introduce two 3D models built from either a 2D Chern insulator or a 2D quantum spin Hall insulator with robust one dimensional states localized at their hinges. We also provide a complete characterization of all the 3D models using the Wannier spectra. Finally, we close with some concluding remarks and discussion in Sec.~\ref{sec:Discussion}.

\section{2D quadrupole insulator and boundary obstructions}
\label{sec:II_DMQI}

\subsection{Review of the quadrupole insulator model}
\label{sec:Review}

\begin{figure}[t]
\centering
\includegraphics[width=\columnwidth]{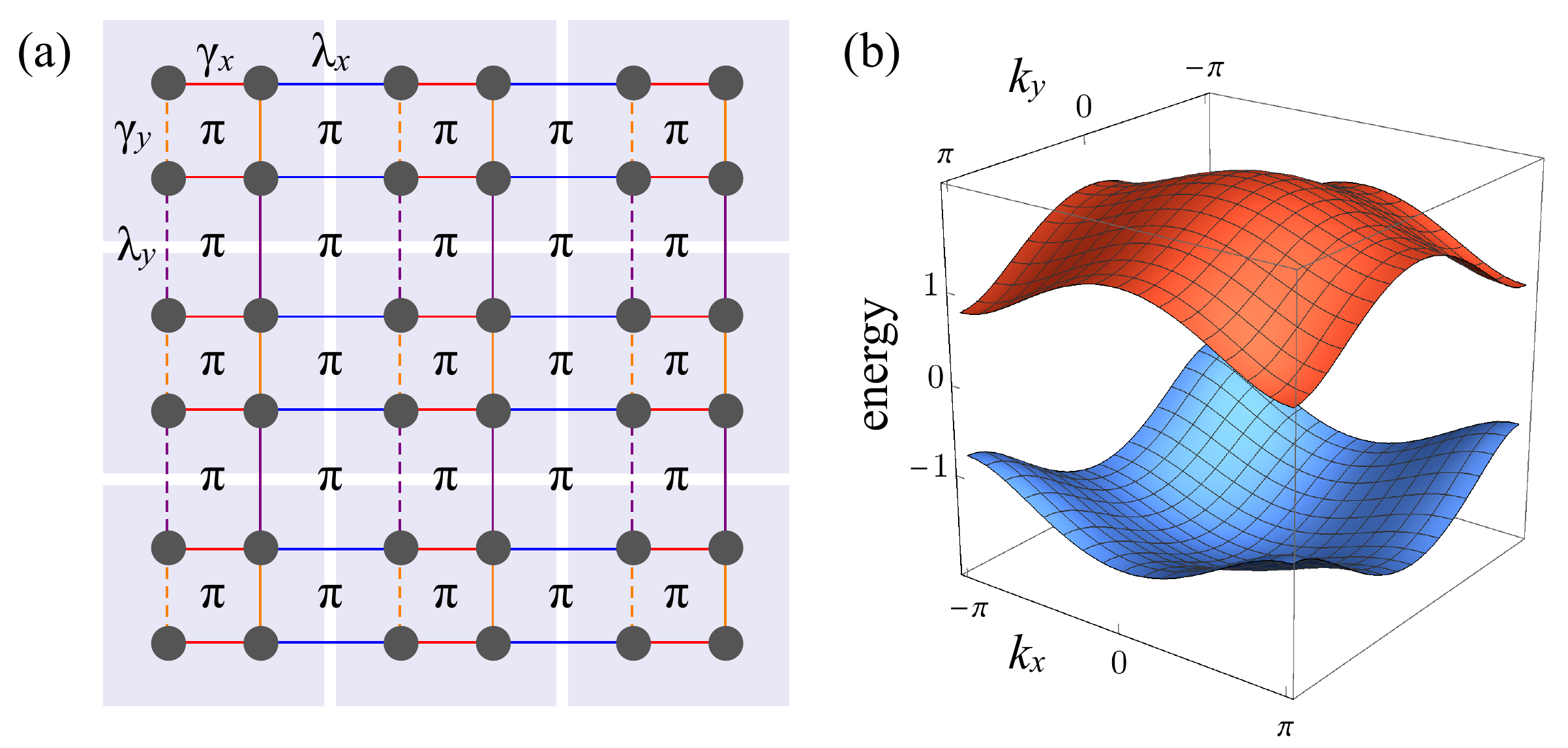}
\caption{Lattice model used to define the quadrupole insulator in Eq.~\ref{eq:QuadHamiltonian}. In (a), dashed lines have a relative negative sign to account for a flux of $\pi$ threading each plaquette. The flux is the origin of the anticommuting reflection operations.  (b) Band structure of the quadrupole insulator with $\gamma_x/\lambda_x=0.5$ and $\gamma_y/\lambda_y=0.4$. Each energy band is twofold degenerate for a total of four bands.}
\label{fig:quad_lattice}\label{fig:WannierBands}
\end{figure}

We start our discussion by reviewing the DMQI model introduced in Ref.~\cite{Benalcazar17}. It consists of four orbitals describing spinless fermions arranged on a two-dimensional rectangular lattice with dimerized hopping amplitudes along the $x$ and $y$ directions, and $\pi$-fluxes threaded through each plaquette as shown in Fig.~\ref{fig:quad_lattice}.
The Bloch Hamiltonian is given by
\begin{align}
\H(\bk)&=\left[\gamma_x + \lambda_x  \cos(k_x)\right] \Gamma_4 + \lambda_x \sin(k_x) \Gamma_3\nonumber\\
&+ \left[\gamma_y + \lambda_y \cos(k_y)\right] \Gamma_2 + \lambda_y \sin(k_y) \Gamma_1,
\label{eq:QuadHamiltonian}
\end{align}
where $\gamma_x$ and $\gamma_y$ are hopping amplitudes within a unit cell along $x$ and $y$ respectively, and $\lambda_x$ and $\lambda_y$ are the inter-cell hopping amplitudes to nearest neighbor unit cells. The negative signs, represented by the dashed lines in Fig.~\ref{fig:quad_lattice}a, correspond to our gauge choice for the  $\pi$-flux threaded through each plaquette. The $\Gamma_{0,\dots,4}$ matrices  satisfy $\{\Gamma_i, \Gamma_j\} = 2 \delta_{i,j}$, and are represented explicitly by $\Gamma_0= \sigma_3  \tau_0$, $\Gamma_k=-\sigma_2  \tau_k$, $\Gamma_4=\sigma_1  \tau_0,$ for $k=1,2,3$, where $\sigma_{0,\dots,3}$ ($\tau_{0,\dots,3}$) denote the Pauli matrices, and the tensor product is implicit.  

The model in Eq.~\ref{eq:QuadHamiltonian} has reflection symmetries along both $x$ and $y$, which for spinless electrons satisfy $M_x^2=M_y^2=1$. They are represented by
\begin{align}
M_x \H(k_x,k_y) M_x^\dagger&= \H(-k_x,k_y), \qquad M_x = \sigma _1  \tau_3, \nonumber\\
M_y \H(k_x,k_y) M_y^\dagger&= \H(k_x,-k_y), \qquad M_y = \sigma_1  \tau_1.
\label{eq:QuadReflectionSymmetries}
\end{align}
The model is also invariant under the combination $M_x M_y=C_{2z}$, where $C_{2z}$ is a two-fold rotation symmetry.
The $\pi$-flux leads to the anticommutation of the two reflection operators $\{M_x,M_y\}=0$. As a result, $C_{2z}$ satisfies $(C_{2z})^2=-1$, as would be the case for spinful fermions. Other than the crystalline symmetries, the model $\H(\vec k),$ as written, {also has (spinless) time-reversal symmetry given by complex conjugation $\T = \K,$ as well as chiral symmetry $\{\H, \Gamma_0\} = 0$. As a result, it} lies in class BDI, though these internal symmetries are not crucial for the conclusions.

{One important aspect to highlight here is that, due to the $\pi$ fluxes, the spatial symmetries of the model as defined above already include some gauge transformations. As a result, the symmetry representation in question is projective. 
Although the implementations of $M_x$ and $M_y$ themselves are gauge dependent, the quantity 
\beq
M_x M_y M_x^{-1} M_y^{-1} = -1
\label{eq:piflux}
\eeq
captures the gauge invariant plaquette flux. We note that since the flux attachment does not affect the translation symmetries ($T_x$ and $T_y$ still commute), and since the spatial symmetries are symmorphic, the full space group is given by a simple product of the point group $F=D^\pi_2$ and the group of 2D translations i.e. $G = F \times T_x \times T_y$. The projective representation considered can be equivalently understood as a $\Z_2$ extension of the point group $D_2$ defined by Eq.~\ref{eq:piflux}. We refer to such $\Z_2$ extension with a $\pi$ superscript, i.e., $D_2^{\pi}$. { The point group $D_2^\pi$ has four one-dimensional, irreducible representations ($A_1$, $A_2$, $B_1$, $B_2$), and one two-dimensional irreducible representation ($\bar E$) (see Appendix \ref{app:charactertable}).} 
 We notice that the full projective representation defined by Eq.~\eqref{eq:piflux} is equivalent to a double (spinful) representation for the space group 25 ($Pmm2$) restricted to the 2D plane or equivalently the layer group 23 ($pmm2$). This aspect was discussed in the original works \cite{Benalcazar17, Benalcazar17b}, and investigated in more detail recently \cite{Wieder2020}, but for the rest of this work we will focus on the \emph{spinless} version to simplify the intuition for the degrees of freedom in the unit cell.}

The energy spectrum of the Hamiltonian (\ref{eq:QuadHamiltonian}) is two-fold degenerate and gapped across the entire bulk Brillouin zone (BZ) (cf.~Fig.~\ref{fig:quad_lattice}) unless both $\vert\gamma_x/\lambda_x\vert=1$ and $\vert \gamma_y/\lambda_y\vert=1$. Hence, we can connect the Hamiltonians at any pair of points in the $(\gamma_x/\lambda_x, \gamma_y/\lambda_y)$-plane without closing the bulk gap. This implies the absence of any bulk topological distinctions in the model. We note that the presence of $C_{4z}$ symmetry alters this conclusion since it forces $\lambda_x = \lambda_y=\lambda$ and $\gamma_x = \gamma_y=\gamma$, thus changing the sign of $\vert\gamma/\lambda\vert - 1$ is necessarily accompanied by a bulk-gap-closing phase transition, which indicates that the model has at least two distinct bulk-protected topological phases.

Despite the absence of bulk topological distinctions in the model protected by mirror symmetries, Refs.~\onlinecite{Benalcazar17, Benalcazar17b} have uncovered a more subtle topological distinction encoded in the topology of the Wannier bands instead. The Wannier bands along the $x$($y$) direction are obtained as the eigenspectrum of a Wilson loop operator along this direction for a fixed  momentum $k_y$ ($k_x$)\cite{Marzari97, Marzari12}. The Wilson loop operator is denoted $\W^\bb(\bk)$, where $\bb$ is the direction along which the operator is taken in the Brillouin zone. $\W^\bb(\bk)$ is a unitary operator whose eigenvalues have the form $e^{2\pi i \nu_\bb(\bk)}$. The values $\nu_\bb(\bk)$ are defined modulo one, and represent the positions of the charge centers (within the unit cell) of hybrid Wannier functions that are maximally localized in the $\bb$-direction, but are delocalized Bloch waves in the perpendicular direction parameterized by the transverse momentum $\bk$ \cite{Marzari97, Marzari12}. For the DMQI model, the two Wannier bands of the occupied states are generally gapped  and symmetrically displaced away from the high-symmetry lines $\nu_\bb=0$ and $\nu_\bb=1/2$ (cf.~Fig.~\ref{fig:numerics2d}b). Among other things, this indicates that there is no spectral flow as would be the case, for example, for a strong topological insulator \cite{qibernevigwilsonloop}. 

Whenever the two Wannier bands are separated by gaps from above and below we can consider the projector onto one of these two bands. This projector is effectively projecting onto the ground state of a 1D reflection-symmetric insulator with one electron per unit cell. Such an insulator can be characterized by a half-quantized polarization $p,$ {measured in integer units of the charge}, distinguishing the cases where the charge center is at the center ($p = 0$), or the edge ($p=1/2$), of the 1D unit cell. Since there are two possible sets of Wannier spectra, i.e., one along the $x$-direction and one along the $y$-direction, there are two distinct quantized Wannier band polarizations $p_{x,y} = 0, 1/2$ and $p_{y,x} = 0, 1/2$ where $p_{i,j}$ denotes the polarization in the $j$-direction for a band taken from the Wilson loop in the $i$-direction. This yields a $\Z_2 \times \Z_2$ invariant capturing the topology of the Wannier bands. Such invariants are protected by the gap in the Wannier spectrum rather than the energy spectrum, which means that their value can be changed without going through a bulk gap closing. As we will explain in detail later, these types of transitions can {\it in some cases} be associated with an energy gap-closing at the {\it edge}, rather than the bulk, when the system is considered with open boundary conditions in both directions. This will be the topic of Sec.~\ref{sec:edge_Wannier}.

For $\bp^\nu = (p_{x,y}, p_{y,x})=(1/2, 1/2)$, the model exhibits mid-gap corner modes when considered on a rectangular geometry with open boundaries in both directions, (and {when the boundary coincides an edge of the unit cell).} 
However, these corner modes are protected by chiral symmetry which can be broken without changing the Wannier band polarization (whose quantization relies on only the mirror symmetries). Thus, these corner modes are not generically associated with the quantized Wannier band polarization. { Instead we can associate the $(1/2, 1/2)$ phase with corner {\it charge} \cite{Benalcazar17,Benalcazar17b, Benalcazar18, Liu18, YouBurnell} as will be discussed in detail in Sec.~\ref{sec:filling_anomaly}. It is worth noting that the fractional corner charge can be removed by adding edge degrees of freedom. Adding fractional corner charge to the trivial phase, will interchange which phase we identify as topologically non-trivial, but the fact that the two phases can be distinguished by the fractional corner charge remains true. This topological distinction will be discussed in Sec.~\ref{sec:filling_anomaly}.} 

Following this review of the DMQI, we will dedicate the remainder of this section to showing that the topology of the DMQI can be captured using the notion of edge topological obstructions; a new type of topological obstruction that is present only when the model is placed on certain geometries with open boundary conditions. In particular,  we will provide an intuitive picture for a topological obstruction that is associated with an edge gap-closing transition rather than a bulk one. Furthermore, we will show how the $\Z_2 \times \Z_2$ topological distinction encoded in the Wannier band polarizations is related to the corner charge, which is determined by only a single $\Z_2$ invariant that distinguishes $\bp^\nu = (1/2,1/2)$ from the other three cases. We note here that picking $\bp^\nu = (1/2,1/2)$ as the non-trivial phase relies on the boundary termination convention {shown in Fig.~\ref{fig:WannierBands} which is chosen such that the vertical (horizontal) edge is invariant under $M_y$ ($M_x$), \emph{and} the edge lies at the border of the unit cell depicted in the figure (denoted by $x,  y = \pm 1/2$). As we will discuss later, the choice of the boundary decides which Wannier phase is distinguished from the others. In all cases, however, the resulting classification for a fixed boundary is $\Z_2$. This is similar to the Su-Schrieffer-Heeger chain where changing the boundary termination by half a unit cell will change the identification of the trivial and obstructed atomic limits.

\begin{figure}[t]
    \centering
    \includegraphics[width=\columnwidth]{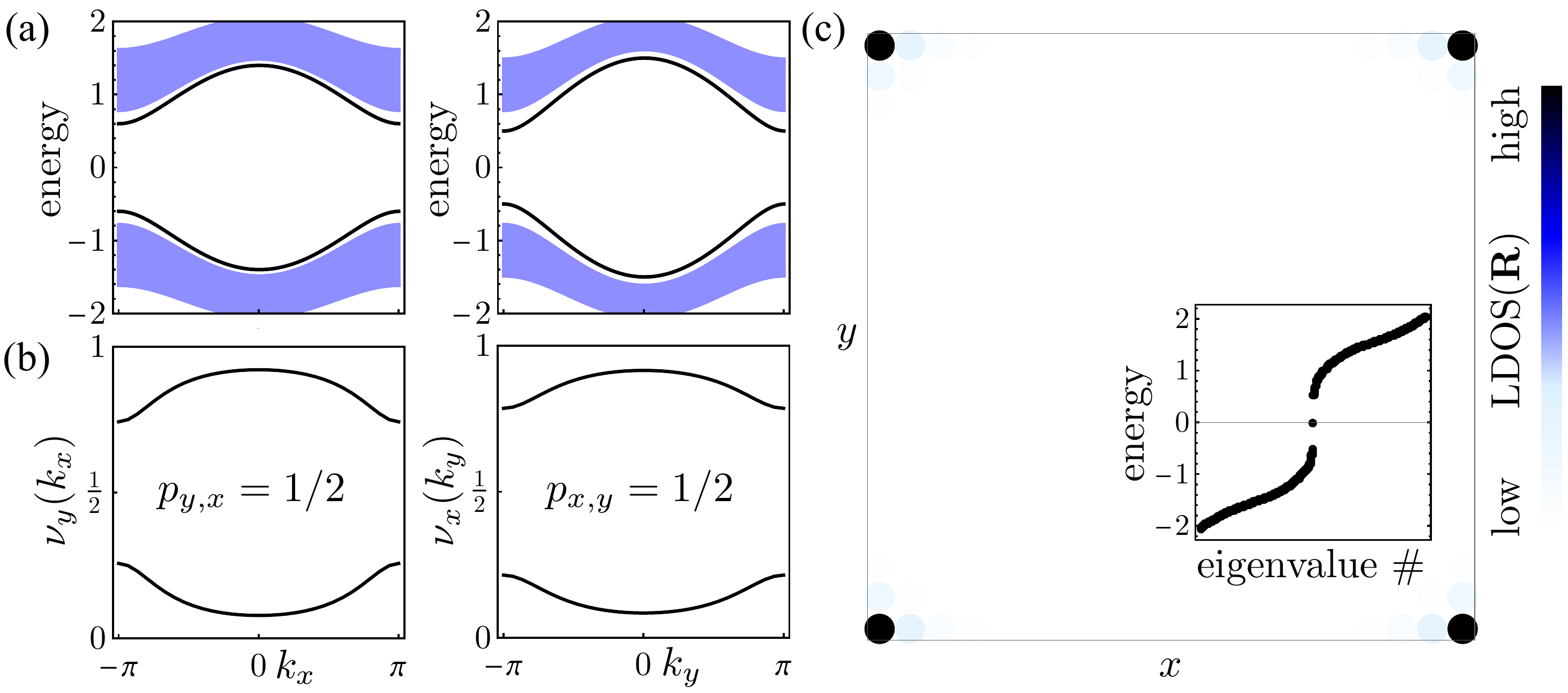}
    \caption{(Color online) (a) Band structure of the two-dimensional quadrupole insulator with open boundaries in a single direction, $y$ (left) or $x$ (right). The blue bands correspond to the bulk band continuum, and the black bands that are slightly separated from the bulk bands are localized at the edges. (b) Wannier bands $\nu_{y}(k_x)$ (left) and $\nu_{x}(k_y)$ (right), with a quantized Wannier band polarization of $p_{x,y}=p_{y,x}=1/2$. The Wannier spectra are topologically equivalent to the surface bands in (a) when we identify the gap around $\nu_i(k)=1/2$ to represent the surface gap, see Sec. \ref{sec:edge_Wannier}. (c) Local density of states at zero energy ${\rm LDOS}(\vec r)=-{\rm Im}\tr (H(\vec r)+i\eta)\inv/\pi$ for a small broadening $\eta=0.1$ in a system with open boundaries in $x$ and $y$. The inset shows the existence of zero energy states inside the gap.}
    \label{fig:numerics2d}
\end{figure}

\subsection{Real space picture}
\label{sec:2DReal}

In the following, we will present an understanding of the edge topological obstruction of the DMQI in terms of a real space picture. At half-filling, the space of filled bands is Wannier representable, i.e., it is possible to find a basis of localized symmetric orbitals that span the subspace of filled bands. In a Wannier representable system, localized orbitals are labelled by their so-called Wyckoff position $\Q = \{\bq_i\}$ which denotes a \emph{set} (symmetry orbit) of spatial positions $\bq_i$ with $i=1,...,N_\Q,$ which is invariant as a whole under the symmetry group $G$. The positions $\bq_i$ can map to each other under some elements of $G$, $\bq_i=g_i\bq$ for some reference $\bq$, but they are invariant under a (site symmetry) subgroup $G_\bq \subseteq G$ \footnote{The site symmetry groups for different elements $\bq_i$ in the same Wyckoff position are equivalent up to conjugation {by an element in $G$}, so we denote them by the same symbol $G_\bq.$}. $N_\Q$ denotes the size of the set of spatial positions $\bq_i,$ and is called the Wyckoff multiplicity of position $\Q$. Wannier representable (atomic) insulators have electronic configurations constructed by decorating the various Wyckoff positions with orbitals that transform under the site-symmetry representations of $G_\bq$. These different Wannier representable phases transform under space group representations that are induced from a local representation of $G_\bq$ \cite{zak80,zak1982band}, and are each associated with a \emph{band representation} (BR) \cite{zak80,zak1982band,michel2000elementary,bradlyn2017topological,vergniory2017graph} that encodes both its transformations under the symmetry group, and its Zak-Berry phases {in momentum space}. A short derivation of the band representation from the local representation of $G_\bq$ is presented in Sec.~\ref{sec:bandreps}.

In the DMQI case, {the full space group is a direct product of the point group $F=D^\pi_2$ and the group of translations of a rectangular lattice in two dimensions, $G=F\times T$. As previously mentioned, it can be thought of a $\Z_2$ extension of the layer group 23 ($pmm2$) using Eq.~\ref{eq:piflux}}. 
This group has the same set of Wyckoff positions as the { original layer group}: (i) four maximal Wyckoff positions with multiplicity 1 corresponding to positions lying at the intersection of the two mirror-invariant lines, i.e., at positions $(0,0)$, $(0,1/2)$, $(1/2,0)$ and $(1/2,1/2)$ (labelled $1a$, $1b$, $1c$ and $1d$, respectively){ with $G_\bq$ coinciding with the full point group $F$}; (ii) four Wyckoff positions with multiplicity 2 corresponding to positions lying along one of the mirror-invariant lines, $(\pm x,0)$, $(\pm x,1/2)$, $(0,\pm y)$, $(1/2,\pm y)$,  (labelled $2e$, $2f$, $2g$, and $2h$, respectively) with $G_\bq = D_1$; and (iii) one general Wyckoff position $4i$ with multiplicity 4 at four generic symmetry-related points $(\pm x,\pm y)$ where $G_\bq=C_1$. {The positions in the unit cell associated with these labels are shown in Fig.~\ref{fig:WyckoffpositionsD2}.} 

\begin{figure}
    \centering
    \includegraphics[width=.6\columnwidth]{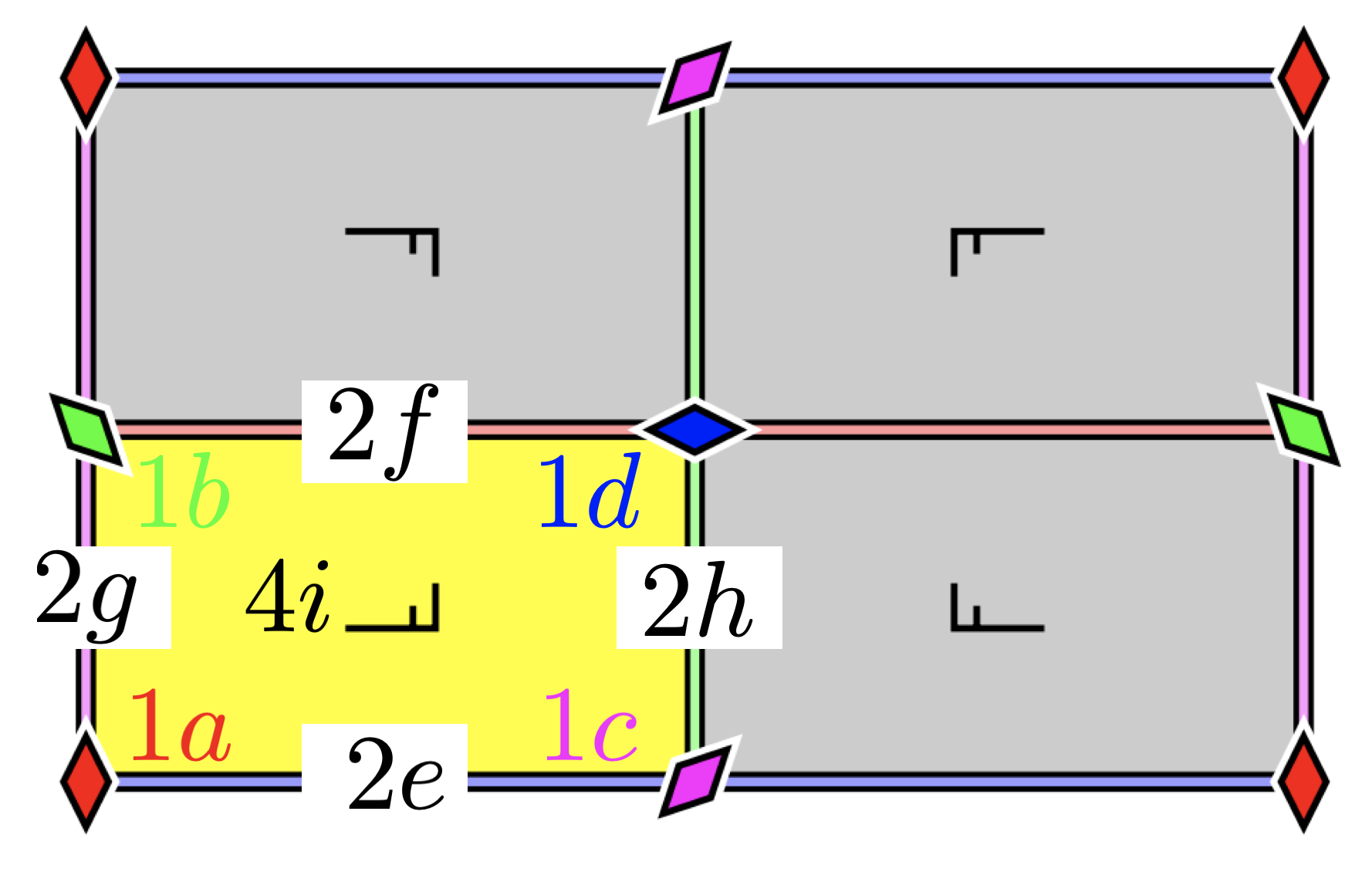}
    \caption{Wyckoff positions for the space group of the DMQI model, with site symmetry groups $D^\pi_2$, $D_1$ and $C_1$, for those Wyckoff positions with multiplicity 1, 2, and 4, respectively. Figure adapted from Ref.~\cite{Wiki}.}
    \label{fig:WyckoffpositionsD2}
\end{figure}

To illustrate the absence of a bulk obstruction in the DMQI model, consider a translationally invariant lattice with $M_x$ and $M_y$ symmetries having periodic boundary conditions \footnote{In general, point group operations act non-locally in real-space, and generically relate orbitals in different unit cells. However, with periodic boundary conditions one can use translation symmetry on a transformed orbital to shift it back to the initial unit cell. This allows us to represent the different point-group operations as an action on Wannier centers inside a unit cell.}. For a filling of two electrons per unit cell, there are only three  allowed options for placing the atomic orbitals: (i) both are placed at the {\it same} maximal Wyckoff position, (ii) they are placed at a Wyckoff position of multiplicity 2, or (iii) the two orbitals are placed at {\it different} maximal Wyckoff positions. Case (iii) has a non-vanishing polarization of $1/2$ along $x$ or $y$ (or both) which rules it out for the DMQI model for which both polarizations vanish. Interestingly, cases (i) and (iii) may be adiabatically deformed into each other in some cases, i.e., they may belong to \emph{equivalent} band representations. This is the case when the two electrons (per cell) transform as a two dimensional representation of the $D^\pi_2$ point group, i.e., when the matrix representations of the two mirror operators anticommute when acting on the electron orbitals. To show this, we note that equivalence between two band representations can be demonstrated if there is a symmetric, adiabatic deformation between the two configurations, see Refs.~\cite{bradlyn2017topological,cano2018building} and App.~\ref{app:BRs}.  In Appendix \ref{app:mirrors} we show that in order to carry out such a continuous deformation between Wyckoff position configurations, the two mirror operators are required to anticommute, which is exactly the case for the DMQI model. The necessity for anticommutation of mirror operators was previously proven in Ref. \cite{Benalcazar17, Benalcazar17b} using the symmetry indicators of the Wannier bands, where it was shown to be linked to the existence of gaps in the Wannier spectra.


In the following, we will present an alternative and more intuitive argument based for the equivalence of these representations based on an illustration of real space deformations. This will be backed by a more rigorous analysis in subsequent sections. In Fig.~\ref{fig:edge_obstruction} we explicitly show how to deform the configuration with both electrons at {$1a$} to the one where both are at $1d$ while preserving the symmetry and the bulk gap. {There are two distinct ways to do this. The first one goes through $1c$ by first moving horizontally through $2e$ then vertically through $2h,$ whereas the second goes through $1b$ by first moving vertically through $2g$ then horizontally through $2f$. Such deformations can be written in terms of deformations of the parameters of the Hamiltonian. For example, the deformation $1a \rightarrow 1b \rightarrow 1d$ starts with the parameters $\gamma_x = \gamma_y \neq 0$ and $\lambda_x = \lambda_y = 0$ which yields Wannier orbitals centered at $1a$. Next, we gradually increase $\lambda_x$ and decrease $\gamma_x$ to get to $\lambda_x \neq 0$ and $\gamma_x = 0$ whose Wannier orbitals are centered at $1b$. Finally, we increase $\lambda_y$ and decrease $\gamma_y$ to get to $\gamma_x = \gamma_y \neq 0$ and $\lambda_x = \lambda_y = 0$ which yields Wannier orbitals centered at $1a$. The center of the Wannier orbitals in these cases can be obtained easily since all of these are fully dimerized limits.} 

It is instructive again to compare to the model with $C_{4z}$ symmetry for which {the movable Wyckoff positions $2e$, $2f$, $2g$, $2h$ are not allowed}. In this case, it is impossible to move two electrons from $1a$ to $1d$ without breaking the symmetry, indicating that the two correspond to distinct bulk phases. {In terms of the Hamiltonian parameters, $C_4$ symmetry forces $\lambda_x = \lambda_y$ and $\gamma_x = \gamma_y$. Under this constraint, it is impossible to go from $\lambda_x = \lambda_y \neq 0$, $\gamma_x = \gamma_y = 0$ (corresponding to $1a$) to $\lambda_x = \lambda_y = 0$, $\gamma_x = \gamma_y \neq 0$ (corresponding to $1d$) without going through the point $\gamma_x = \gamma_y = \lambda_x = \lambda_y$ where the bulk Hamiltonian is gapless.}

\begin{figure}
\center
\includegraphics[width=\columnwidth]{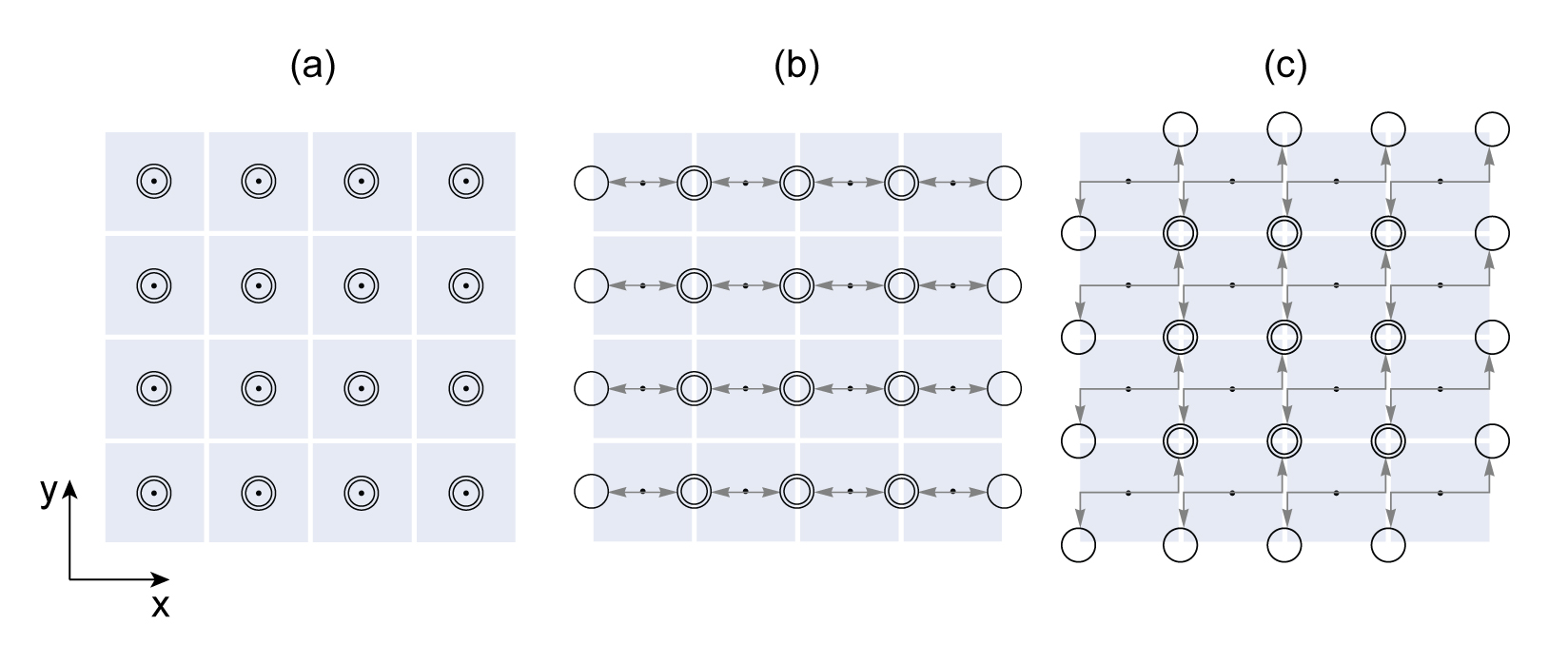}
\caption{ Deformation of the Wannier centers (open circles), in the double-mirror quadrupole model, from the $1a$ to the $1d$ Wyckoff position along movable Wyckoff positions. We see that from panel (b) to panel (c) we cannot preserve the symmetry at the edges. }
\label{fig:edge_obstruction}
\end{figure}

{Let us now see what happens in the presence of a boundary termination which we take, for definiteness, to be a rectangle whose horizontal (vertical) edges are parallel to the $x$ ($y$) axis and lie at a position $x, y =1/2$. 
This means that the edges of the sample coincide with the the edge rather than the center of our chosen unit cell. We then imagine how the bulk deformations explained in the previous paragraph, namely the two possible deformations connecting $1a$ to $1d$ either through $1b$ or $1c$, are affected by the presence of the vertical and horizontal edges. Such deformations are carried out by changing the parameters $\lambda_{x,y}$ and $\gamma_{x,y}$ of the Hamiltonian while keeping the boundary fixed. For definiteness, we consider a trajectory of bulk Hamiltonians $\H(t)$ such that $\H(0) = \H_{1a}$ and $\H(1) = \H_{1d}$ which, upon choosing a boundary termination, induces a trajectory of Hamiltonians in the open system $\H_{\rm obc}(t)$ (we provide a precise description of this in Section \ref{sec:GenDef} ). Deep inside the bulk, $\H(t)$ and $\H_{\rm obc}(t)$ are the same so that the picture of sliding the Wannier centers explained above makes sense inside the bulk of the open system. In addition, if the Hamiltonian $\H_{\rm obc}(t)$ is gapped for all $t$, we may be tempted to assume that the position of the charge centers remain well-defined also close to the edge as we change $t$. However, the schematic illustration of Fig.~\ref{fig:edge_obstruction} suggests that this is impossible, as we now discuss in detail.}

To explain the intuitive picture behind the boundary obstruction depicted in Fig.~\ref{fig:edge_obstruction}, let us consider the trajectory connecting $1a$ to $1d$ through $1c$. We note that in the presence of the vertical boundary at $x = 1/2$, the electron filling of position $1c$ \emph{on the edge} is half of its filling in the bulk (since such a site is shared by two unit cells when in the bulk, but one of them is now absent on an edge). This means that there is only one electron at the $1c$ position at the edge, instead of two. This makes it impossible to vertically move it to position $1d$ while preserving the symmetry and energy gap.  Similarly, the presence of the horizontal edge at $y = 1/2$ prohibits the deformation $1a \rightarrow 1b \rightarrow 1d$ as position $2f$ is unavailable at the edge. Thus, with open boundaries we can pass from $1a$ to $1b$ or $1c$, but not to $1d,$ which distinguishes $1d$ from the other three configurations in the presence of open, symmetric boundaries.

To summarize, Fig.~\ref{fig:edge_obstruction} suggests there are two ways to move two electrons from $1a$ and $1d$: one of which goes through position $1c$ and one through $1b$. Importantly, each of the two edges ($x$ or $y$) prohibits only \emph{one} of the two trajectories. Thus, if we open the boundaries along only one direction by considering the system on a cylinder, it is still possible to deform $1a$ to $1d$, and thus they are not distinguished. However, once we take open boundaries along both the $x$ and $y$ directions (with an edge termination consistent with the unit cell), these two atomic configurations cannot be continuously connected while preserving the symmetry. This picture suggests the DMQI exhibits what we will call an \emph{edge obstruction} 
wherein two bulk Hamiltonians which are smoothly deformable in a periodic system (i.e. in the bulk) are not continuously deformable in the open system, i.e., any trajectory connecting them will involve a gap closing on the edge or breaking the symmetry. In the following two subsections, we will expand on the intuitive understanding presented here by presenting two rigorous formulations of the boundary obstruction in the system: one in terms of edge phase transitions, and the other in terms of a Wannier sliding argument in the open system.

\subsection{Wannier bands and edge spectrum}
\label{sec:edge_Wannier}

 We now switch our attention to the diagnosis of the edge obstruction in terms of the Wannier spectrum. As discussed in Sec.~\ref{sec:Review}, and Refs.~\onlinecite{Benalcazar17, Benalcazar17b}, the Wannier bands of the DMQI model can be characterized by a $\Z_2 \times \Z_2$  invariant $\bp^\nu = \tfrac{1}{2}(\Theta(1-|\gamma_x/\lambda_x|), \Theta(1-|\gamma_y/\lambda_y|))$ (where $\Theta(x)$ is the step function). Using the correspondence between the Wannier spectrum and the edge spectrum \cite{Fidkowski11}, we may naively think that the DMQI has four distinct phases that cannot be smoothly connected to each other without closing the energy gap at the edge. This, however, contradicts the analysis of the previous sections {as well as the result of Refs.~\cite{Benalcazar17, Benalcazar17b} which} found two, rather than four, distinct phases {distinguished by the corner charge (or alternatively the quadrupole moment)}.

The resolution to this puzzle lies in the observation that the edge spectrum and the Wannier spectrum differ in one fundamental aspect, despite being {continuously} deformable to each other. Namely, the Wannier spectrum is periodic, whereas the edge spectrum is not. This means that of the two gaps of the Wannier spectrum, one at $\nu = 0$ and one at $\nu = 1/2$, only one corresponds to the actual energy gap at the edge. Thus, only Wannier transitions which involve closing this particular gap correspond to actual edge obstructions even though gap closings at either $\nu$ can act to change Wannier topology; this is an important subtlety that we treat in detail below. The determination of which Wannier gap corresponds to the edge energy gap depends on details, including the surface termination as we will see below.

To understand the relationship between a surface spectrum and a Wannier spectrum, let us follow Ref.~\onlinecite{Fidkowski11} in implementing the edge via the replacement 
\beq
\H \rightarrow \H_{\rm obc}(x) = P \phi_M(x) P + M (1 - P)
\label{Hphi}
\eeq
where $P$ is the projector onto the filled bands of the periodic Hamiltonian $\H_{\rm pbc}$ 
and $\phi_M(x)$ is a regularized linear potential given by
\beq
\phi_M(x) = \begin{cases} x, & |x|<M \\
\sgn(x) M, & |x|>M.
\end{cases}
\label{VM}
\eeq
For a given filling specified by a chemical potential $\mu$ satisfying $-M < \mu < M$, this Hamiltonian reduces to the spectrally-flattened bulk Hamiltonian with filled (empty) states at energy $-M$ ($M$) in the bulk. Thus, $\H_{\rm obc}(x)$ has the same filled states as $\H_{\rm pbc}$ deep within the sample, and it implements the vacuum Hamiltonian, where all eigenstates are empty (trivial projection operator), far outside the sample. {Although the existence of an edge obstruction does not depend on the details of the chosen potential $\phi(x)$, the choice (\ref{VM}) makes the connection to Wannier bands manifest since the linear potential
sets a direct proportionality between distance and energy\footnote{To match the units between energy and distance, we note that the Hamiltonian $\H_{\rm obc}(x)$ is a dimensionless band-flattened Hamiltonian where filled (empty) bands correspond to $-M$ ($+M$). Similarly, the position $x$ is measured in units of the lattice constant and can be taken to be dimensionless.}.

 \begin{figure}
    \centering
    \includegraphics[width=.7\columnwidth]{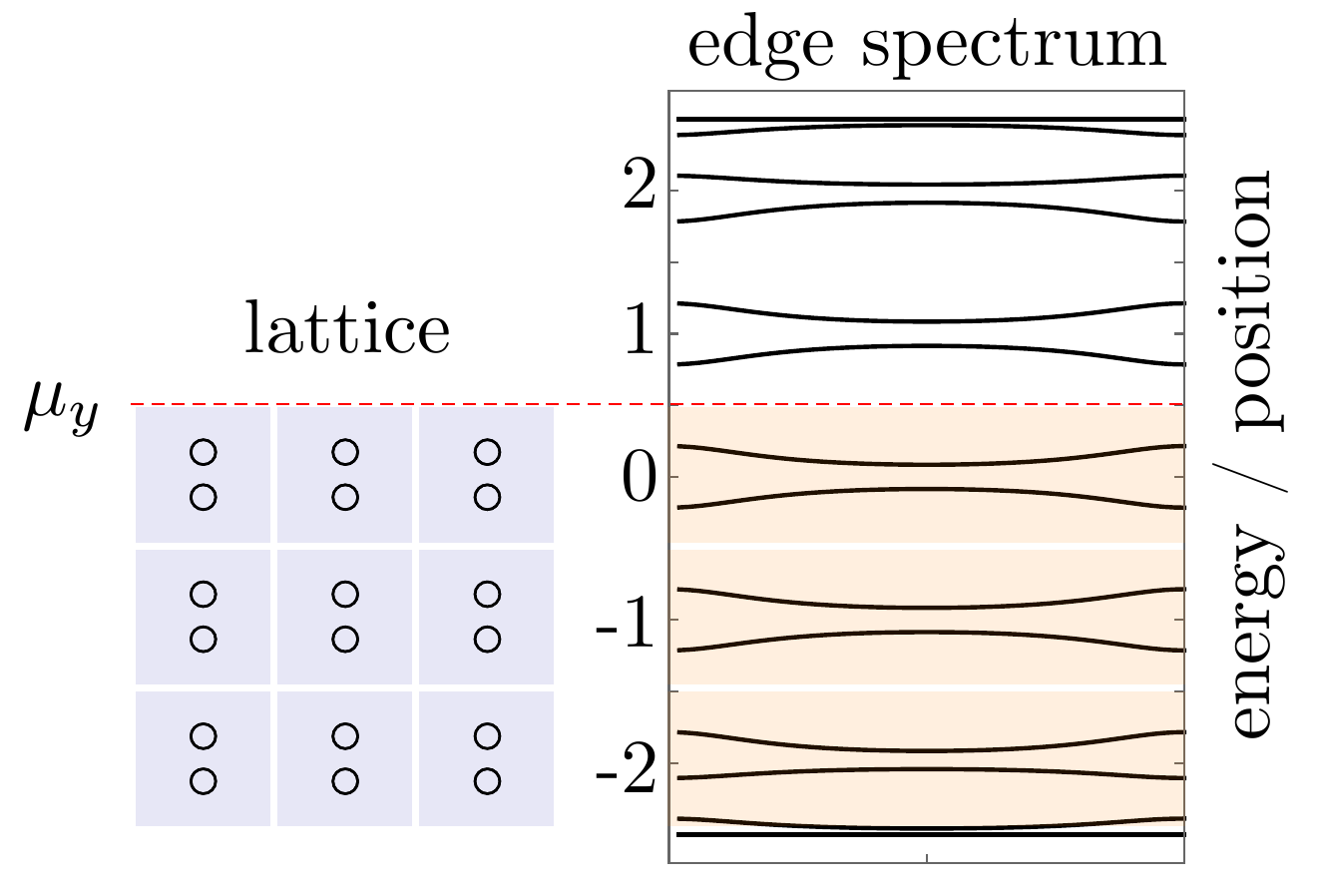}
    \caption{Spectrum for the Hamiltonian $\H_{\rm obc}(x)$ implementing the edge termination with the linear potential $\phi_M(x)$ (Eq.~\ref{VM}) truncated at $M = 2.5$. We can see that the spectrum for energies localized close to the edge matches exactly the Wannier spectrum. Bulk states are all accumulated at energies $\pm M$.}
    \label{fig:edge_spectrum}
\end{figure}

The resulting spectrum of $\H_{\rm obc}$ is shown in Fig.~\ref{fig:edge_spectrum}. We can see that the spectrum for the states localized close to the edge consists of several copies of the Wannier spectrum shifted relative to each other by some integer. The rest of the spectrum far away from the edge accumulates close to the energies $\pm M$ and represent states in the bulk ($-M$) or outside of the sample ($+M$). It is clear from Fig.~\ref{fig:edge_spectrum} that, in this potential,  the edge termination is decided by choosing an energy that separates the filled states inside the sample from the empty states outside, i.e., an ``edge chemical potential." The linear form of the potential $\phi_M(x)$ gives a direct relation between the edge chemical potential and the ``position" of the edge termination. Since the edge energy spectrum is just repeated copies of the Wannier spectrum, the edge chemical potential determines a corresponding \emph{``Wannier chemical potential"} (WCP). The key implication of this analysis is that only Wannier gap-closings that happen at a position corresponding to the WCP indicate a gap closing in the edge energy spectrum. \emph{This establishes the important link between gap closings in the Wannier spectrum and at the edge.}\footnote{Throughout this work, a gap closing at the edge means that the states for which the gap closes are localized close to the edge with localization length much smaller than the system size.}

Similar to the previous subsection, we take a unit cell centered at $1a$ (modulo integers) with edges lying on the lines $x = 1/2$ and $y = 1/2$ (positions $2h$ and $2f$) that intersect at the position $1d$ where the corner is located. This termination corresponds to the WCP $\nu_x = 1/2$ and $\nu_y = 1/2$. Consequently, a Wannier gap-closing at $\nu_{x} = 1/2$ or $\nu_y=1/2$ represents a genuine edge gap closing in our convention, whereas a gap-closing at $\nu_{x} = 0$ or $\nu_y=0,$ does not. Instead the latter type of gap closing indicates mixing between different states at the edge, or states at the edge mixing with the bulk, without closing the edge gap.

\begin{figure}
\includegraphics[width=\columnwidth]{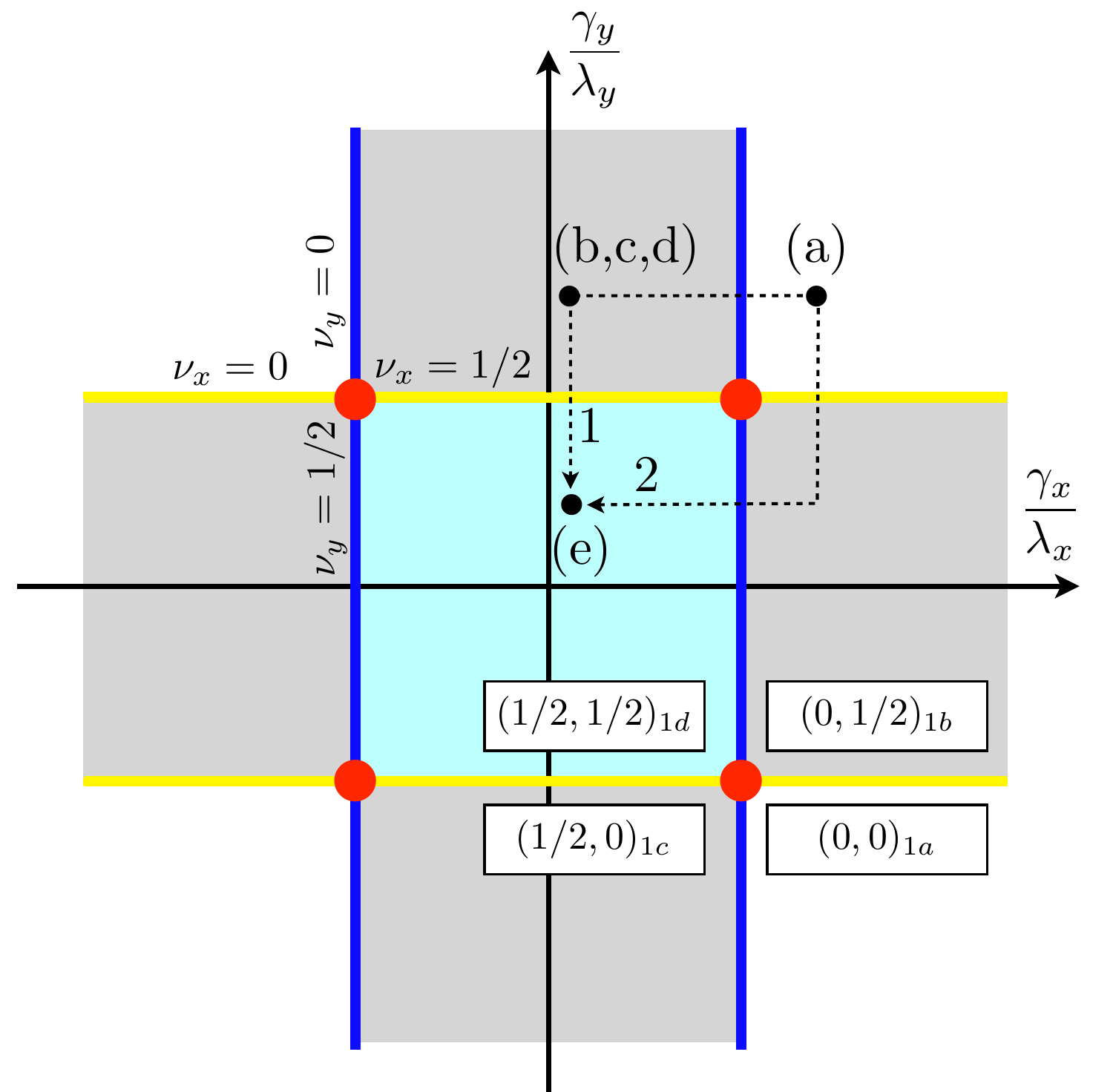}
\caption{ Diagram of the phases of the Wannier bands of the DMQI model, Eq.~\ref{eq:QuadHamiltonian}. On the yellow and blue lines a Wannier gap-closing transition occurs in the Wannier spectra $\W^x$ and $\W^y$, respectively. The gap closing occurs either at $\nu_i=0$ or $\nu_i=1/2$ as indicated. At the intersection of the two lines, marked with a red dot, the bulk spectrum is gapless. To each region, we can associate the Wannier band (nested) polarization $\bp^\nu_\Q$, with $\Q$ the associated maximal Wyckoff position. Black dots indicate specific configurations corresponding to the plots (a-e) in ‚ where we show the evolution of $\W^x$ and $\W^y$ along the path 1, indicated with a dashed line. The phase $(1/2,1/2)_{1d}$ is always separated from the remaining ones by a Wannier gap closing at the Wannier chemical potential $\mu=(1/2,1/2),$ and hence is separated by a gap closing at the physical boundary for our choice of boundary termination. Choosing a different path from (a) to (e), for example dashed line 2, would still imply a boundary gap closing although at a different edge.}
\label{fig:QuadPhaseDiagram}
\end{figure}

Using the preceding analysis, we can now show that for our choice of lattice termination, the state having $\bp^\nu = (1/2, 1/2)$ is distinct from the other three Wannier polarization configurations. The latter three can all be continuously connected while preserving symmetry, but there is an edge obstruction when trying to go from one of $(0,0), (1/2,0), (0,1/2)$ to $(1/2,1/2).$ We can use Fig.~\ref{fig:QuadPhaseDiagram} to illustrate this conclusion. This figure  provides a diagram of all possible Wannier band phases as a function of the ratios $\gamma_x/\lambda_x$ and $\gamma_y/\lambda_y$. The blue central square in the diagram corresponds to the Wannier polarization ${\bf p}^\nu=(1/2,1/2)$; whereas other regions correspond to Wannier polarization equal to either ${\bf p}^\nu=(1/2,0)$ (orange regions), ${\bf p}^\nu=(0,1/2)$ (red regions), or ${\bf p}^\nu=(0,0)$ (white regions). As we can see from the figure, going from the $(1/2,1/2)$ phase to any other phase involves \emph{at least one} Wannier transition at $\nu_x=1/2$ or $\nu_y=1/2$, whereas going between any of the three other phases can be achieved without a Wannier transition at $\nu_{i} = 1/2$. We can conclude that, in our convention, the $(1/2,1/2)$ phase is separated from the other three phases by an edge gap-closing, whereas the three other phases are continuously connected, i.e., any pair of them can be smoothly connected without closing the edge gap or breaking symmetry. This is the resolution to the discrepancy between the topological characterization in terms of edge spectrum vs. Wannier spectrum. We note that if we changed our edge termination, e.g., choosing to terminate the unit cell at $x=0$, $y=1/2$,  the only effect would be to permute which of the four polarization configurations should be distinguished from the other three. 

It is also worth pointing out that the the equivalence of edge and Wannier spectra concerning the gaps only applies for a linear potential. In general, the two are continuously rather than smoothly connected such that a gap in one does not imply a gap in the other. This was pointed out in a recent work \cite{TypeIIQI} which studied a modification of the DMQI with further neighbor hopping. This means that, in general, different choices of the function  $\phi_M(x)$ in (\ref{Hphi}) can move the `phase boundaries' between different obstructed phases. On the other hand, the overall number of phases does not depend on such a choice since they can be diagnosed by discrete features such as filling anomalies or symmetry representations. We will discuss the detailed dependence on the boundary termination in Sec.~\ref{sec:GenDef}.

\begin{figure*}[t]
\centering
\includegraphics[width=2.\columnwidth]{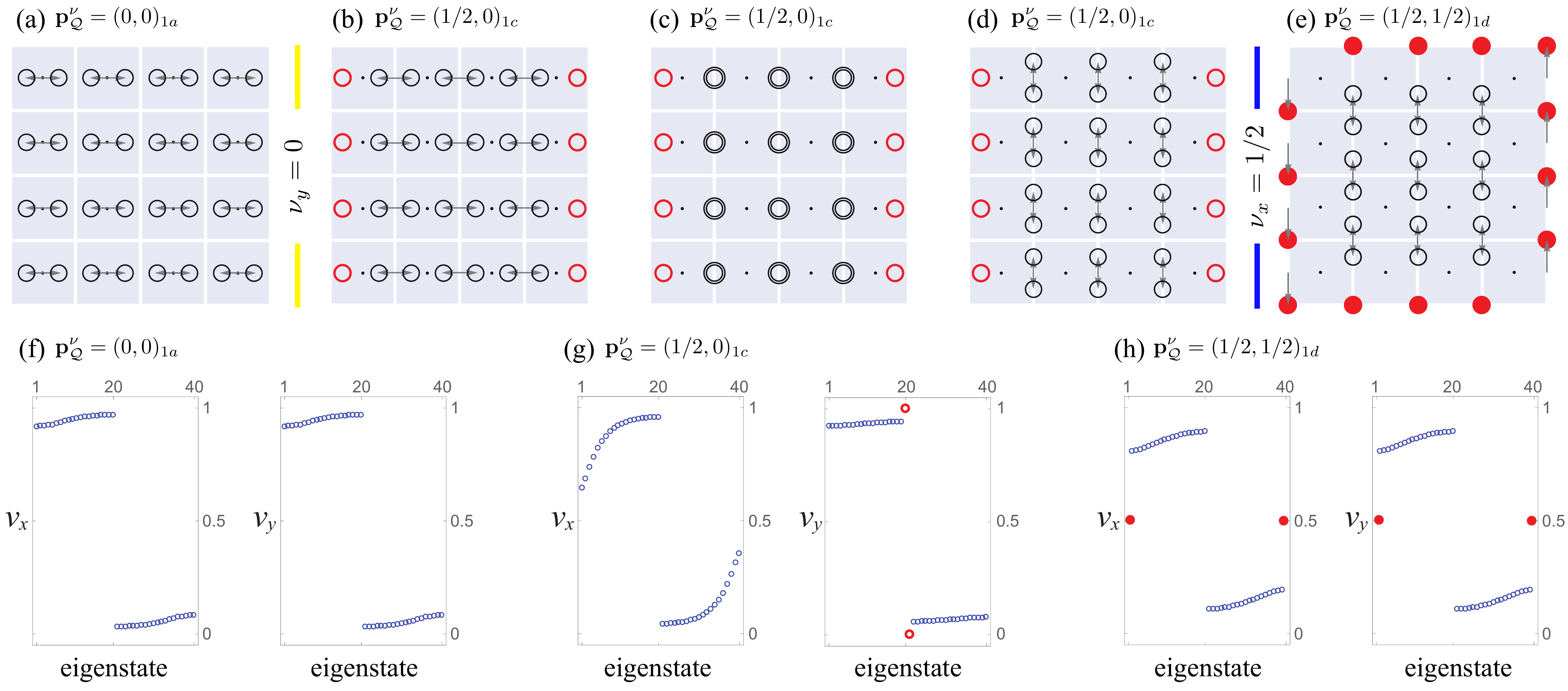}
\caption{Correspondence between real-space configurations and Wannier bands in the DMQI model, Eq.~\ref{eq:QuadHamiltonian}. (a-e) Wannier center location and nested (Wannier band) polarizations along a path illustrated schematically in Fig.\ref{fig:QuadPhaseDiagram} (path 1). A Wannier gap closing transition happens between panels (a) to (b) at $\nu_y=0$, and between (d) and (e) at $\nu_x=1/2$. The latter happens at the Wannier chemical potential and corresponds to a boundary gap closing, and consequently to a topological phase transition.  (f-h) Wannier spectra for cylindrical geometries for the three Wannier phases in (a-e). Wannier eigenvalues in red have eigenstates that are localized on the edges of the cylinder. These results exactly match the configurations of the real space pictures in (a-e). The panels are labelled by the Wannier band polarization $\bp^\nu_\Q$ where the subscript is an indication of the maximal Wyckoff position $\Q$ that is associated to $\bf p^\nu$. The arrows in (a-e) are drawn with their center at $\Q$ as visual guide.}
\label{fig:QuadWPAssignments}
\end{figure*}

Having set up this background, we can now study the connection between Wannier polarization and the real space picture of the edge obstruction. Fig.~\ref{fig:QuadWPAssignments} shows a detailed version of the deformation process in Fig.~\ref{fig:edge_obstruction}, and  illustrates the different Wannier polarization configurations associated with each intermediate state. In Fig.~\ref{fig:QuadWPAssignments}(a), we have a phase characterized by ${\bf p}^\nu=(0,0)$. As the Wannier centers become separated horizontally away from $1a$, a first Wannier transition occurs between Fig.~\ref{fig:QuadWPAssignments}(a) and (b) where the  $y$-Wannier gap closes at $\nu_y(\pi)=0$. If we interpret the edges of the geometry to be open instead of periodic then this transition, which changes ${\bf p}^\nu$ to $(1/2,0),$ leaves the outermost Wannier centers (those on the left an right edges) unpaired [red open circles in Fig.~\ref{fig:QuadWPAssignments}(b,c)]  and pinned to the position $\nu_y=0$. In contrast, the bulk Wannier centers are still free to move, and thus can adopt Wannier eigenvalues that come in pairs $\nu,-\nu$ along both $x$ or $y.$  

We can verify this piece of the deformation through the explicit calculation of the Wannier spectra $\nu_x$ ($\nu_y$) of the DMQI on a cylindrical geometry in which boundaries are closed along $x$ ($y$) and open along $y$ ($x$). In Fig. \ref{fig:QuadWPAssignments}(f) we see that all of the Wannier eigenvalues come in $\pm \nu_i$ pairs, which matches our picture of having bulk Wannier functions with none isolated on the boundaries. In comparison we can look at  Fig.~\ref{fig:QuadWPAssignments}(g), in which there is a pair of isolated Wannier eigenvalues (for every unit cell along the $y$-direction) exactly pinned at $\nu_y=0$. If we look at the eigenstates associated to these pinned Wannier eigenvalues, we find that one set is localized at one of the open, vertically oriented edges, and the other set is on the other vertical edge. Notice, however, that since their eigenvalues are pinned to $\nu_y=0,$ these edge Wannier states lie at the center of the unit cell in the direction parallel to the edge, and thus amount to a trivial edge polarization. Also, notice that such edge isolated Wannier states do not exist for the $\nu_x$ spectrum. 

Now we can  proceed with the deformation process along $y.$ A second Wannier transition occurs between Fig.~\ref{fig:QuadWPAssignments}(d) and (e). At this transition, there is a gap closing in the  $x$-Wannier bands at $\nu_x(\pi)=1/2$, so that in Fig.~\ref{fig:QuadWPAssignments}(e) ${\bf p}^\nu=(1/2,1/2)$. Notice the appearance of unpaired Wannier centers at the horizontal edges with Wannier eigenvalues fixed to $\nu_x=1/2$ [red solid circles in Fig.~\ref{fig:QuadWPAssignments}(e) along horizontally oriented edges]. {Since} the vertical {edge has reflection symmetry, the values $\nu_y$ for states on this edge are quantized to 0 or 1/2. Being initially pinned at $\nu_y=0$ after the first Wannier transition, they} undergo an \emph{edge phase transition} that changes their value to $\nu_y=1/2$ [red solid circles in Fig.~\ref{fig:QuadWPAssignments}(e) along vertical edges]. This is further confirmed by the explicit calculation of the Wannier spectrum on a cylinder shown in Fig.~\ref{fig:QuadWPAssignments}(h), which shows pairs of isolated Wannier eigenvalues (per unit cell along the periodic direction) at both $\nu_x=1/2$ and $\nu_y=1/2.$ If we look at the eigenstates corresponding to these isolated Wannier eigenvalues we find that they are localized at the horizontal or vertical edges respectively. We emphasize that this last transition is the only one during this process that is accompanied by an edge phase transition which closes the edge energy gap, and that this transition is unavoidable due to the fact that the isolated states (red) on the vertical edge constitute effective one-dimensional obstructed atomic limits protected by $M_y$ symmetry.

A final remark is due regarding Fig.~\ref{fig:QuadWPAssignments}(e). The overall configuration of Wannier centers is such that reflection symmetries are preserved in the bulk, but not at the edges. Indeed, it is not even possible to maintain a symmetric configuration of Wannier centers for this system in an open boundary given the number of occupied states we need to fill the lower energy bands in periodic boundary conditions. This is an example of a \emph{filling anomaly}, i.e., a condition where the constraint of charge neutrality (integer band filling) is not compatible with the spatial symmetry~\cite{Benalcazar18}. We will provide a more detailed discussion of this concept in the next subsection. 

\subsection{Corner charge and filling anomaly}
\label{sec:filling_anomaly}
As discussed briefly in Sec.~\ref{sec:Review}, the $\bp^\nu = (1/2, 1/2)$ phase exhibits zero energy corner modes, but only when particle-hole or chiral symmetries are present. In the absence of these symmetries, these corner states are not pinned to zero energy and, as a result, can be moved up and down in energy, though they remain degenerate because of the spatial symmetry \cite{Goldstone1981, AhnJung}. Once they are pushed to the conduction or valence band they can hybridize with the delocalized states there, and do not necessarily remain exponentially localized at the corners. Thus, the edge obstructed phase is not associated with corner modes in general. 

The absence of zero-energy corner modes, however, does not mean that other signatures such as corner {\it charges} (rather than states) are absent. To investigate this possibility, let us start with the particle-hole/chiral symmetric case and then add terms that break these symmetries. We note that the DMQI has $4N$ states in total, where $N$ is the number of unit cells. In the particle-hole/chiral symmetric case, we know there are 4 zero-energy corner states leaving $2N-2$ states in both the conduction and valence bands. Once particle-hole/chiral symmetry is broken, the corner states can move in energy but they remain degenerate due to mirror symmetries, leading to two possibilities: they either remain inside the gap or move into the valence or conduction band. In the first case, they remain localized eigenstates of the Hamiltonian having degeneracy protected by the two mirror symmetries. As a result,  the system can not be gapped \footnote{A gapped system is one with no charge excitations at zero or infinitesimally small energy.} at half-filling since there are two electrons that need to occupy four degenerate states.\footnote{We note that the usual paradigm of slightly breaking the symmetry so that two out of four modes could be filled is unavailable as we are strictly enforcing the symmetry. Indeed filling the modes this way essentially results in spontaneous symmetry breaking in the thermodynamic limit.} In the second case, these states hybridize with the states in the valence (conduction) band forming new and more extended eigenstates. As a result, the number of states in the valence (conduction) band is changed to $2N+2,$ and the chemical potential will lie in the valence (conduction) band at half filling which implies that the system is again not gapped. This phenomenon, in which it is impossible for a system to be simultaneously gapped, symmetric, and charge-neutral at a certain filling has recently been referred to as ``filling anomaly" \cite{Goldstone1981, AhnJung, Benalcazar18, Wieder18, MiertOrtix, rhim2017}. Crucially, if we impose the conditions that the system is a symmetric gapped insulator, then the filling anomaly will manifest in excess or deficient charges distributed symmetrically in the four corners. For the DMQI, such charge is equal to $(n+ e/2)$  per corner for some integer $n$\footnote{Corner charges associated with filling anomalies are only defined modulo an integer charge since we are free to add integer charges to the corners while keeping the system a symmetric insulator.}.


Before considering the general case in the next section, let us build some intuition about filling anomalies in bulk and boundary obstructed phases by contrasting two important examples: the bulk obstructed SSH chain in one dimension protected by inversion symmetry; and the boundary obstructed DMQI. First, imagine we take an SSH chain having $N$ unit cells (with two orbitals per unit cell) at half-filling so that the total number of electrons is $N$. Following \cite{MiertOrtix, rhim2017}, we will now show that, in some cases, it is impossible for this system to be gapped, inversion-symmetric, and charge-neutral when considered with open boundaries. Since every 1D system is Wannierizable, we can assign real space position  corresponding to the centers of the atomic orbitals in the gapped, periodic system, i.e., Wannier centers. We now open the boundaries and denote the inversion center of the open system by $\O$. For a sufficiently large system, the filling at a small inversion-symmetric region surrounding $\O$ , which we denote by $\nu_\O$, should not be affected by opening the boundary and thus should be the same in the open and periodic systems. 
The assumption of charge neutrality means that the number of electrons in the open system is $N$, the same as the periodic system. We now note that it is impossible to find an inversion-symmetric arrangement of real space positions away from the inversion center $\O$ if $N - \nu_\O$ is odd. Thus, one of the three assumptions -- gap, inversion symmetry, charge neutrality -- must fail in the open system. For example, if $N - \nu_\O$ is odd and  we require the gap and neutrality, then there will be a charge imbalance on the left and right halves of the system such that inversion symmetry is broken. If we instead balanced the charge to preserve inversion symmetry then the system would no longer be neutral.

Another way to intuitively understand the filling anomaly is to see that at a filling corresponding to neutrality an electron would be forced to fractionalize into two pieces to preserve the inversion symmetry, as illustrated in Fig. \ref{fig:sshobstruct}. If we require that the system be neutral then the filling anomaly in the SSH chain generates the bulk polarization, i.e., it implies that the inversion symmetry will be broken spontaneously and a dipole moment will appear that creates opposite fractional charges of $\pm e/2$ bound at the two edges. If we require that the system remains symmetric then it will either be gapless since there will be degenerate low-energy modes, or we will need to add an extra electron (at the minimum), which will violate neutrality.

\begin{figure}[t!]
    \centering
    \includegraphics[width=\columnwidth]{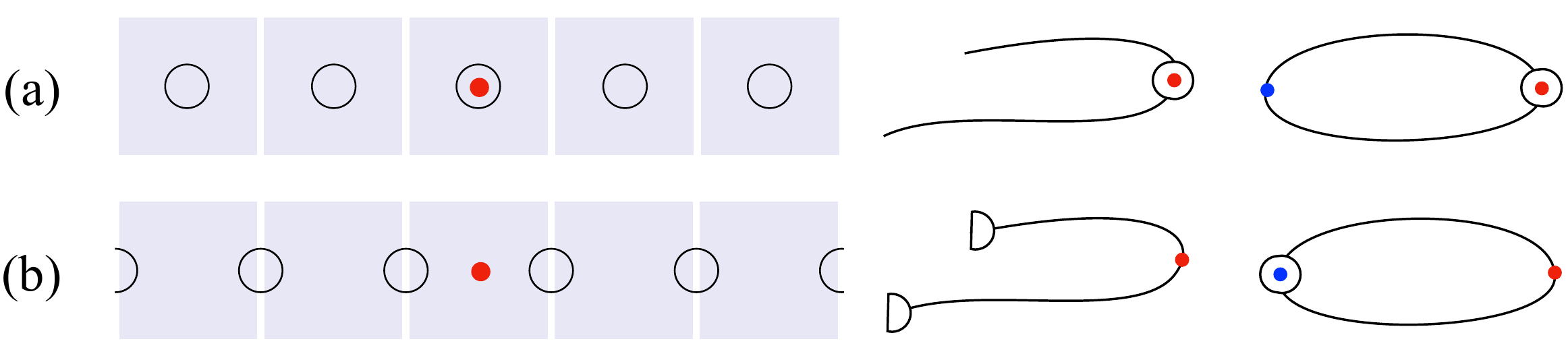}
    \caption{Bulk obstruction and filling anomaly in an SSH chain protected by inversion symmetry. With open boundaries there is a single inversion center (red dot), while when the boundaries are identified an antipodal inversion center is created at the stitching point (blue dot). (a, left) Trivial state: the filling of the inversion center (red dot) is odd just as the total filling of the chain is odd, which implies the electrons can be filled while preserving inversion. (a, right) When the end points are identified only the red inversion center is filled. (b, left) Nontrivial state: at an odd filling, if the red inversion center is not filled there is no way to fill the chain while preserving inversion, and there is a filling anomaly associated with end charges that need to be fractional at neutrality to preserve the symmetry. (b, right) Identifying the end points cures the anomaly, i.e., we no longer require fractionalization at neutrality, and only the blue inversion center is filled. With open boundaries the two phases can be distinguished by the endpoint charges, while with closed boundaries a topological distinction remains since the charge at the blue site cannot be moved to the red site while preserving both the symmetry and the bulk gap. This is an example of a bulk obstruction.}
    \label{fig:sshobstruct}
\end{figure}

As an aside it is instructive to relate filling anomalies to the recently developed framework of layer construction of topological crystalline phases \cite{song2017topological, Huang17} since we can use it for the description of the filling anomaly of the DMQI model. Within the layer construction, any $D$-dimensional topological crystalline phase can be built as follows. We start with some $d$-dimensional region $\Gamma$, with $d<D$, that is invariant under the spatial symmetry group. We then decorate this region with some topological phases whose dimension is smaller than or equal to $d$. Finally, we build the $D$-dimensional system by symmetrically attaching or ``adjoining" lower dimensional units to $\Gamma$ until the whole $D$-dimensional space is filled. The topological invariants of the $D$-dimensional topological phase will be given by those of the lower dimensional topological phase modulo those related by the adjoining procedure. The layer construction was recently employed for the classification of higher-order topological phases in Refs.~\cite{Song2018, Khalaf18}. It was also employed to study obstructed atomic insulators and fragile phases which can be built by layering 0D units, i.e., orbitals, in Ref.~\cite{Liu18}. For example, in the SSH chain example considered above, $\Gamma$ is the origin $\O$. Its topological invariant is the filling $\nu_\O$ which is an integer invariant protected by ${\rm U}(1)$ charge conservation. The adjoining operation is the symmetric addition of zero-dimensional inversion-related units (which are just a pair of atomic orbitals) which changes the filling $\nu_\O$ by an even number. Thus, after modding out by this operation, we end up with a $\Z_2$ invariant diagnosing the filling anomaly in the system.

A similar analysis can be performed for the filling anomaly of the DMQI model. We start by considering $N_x N_y$ unit cells at a filling of two electrons per unit cell. With open boundaries, the symmetry group of the model is the point group $D^\pi_2$ relative to $\O$, with irreducible representations given in Appendix \ref{app:charactertable}, Table \ref{tab:CharacterQuaternion}.
Similar to the discussion on the SSH model, we will now show that sometimes it is impossible for the  system to be a symmetric charge-neutral atomic insulator\footnote{Note that here we need the extra assumption that the system is an atomic insulator, i.e., it admits a Wannier representation, which was not needed in the 1D case since all 1D systems are Wannierizable.}.

Our argument proceeds very similarly to the 1D case. We first select a symmetry invariant region $\Gamma$ consisting of the union of the $x = 0$ and $y = 0$ lines. If the filling of this region, $\nu_\Gamma$, is the same in the open and periodic system (this assumption will be justified below), then there are $2N_x N_y - \nu_\Gamma$ remaining electrons that should be symmetrically arranged throughout the rest of the open system (excluding the region $\Gamma$) to obtain a symmetric charge neutral atomic insulator. However, this is only possible if $2N_x N_y - \nu_\Gamma$ is a multiple of 4 since every point in the open system that does not lie inside the region $\Gamma$ is mapped by $M_x$, $M_y$, and $M_x M_y$ to three other distinct points, hence leading to a symmetry orbit of size 4. This can be used to establish the existence of a filling anomaly. Notice that $\nu_\Gamma$ is always even. This can be seen by noting that it consists of (i) the filling at the origin, (ii) the filling at the line $x = 0$ excluding the origin, and (iii) the filling at the line $y = 0$ excluding the origin. The fillings (ii) and (iii) are manifestly even due to mirror symmetries, whereas (i) is even since the bulk electrons at maximal Wyckoff positions always transform as a 2D $\bar E$ irrep in the DMQI model, regardless of the phase. Thus, $\nu_\Gamma$ is always even and the filling anomaly is determined by the $\Z_2$ invariant $N_x N_y - \nu_\Gamma/2 \mod 2$.

The previous argument has one missing ingredient: how can we show that $\nu_\Gamma$ is the same in the periodic and open systems? In the 1D case, this was simple to show since opening the boundary did not affect the vicinity of the inversion center $\O$ (which corresponds to the region $\Gamma$ in that case). For the 2D case, this can be established only if we make the extra assumption that the polarization per unit cell in both directions vanishes: $p_x = p_y = 0$, which holds for the DMQI model. This assumption means that if we consider a cylinder geometry  where translation symmetry is retained along one of the directions, there is no filling anomaly and we can have a symmetric insulator whose filling per unit length along the periodic directions is the same in the open and periodic systems. Thus, for a cylinder with open boundaries in the $x$-direction and PBC in the $y$-direction, the filling of the $x = 0$ line is the same as in the periodic system. Obviously, this filling is unaffected by opening the boundary in the $y$-direction which takes place far away from the $x = 0$ line. Thus, by opening the boundary in two steps, first in $x$ then in $y$, we can show that the filling at the $x = 0$ line is the same in the open and periodic systems. Similarly, we can open the boundary first in $y$ then in $x$ and show that the filling at $y = 0$ is the same in the open and periodic systems. This establishes that the filling at $\Gamma$, the union of $x = 0$ and $y = 0$, is the same in the open and periodic systems.

We remark that for the point group $D_2^\pi$, the $\Z_2$ filling anomaly explained above is the only filling anomaly possible. It is instructive to compare this to the point group $D_2$ of spinless electrons without the $\pi$ flux. In this case, the filling at the origin $\nu_\O$ is not necessarily even, and can lead to a different filling anomaly diagnosed by the parity of the full filling at the origin $\nu_\O$. This type of filling anomaly reflects a bulk obstruction, just like it does for the 1D SSH chain, and diagnoses the sum of the polarizations along the $x$ and $y$ directions, each of which must be quantized by the mirror symmetries. To identify the individual $x$ and $y$ polarizations, we would need to consider cylindrical geometries with an open boundary along one of the directions and a periodic boundary along the other.

Though we have now identified the filling anomaly for the DMQI mode, the previous discussion 
has not yet clarified the distinction between filling anomalies in boundary-obstructed models like the DMQI, compared to bulk-obstructed models such as the SSH chain. Our goal now 
is to elucidate this difference. The distinction can be understood by studying what happens when the anomaly is resolved, i.e., removed, by identifying some of the edges of the open system. What we mean by ``resolved" is that the change of geometry to a periodic system now allows for a symmetric, gapped, and neutral configuration at our filling of interest. For example, in the SSH case, we can resolve the anomaly by identifying the two endpoints, which generates a new inversion center $\O'$. In this case, it is possible to have a symmetric, charge-neutral insulator whose filling $N$ has different parity from $\nu_\O$ since there is now a new inversion center that can compensate for the parity mismatch. In the phase with (without) the filling anomaly, the new inversion center $\O'$ will have an odd (even) filling whose parity cannot be changed by symmetrically adding or removing electrons. Thus, the filling anomaly in the open system corresponds to a bulk topological distinction in the periodic system. 
That is, while the anomaly can be resolved by identifying the edges, the topological distinction between the two phases still remains with periodic boundaries; they are distinct atomic limits.

\begin{figure}
    \centering
    \includegraphics[width=\columnwidth]{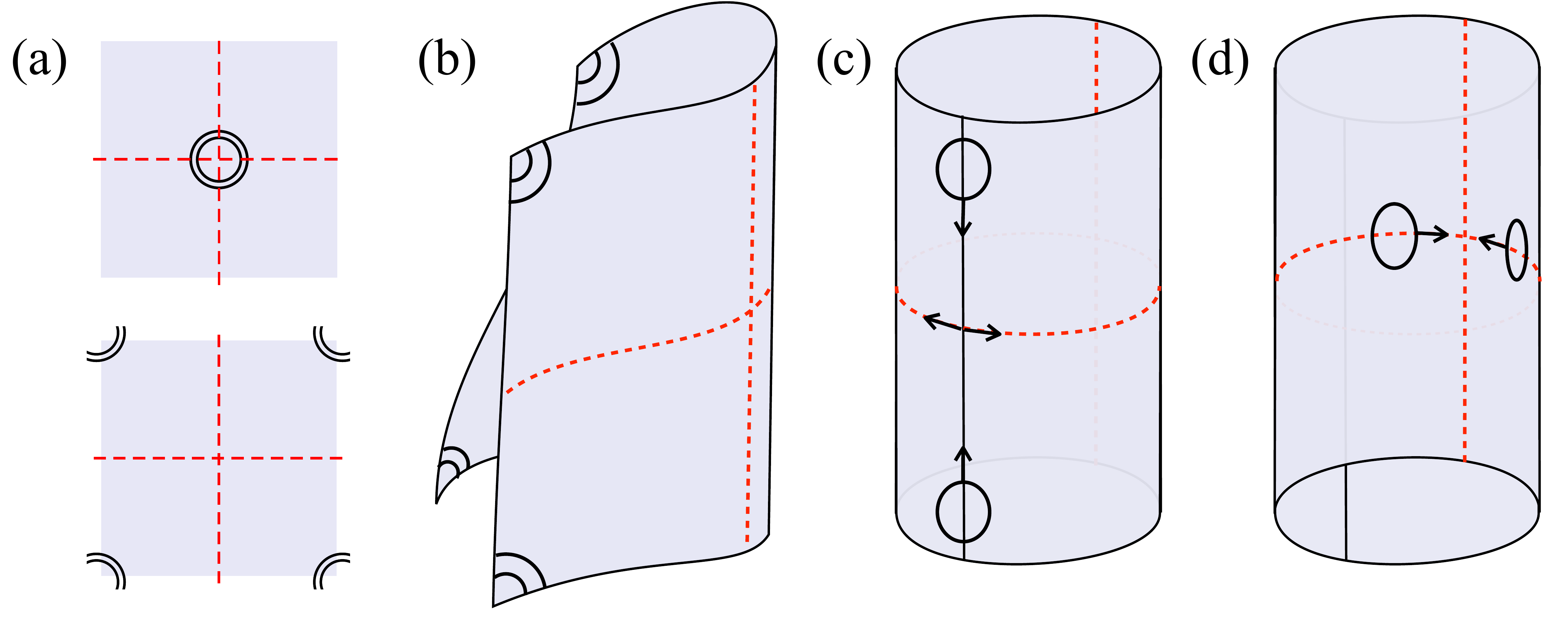}
    \caption{Sewing the boundaries in a boundary obstructed atomic insulator (e.g., the DMQI quadrupole insulator). When the boundary is stitched along one direction, the filling anomaly is lifted since the two corners contribute with half a charge so that all sites will now have integer filling. Interestingly, in this case the bulk obstruction is also lifted: the seam line is a high-symmetry line where the two states can freely move towards each other and symmetrically reach all maximal Wyckoff positions, including the original symmetry origin.}
    \label{fig:boundary_sewing}
\end{figure}

We can similarly resolve the filling anomaly of the DMQI by identifying either pair of opposite edges. When the boundary is stitched together, the symmetry of the corner and the seam line is enhanced compared to the open geometry. The filling anomaly is resolved following this boundary identification since only two electrons
are required to fill the edge states after the stitching, as illustrated in Fig. \ref{fig:boundary_sewing}. Another way to say this is that there is a new high symmetry line (invariant under $M_x$ or $M_y$ depending on the identified edges) whose filling can compensate for the mismatch between $\nu_\Gamma$ and $2 N_x N_y$ modulo 4. However, in contrast to the SSH chain, the resulting system is completely trivial since the two electrons at {the new high symmetry line} can now be symmetrically moved to the original center $\O$. The reason is that, unlike the SSH case, the high symmetry positions at the center $\O,$ or at any mirror line, are all connected. Thus, the difference between filling anomalies associated with boundary and bulk obstructions is that when the anomaly is resolved by identifying some of the edges, the former leads to completely trivial bulk phases, whereas the latter leads to topologically non-trivial bulk phases (e.g., obstructed atomic limits).

One final remark concerns the effect of adding some extrinsic degrees of freedom at the edge of the DMQI. For example, we can add an SSH chain with $N_{\rm SSH}$ lattice sites to one edge (say the one parallel to the $x$-axis), and its mirror image to the opposite edge. This process changes the total number of states by $2N_{\rm SSH}$. It also introduces $2\nu_{\O,\rm SSH}$ states at the $y=0$ mirror line, where $\nu_{\O,\rm SSH}$ is the filling at the center of each SSH chain. This changes $\nu_\Gamma$ as $\nu_\Gamma \mapsto \nu_\Gamma + 2\nu_{\O,\rm SSH}$, and as a result, if the SSH chain has a filling anomaly indicated by parity mismatch between $N_{\rm SSH}$ and $\nu_{\O,\rm SSH}$, then its addition to the DMQI in the anomalous (non-trivial) phase will cancel the filling anomaly of the DMQI. Such a process will, however, induce a filling anomaly in the non-anomalous (trivial) phase. Thus, the effect of \emph{symmetrically adding SSH chains at the edge essentially exchanges what we would identify as the trivial and non-trivial phase, but keeps the distinction between them}. As we will see later, this is a general feature for {boundary}-obstructed phases.

\subsection{Symmetry representations}
\label{sec:symmeigsDMQI}

In the previous discussion we clarified that a topological obstruction in the DMQI model exists if and only if the system has open boundaries. This statement can directly be reiterated as a statement about how the filled electrons transform under the space group $G$ with periodic boundaries, versus how it transforms under the point group $F$ with open boundaries. Namely, the representations $\rho$ of $G$ for the filled electron states corresponding to different parameters of the DMQI model are isomorphic with periodic boundary conditions,
while upon opening boundaries, the filled electron states may transform under distinct, i.e., non-isomorphic, representations of the point group $F$, depending on a choice of boundary. In the following we explicitly show this symmetry distinction by computing the $G$ and $F$ representations for this model as a means to build intuition for the general formulation that will be presented in Sec.~\ref{sec:bandreps}.

\subsubsection{Equivalence between band representations in the bulk periodic system}
\label{sec:bandrepsDMQI}

Now let us show explicitly using symmetry representations that with periodic boundaries there are no distinct topological phases in the DMQI model. As described in detail in Refs.~\cite{zak80,cano2018building,Bradlyn17}, as well as in Sec.~\ref{sec:bandreps}, all atomic bands transform under representations of a space group $G,$ which are obtained from local representations of the site-symmetry group $G_\bq$ of a maximal Wyckoff position. In the DMQI case, these correspond to the four maximal Wyckoff positions $1a,~1b,~1c,$ and $~1d$ of multiplicity one whose site-symmetry group is the point group $F$. Although they have isomorphic site symmetry groups, the four position centers correspond to distinct coset decompositions of the space group under translations $F=G/T$, which implies the groups $F_{1a}$, $F_{1b}$, $F_{1c}$ and $F_{1d}$ are isomorphic but not related by conjugation under any element of $G$. The representations $\pi$ of $F$ in the DMQI model with periodic boundaries, are given by the two-dimensional representation $\bar E$, for all choices of parameters, guaranteeing the anti-commutation of $M_x$ and $M_y$.

To prove that all Wannier configurations in the DMQI are topologically equivalent with periodic boundaries it suffices to find a symmetry-preserving unitary transformation that acts in the space of filled bands and transforms between the distinct representations $\rho$ of $G$. In momentum space, the diagonal entries of these matrix representations form the band representations $\rho_\bk$. We discuss how to find $\rho_\bk$ in Sec.~\ref{sec:bandreps} following the standard references \cite{zak80,zak1982band,bradlyn2017topological,cano2018building}. The possible representations induced by the local representation $\bar E$ at the four maximal Wyckoff positions $\Q$ have distinct momentum dependencies $\rho_\bk^\Q,$ which we will now tabulate. 
First, we see that, because of the $\pi$ flux, 
\beq
\rho_\bk(C_{2z}^2) \!=\!-1,
\label{RhoC2}
\eeq
for all Wyckoff positions. Next, let us focus on position $1a$ and consider the representations of $\rho_\bk(M_x)$ and $\rho_\bk(M_y)$. Since this position is chosen as the center of the unit cell, the representation matrices $\rho_\bk$ will be $\bk$-independent. One possible choice for $\rho^{1a}_\bk(M_{x,y})$ is $\rho^{1a}_\bk(M_x)\!=\!\sigma_1$ and $\rho^{1a}_\bk(M_y)\!=\!\sigma_3$ which corresponds to the eigenbasis of $M_y$. Alternatively, we can choose the eigenbasis of $M_x$: $\rho'^{1a}_\bk(M_x)\!=\!\sigma_3$ and $\rho'^{1a}_\bk(M_x)\!=\!\sigma_1$. Clearly, the two BRs are equivalent with the momentum-independent Hadamard transformation $U_\bk\!=\!(\sigma_x+\sigma_z)/\sqrt{2}$ relating the two. 

Next, we can compare the band representations induced from distinct maximal Wyckoff positions, located at parallel high symmetry lines. Inducing a representation from $1b$ we find that $\rho^{ 1b}_\bk(M_x)=\sigma_1$ and $\rho^{ 1b}_\bk(M_y)=\exp(ik_y)\sigma_3$. The unitary transformation that relates $\rho^{1a}_\bk(M_y)$ and $\rho^{1b}_\bk(M_y)$ is %
\begin{align}
    U_\bk=\half\begin{pmatrix}1+\exp(ik_y)&1-\exp(ik_y)\\1-\exp(ik_y)&1+\exp(ik_y)\end{pmatrix}.
\end{align} 
Finding the other unitary transformations follows analogously. Since the band representations are used to distinguish the different atomic limit phases, we conclude from this argument that they are all equivalent to the trivial atomic limit when the system has periodic boundary conditions. We remark that since the unitary transformation, acting on the filled bands alone, changes the nested polarization $p_{x,y}\to p_{x,y}+1/2$, it explicitly highlights that these values \emph{cannot} be naively used as a bulk topological invariant.

\subsubsection{Symmetry representations of the open system}
Now let us consider the open system without periodic boundary conditions. Assuming the system has open boundaries implies that the symmetry group is restricted from $G$ to $F$, i.e., translation symmetry is no longer a symmetry of the system, and the number of high symmetry lines in the lattice is substantially reduced to two perpendicular lines crossing a single point $\O$ which is the center of the lattice. Above, in the layer construction discussed in Sec.~\ref{sec:filling_anomaly}, this symmetric region was labelled by $\Gamma$. 

We want to consider restricting the representation $\rho$ of the full space group $G$ of the periodic system to a (reducible) representation $\pi$ of the point group $F$ of the open system. This implies selecting only group elements that belong to $F$, which is symbolized by $\pi=\rho\downarrow F$. Thus, starting from a representation $\rho_\bq$ of the site symmetry group $G_\bq$, we first generate a representation $\rho$ of the full space group $G$ which is then restricted to a representation $\pi$ of the point group $F$ of the open system. In the language of representations a boundary obstruction is the statement that different site symmetry representations $\rho_\bq$ which generate the space group representation $\rho$ can generate different representations $\pi$ of the open system despite being equivalent in the bulk periodic system (as we just showed above). This provides a symmetry-based characterization of boundary obstructions which we will illustrate for the DMQI below and discuss in more general terms in Sec.~\ref{sec:bandreps}. In fact, we will prove a stronger statement in what follows: not only are the representations $\pi$ of the two DMQI `phases' distinct, but only one of them is a local representation of the point group $F$. We say a representation of the point group $F$ is local if it can be generated by the action of $F$ on localized atomic orbitals in real space. This means that it is generated as a sum of site symmetry representations of the Wyckoff positions in the open system (the Wyckoff positions of the open system are defined in analogy to the periodic system, as a set of points forming a closed orbit under the action of the point group $F$). The notion of a local representation thus provides a definition for a symmetric atomic insulator in the open system.

{Let us now see how this explicitly applies to the DMQI. The Wyckoff positions consist of the center $\O$ with multiplicity 1, the $x$ and $y$ axes each with multiplicity 2, and the general position with multiplicity 4. We begin by assuming that the representations generated from both $1a$ and $1d$ are local and show this leads to a contradiction. First note that restricting a representation from $G$ down to $F$ cannot change the dimension of the representation. Thus, since $\pi^{1a}$ and $\pi^{1d}$ are obtained from restricting equivalent bulk representations $\rho$ to the point group $F$, we have
\beq
\tr \pi^{1a}(1) = \tr \pi^{1d}(1),
\label{Dim1a1d}
\eeq which uses the fact that the dimension of the representation is the trace of the representation of the group identity element.
Second, note that the dimension of a symmetry representation acting on Wyckoff position $\Q$ is necessarily a multiple of its multiplicity, thus the contribution from the general Wyckoff position to $\tr \pi(1) \!\!\mod 4$ drops out. If the representation $\pi$ is local, then it is a sum of representations acting on the different Wyckoff positions of the open system, and furthermore it implies that $\tr \pi(1) \!\!\mod 4$ receives contributions only from the Wyckoff positions lying on high symmetry lines, i.e.,the region denoted by $\Gamma$. As a result, we find
\beq
\nu_\Gamma = \tr \pi(1) \!\!\mod 4.
\label{nuGammapi}
\eeq
On the other hand, we can read off $\nu_\Gamma$ from the site filling of the periodic system assuming vanishing polarization in the $x$ and $y$ directions (see Sec.~\ref{sec:filling_anomaly}).  For definiteness, let us assume both $N_x$ and $N_y$ are odd such that the boundary of the open system relative to the center of the unit cell coincides with the edge of the unit cell. Clearly for $\rho^{1a}$, there are two orbitals at the center of the unit cell which lie inside $\Gamma$ leading to $\nu_\Gamma = 2(N_x + N_y - 1) = 2 \!\!\mod 4$. On the other hand, for position $1d$, there are clearly no orbitals lying on $\Gamma$ leading to $\nu_{\Gamma}^{1d} = 0$. Thus,
\beq
\nu_{\Gamma}^{1d} \neq \nu_{\Gamma}^{1a}
\label{nuGamma1a1d}
\eeq
We can see that Eq.~\ref{Dim1a1d}, \ref{nuGammapi}, and \ref{nuGamma1a1d} lead to a contradiction which implies that the assumption that both $\pi^{1a}$ and $\pi^{1d}$ are local must be wrong. For the chosen boundary, it is easy to see that $\pi^{1a}$ is local since none of the filled sites lie at the boundaries, thus we conclude that $\pi^{1d}$ cannot be local, i.e., it cannot be described by a symmetric atomic insulator with the same filling as the bulk, which precisely defines a BOTP with a filling anomaly.  Once we have established the existence of two distinct phases in the open system, we can classify representations obtained from any other site symmetry representation $\rho_\bq$ which will be equivalent to either $\pi^{1a}$ or $\pi^{1d}$ depending on whether $\nu_{\Gamma} = 2, 0$. These results are summarized in Table \ref{tab:WannierBandsWyckoffPositions}, from which we see that $\pi^{1d}$ is distinct from $\pi^{1a}, \pi^{1b}, \pi^{1c},$ which are equivalent to each other.

\subsubsection{Wannier band representations}

We have seen in Sec.~\ref{sec:edge_Wannier} that a natural diagnosis tool for boundary topology is  the Wannier spectrum. The Wannier spectrum, with eigenstates spanning the same space as the periodic Hamiltonian, transforms under representations related to the periodic and open boundary representations $\rho$ and $\pi$, respectively. By constructing hybrid Wannier functions localized along the $\ba$ direction dual to $\bb$ ($\ba\cdot\bb=2\pi$), each individual band in the Wannier spectrum $\W^\bb(\bk_\perp)$ preserves a subgroup $G_\bb$ of the space group $G$. Therefore, we can associate a lower dimension band representation to these bands, identified by the symmetry character of the Wannier states at high symmetry momenta in the (Wannier) Brillouin zone transverse to $\bb$ spanned by the momenta $\bk_\perp$. The representation of each Wannier band $w_a^\bb(\bk_\perp)$ is related to the representation $\rho$ of the periodic boundary Hamiltonian  by restricting this representation to the group elements of $G_\bb$; this process is denoted as $w^\bb=\rho\downarrow G_\bb$. Since the Wannier spectrum can be decomposed into a set of disconnected bands, we can additionally write $w^\bb=\bigoplus_aw^\bb_a$, where $a$ is a band index. The explicit form of the representation is given in further detail in Sec. \ref{sec:bandreps}. If we look at the Wannier band representations in different directions $\bb$ such that $G=\bigcup_\bb G_\bb$, then one can unambiguously determine from which original bulk representation they are restricted, as well as the specific coset decomposition $G/T$ that identifies the maximal symmetry point in the unit cell that serves as its center.


In the DMQI model, the four coset decompositions, labelled by the unit cell centers $1a$ to $1d$, can therefore be distinguished by the Wannier band representations. In this case, the Wannier band representations are given by $\rho$ restricted to the subgroups $G_x=\{T_x,M_x\}$ and $G_y=\{T_y,M_y\}$. We can now illustrate how to identify the bulk representation $\pi$ from the set of Wannier band representations. Let us choose the bulk Wannier functions to be located at each of the possible four centers, where they transform under the $\bar E$ representation of $F=D_2^\pi$. When the center is $1a$, the hybrid Wannier functions are constructed out of linear combinations of localized Wannier functions at the center of the unit cell in the perpendicular plane. For example, if we choose to keep $G_y$, the two states in the unit cell have opposite eigenvalues of $M_y,$ and these eigenvalues have no momentum dependence. This is analogous to the SSH chain where the representation of the (energy) bands  do not have a momentum dependence when the Wannier centers occupy the inversion center in the unit cell. The Wannier bands transform under the band representations $[w^x_{\pm}]_{k_y}(M_y)=\pm1$. Alternatively, when keeping $G_x$ the hybrid Wannier functions transform under the band representation $[w^y_{\pm}]_{k_x}(M_x)=\pm1$. If instead the bulk Wannier centers  occupied the $1b$ position, which is located at $y=1/2$ in the unit cell, then the restriction of $\rho$ to $G_y$ would reflect this  by acquiring a phase of $\exp\{ik_y\}$ under the action of $M_y$. That is $[w^x_{\pm}]_{k_y}(M_y)=\pm \exp\{ik_y\}$. Indeed, all four coset decompositions $G/T$ correspond to $1a, 1b, 1c, 1d$ generate distinct Wannier band representations, and are shown in Table \ref{tab:WannierBandsWyckoffPositions}. These representations exhaust the possible restrictions of $\rho$, and do not depend on the possible symmetric displacement of the Wannier centers away from the high symmetry points. We see that all bulk representations of this model are isomorphic, and the different Wannier representations reflect a choice of basis.  

As we have seen in the previous section, the difference between the basis choice becomes important with open boundaries. Namely, it is important to know if the center of the unit cell occupies a high symmetry region $\Gamma$ in the open boundary system. This depends on the choice of boundary. In Table \ref{tab:WannierBandsWyckoffPositions} we compare the representations of the Wannier spectrum, to the filling $\nu_\Gamma$ of the open boundary system, where the corner position coincides with the high symmetry position $1d$, and the center $\O$ with $1a$ with periodic boundaries. This is the conventional boundary termination, and the direct comparison of the Wannier spectrum representations obtained from the Hamiltonian for a given set of parameters can be compared with this table to predict if $\pi$ is a local representation of the open boundary symmetry group $F$ characterized by a filling anomaly.


\begin{table}[h!]
\centering
\begin{tabular}{c|cc|cc|c}
 & $k_x=0$ & $k_x=\pi$ & $k_y=0$ & $k_y=\pi$ &   \\
\hline
$\rho$ & $w^{y}_+(M_x)$ & $w^{y}_+$ & $w^{x}_+(M_y)$ & $w^{x}_+$ & $\nu_\Gamma$  \\
\hline
$(1a,\bar E)$ &$+$ & $+$ & $+$ & $+$ & $2$\\
$(1b,\bar E)$ &$+$ & $+$ & $+$ & $-$ & $2$\\
$(1c,\bar E)$ &$+$ & $-$ & $+$ & $+$ & $2$\\
$(\vec{1d},\bar{\vec E})$ &$\vec +$ &  $\vec-$ &  $\vec+$ & $\vec-$ & $\vec 0$ 
\end{tabular}
\caption{Diagnosis of boundary obstructions through Wannier band representations $[w^{\bb}_{a}]_k(g)$ for the band $a=+$ (selected to have a positive eigenvalue at $k=0$) in the DMQI model evaluated for the mirror operator preserved by the perpendicular direction to $\bb$ at different high symmetry momenta. The bold lines correspond to the phases with a boundary obstruction when the conventional Wannier chemical potential is chosen, i.e., $\mu=(1/2,1/2)$ (obstructed high symmetry site $1d$), and the center of the sample $\mathcal O=(0,0)$ (high symmetry site $1a$). In the last column we show the character $\nu_\Gamma$ which differentiates between the anomalous and the non-anomalous configurations with this boundary choice.
}
\label{tab:WannierBandsWyckoffPositions}
\end{table}

\section{Boundary obstructions: Generalities}

\subsection{General definition}
\label{sec:GenDef}
{After discussing the properties of the DMQI model in detail, our goal now is to introduce a general definition to capture topological distinctions encoded in a boundary obstruction rather than a bulk obstruction. As we have seen for the DMQI, the distinguishing feature between the two phases of the model on an open boundary is a filling anomaly manifesting as a fractional corner charge of $e/2,$ which is present in one of the phases, but absent in the other. However, unlike a standard SPT, e.g., a higher-order TI, the boundary of the DMQI is not anomalous since the corner charges can be removed by the addition of two SSH chains on opposite edges while preserving the mirror symmetries. This means that we should be more careful when defining an edge termination to distinguish surface features which arise from the bulk compared to those arising from the edge termination itself.}

\subsubsection{Boundary termination}
Let us begin by recalling some relevant concepts. Given a Hamiltonian $\H$ and a chemical potential $\mu$, we can define the projector on the filled bands via
\beq
P:=p(\H,\mu) = \sum_{n} \theta(\epsilon_n - \mu) \ket{\psi_n} \bra{\psi_n}, \,\, \H \ket{\psi_n} = \epsilon_n \ket{\psi_n}.
\eeq
 The projector $P$ is well-defined if and only if there are no eigenstates of $\H$ at the chemical potential, i.e., only if $\H$ is gapped at $\mu$.  {Note that the definition above does not assume translation symmetry. Since we are interested in topological properties rather than dynamics, we will consider two pairs $(\H,\mu)$ and $(\H',\mu')$ equivalent if $p(\H,\mu) = p(\H', \mu')$. In the following, unless otherwise stated, we will choose $\mu = 0$ such that $p(\H) = p(\H,0)$. 
 Given an operator $O$ ($\H$ or $P$), we can write it in the position basis as
 \beq
 O(\br,\br') := \langle \br| \hat O| \br' \rangle 
 \eeq
 For example, the position-dependent projector is given by
  \beq
 P(\br,\br') = \sum_n \theta(\epsilon_n - \mu) \psi_n(\br) \psi_n^\dagger(\br'),
 \eeq
 which is a matrix-valued function of two spatial positions $\br$ and $\br'$ (for a model with $N_o$ orbitals, this will be an $N_o \times N_o$ matrix).}
 
We start by considering a periodic system with symmorphic space group $G$ decomposed as a direct product of a group of translations $T,$ 
 and a point group $F$ relative to a fixed point $\O$.
 An $F$-symmetric boundary in $d$-dimensions is defined by specifying a $(d-1)$-dimensional surface $\Sigma$ dividing the space into an interior region $\Lambda_{\rm in}$ and an exterior region $\Lambda_{\rm out}$ with a ``thickness" parameter  $\xi$ such that for distances larger than $\xi$ boundary effects are negligible. We can now define a surface termination as a continuous map $\sigma$ which takes a translationally symmetric Hamiltonian $\H$ (with the corresponding projector $P=p(\H)$) to a Hamiltonian in the open system $\H_{\rm obc}$ (with the corresponding projector $P_{\rm obc} = p(\H_{\rm obc})$), such that $P_{\rm obc}(\br,\br') = P(\br,\br')$ deep inside the sample, and $ P_{\rm obc}(\br,\br') = P_0(\br,\br')$ far outside, with $P_0$ denoting some reference "trivial" projector. For simplicity, we will take $P_0$ to denote projector where all states are empty. What we mean by deep inside and far outside the sample can be made precise by defining
 \beq
 d_{\br, \Sigma} = {\rm min}_{\br' \in \Sigma} |\br - \br'|,
 \eeq
 which simply denotes the shortest distance from a point $\br$ to the surface $\Sigma$. Thus, we can define the most general boundary termination as

 \begin{multline}
 \H_{\rm obc} = \sigma(\H) \text{ such that for } d_{\br,\Sigma}, d_{\br',\Sigma} \gg \xi \\
 \begin{cases}
 P_{\rm obc}(\br,\br') = P(\br,\br') &: \br,\br' \in \Lambda_{\rm in} \\
 P_{\rm obc}(\br,\br') = P_0(\br,\br') &: \br,\br' \in \Lambda_{\rm out} \\
 P_{\rm obc}(\br,\br') = 0 &: \text{otherwise}
 \end{cases}.
 \end{multline}
Here, we also made the assumption that $P(\br,\br')$ vanishes when one of $\br, \br'$ is deep inside and the other is far outside the sample. This means that there are no states which ``propagate'' between the inside and outside of the sample.}
 
 {We notice that in practice, boundary terminations are typically implemented by a sharp edge where the lattice terminates, or more generally by a potential well which separates the filled states inside the sample from the empty states outside. An example of the latter is ultracold atoms \cite{ColdAtoms} where the lattice termination is usually implemented by a potential trap which confines the particles within a certain region. This motivates the consideration of a particularly simple  boundary termination implemented by  making the chemical potential spatially dependent such that it lies in the gap inside the sample, and below the smallest energy eigenvalue outside. 
  This is equivalent to choosing the open system Hamiltonian considered in Ref.~\cite{Fidkowski11}
 \beq
 \H_{\rm obc} = \sigma(\H) = p(\H) \phi(\br) p(\H) + \phi_{\rm out} (1 - p(\H)),
 \label{eq:Hopen}
 \eeq
 where $\phi(\br) = \phi_{\rm in} < 0$ deep inside the sample, and $\phi(\br) = \phi_{\rm out} > 0$ far outside.\footnote{We will usually also impose the condition that $\phi(\br) = 0$ only at the boundary surface $\Sigma$, but this is not crucial for what follows.} This boundary termination, implemented by changing a single scalar parameter (the chemical potential), is the simplest way to interpolate between the bulk Hamiltonian and a fixed trivial reference without introducing any extra degrees of freedom at the boundary. It includes most surface terminations needed in practice, e.g., a linear potential, a sharp edge, etc. and has the advantage of connecting directly to the Wannier spectrum. 
 
To make the notion of boundary obstructions introduced in the next section as general as possible, we will also find it useful to consider an augmented class of boundary terminations by allowing for the addition of gapped degrees of freedom at the boundary. Such additional degrees of freedom do not induce any additional boundary gap-closing, thereby preserving the notion of boundary obstructions. To define this more general class of boundary terminations rigorously, we note that since the open system Hamiltonian for \emph{any} boundary termination is equivalent to (\ref{eq:Hopen}) away from the boundary, we can write the Hamiltonian for the most general boundary termination we consider as
 \beq
 \sigma(\H) = p(\H) \phi(\br) p(\H) + \phi_{\rm out} (1 - p(\H)) + \H_\Sigma,
 \label{eq:HBoundary}
 \eeq
 where $\H_\Sigma$ vanishes away from the boundary. $\H_\Sigma$ is defined on a $(d-1)$-dimensional shell which is topologically equivalent to $S^{d-1}$. We say a boundary termination $\sigma$ is trivial if there exists $\phi(\br)$ such that $\H_\Sigma$ defined by (\ref{eq:HBoundary}) is symmetric and gapped for all $\H$. A trivial termination allows us to compare distinct Hamiltonians with open boundaries by excluding terminations which introduce low energy (gapless) modes at the boundary. Such a restriction is necessary since BOTPs are extrinsic HOTIs, hence their surface states can always be removed through the addition of boundary degrees of freedom and it enables a direct correspondence between the bulk and boundary degrees of freedom.

\subsubsection{Boundary obstructed phases}
{We are now ready to define the notion of boundary obstructions. Given a trivial termination $\sigma$, we say there is a boundary obstruction between two gapped translationally symmetric Hamiltonians $\H_1, \H_2$ with periodic boundary conditions if: 
\begin{itemize}
    \item[(i)] there is a smooth trajectory in the space of symmetric gapped Hamiltonians $\H(t)$ for $t \in [0,1]$ connecting $\H_1$ and $\H_2$ such that $\H(0) = \H_1$ and $\H(1) = \H_2$; 
    \item[(ii)] the trajectory induced by the boundary termination $\sigma(\H(t))$ necessarily involves closing the energy gap at a high-symmetry surface, i.e. there exists a $t \in [0,1]$ such that $\sigma(\H(t))$ has a zero energy (remember we use a convention that sets the chemical potential to 0) eigenstate localized in the direction perpendicular to a high symmetry surface.
\end{itemize}}

A high symmetry surface (HSS) in $d$ dimensions denotes any $D$-dimensional hyperplane ($D<d$) on the boundary that is left invariant by at least one non-trivial point group symmetry operation in $F$. For example, in the simple rectangular termination considered for the DMQI, both edges are high symmetry surfaces since each one is invariant under one of the mirror symmetries. For more complicated boundaries, with arbitrary shape, we can still define high symmetry surfaces by first ``rounding the corners" to make the curvature finite everywhere, and then considering the tangent plane to the boundary at any given point. We then define the surface spectrum as the energy spectrum of the states which are localized in the direction perpendicular to the surface. The surface spectrum is a function of the momentum parallel to the surface which is a vector in the tangent plane at this point \cite{Teo10, Khalaf17, Khalaf18, AhnJung}. See Appendix \ref{app:SSDMQI} for an illustration of the edge spectrum for a general boundary termination of the DMQI.

We can now define boundary-obstructed phases as equivalence classes of Hamiltonians under boundary obstructions defined earlier. That is, we consider a space of gapped Hamiltonians in the translationally-symmetric system $\Xi_{\rm pbc}$, which we take to be topologically trivial, i.e., all gapped phases can be continuously deformed to each other without closing the gap or breaking the symmetry. \emph{Given a trivial boundary termination $\sigma$, we define the group of boundary-obstructed topological phases to be the group of equivalence classes of boundary-terminated Hamiltonians $\sigma(\Xi_{\rm pbc})$ under continuous deformations which do not close the gap at any high symmetry surface.} Intuitively, this definition distinguishes different classes of Hamiltonians based on boundary obstructions, and enables a finer distinction between phases that are topologically identical from the perspective of bulk topology. 
   
The above definition introduces a relative distinction between a pair of Hamiltonians given a certain trivial boundary termination. Such distinctions are generally termination dependent even for trivial terminations. For example, recall from the analysis of the DMQI in the previous section that the termination (which determines the Wannier chemical potential) determines which Hamiltonians will exhibit a filling anomaly. On the other hand, for any symmetric termination, there are always two distinct phases in the open system distinguished by the $\Z_2$ filling anomaly. This suggests that although the boundary determines which BOTP class a given Hamiltonian ends up in, the topological classification (homotopy group) of BOTPs itself is independent of the boundary termination $\sigma$ for sufficiently general topologically trivial boundaries. {Here, by ``sufficiently general", we mean that the distinct topological sectors are stable to symmetry-preserving perturbations of the boundary which keep it gapped and Wannierizable. In particular, this includes the possibility of adding degrees of freedom on the surface which can trivialize the surface bands if they possess fragile topology \cite{PoFragile}}. While a general proof of this statement is beyond the scope of this work, we can directly verify it for all the examples of BOTPs considered in this work (the DMQI considered in Sec.~\ref{sec:filling_anomaly} and the 3D examples of Sec.~\ref{sec:3D}) by showing that the different BOTPs can be distinguished by a physical signature: either filling anomalies or gapless surface states, on any symmetric boundary.
   
In general, boundary obstructions may co-exist with bulk obstructions. For instance, a non-trivial bulk phase, such as a 3D higher-order topological insulator with chiral hinge states, can have several distinct patterns of hinge state. These patterns effectively differ by the addition of a BOTP, i.e., we can always move between such different surface configurations by adding a BOTP. Such a BOTP, which is manifestly trivial in the bulk, can be identified with the \emph{difference} between the Hamiltonians corresponding to the two distinct patterns of surface states. Thus, when studying boundary obstructions, we can restrict ourselves to phases that are completely trivial in the bulk.

One important aspect in our definition is that only gap-closings at high-symmetry surfaces are relevant. {Physically, this is a natural restriction since most simple realistic boundaries tend to be along high symmetry surfaces. One way to see this is by recalling the discussion of  Sec.~\ref{sec:filling_anomaly}, in which the filling anomaly in the DMQI was diagnosed by the mismatch between the total filling and the number of states lying at the two mirror lines modulo 4. Thus, an edge gap-closing can change the filling anomaly by changing the number of states at the mirror lines if and only if this gap-closing occurs at one of the mirror invariant points at the edge. This means that our definition captures the same topological distinction described by the filling anomaly for the DMQI.} 

Another way to see the importance of distinguishing high symmetry points/surfaces versus generic points on the surface is to consider the DMQI on a disk geometry whose boundary is a circle. If chiral/particle-hole symmetry is unbroken, the non-trivial phase of the model is characterized by four symmetry-related zero-energy states. These states can be associated with domain walls for a surface Dirac mass \cite{Khalaf17, Khalaf18, Fang17, Fukui18, AhnJung, Wieder18, Wieder2020} that is invariant under both mirrors and changes sign four times as we go around the circle (cf.~Fig.~\ref{fig:DMQIcircle}) {as shown in Appendix \ref{app:SSDMQI}}. 
In this case, we can see that the gap at any generic point on the edge can be closed by simply moving one of the domain walls through this point. Such a gap-closing cannot be associated with a topological distinction since that would imply the existence of uncountably-many distinct topological phases associated with all possible (continuously varying) positions of the domain walls. However, by restricting ourselves to gap-closing transitions occurring at the HSSs, i.e.,  at $x=0$ or $y=0$, we can distinguish between the cases with and without corner states. This follows because the only way to get rid of the zero-energy domain wall states while preserving symmetry is by annihilating them pairwise at one of the mirror invariant points at the boundary. 

\begin{figure}
    \centering
    \includegraphics[width=0.6\columnwidth]{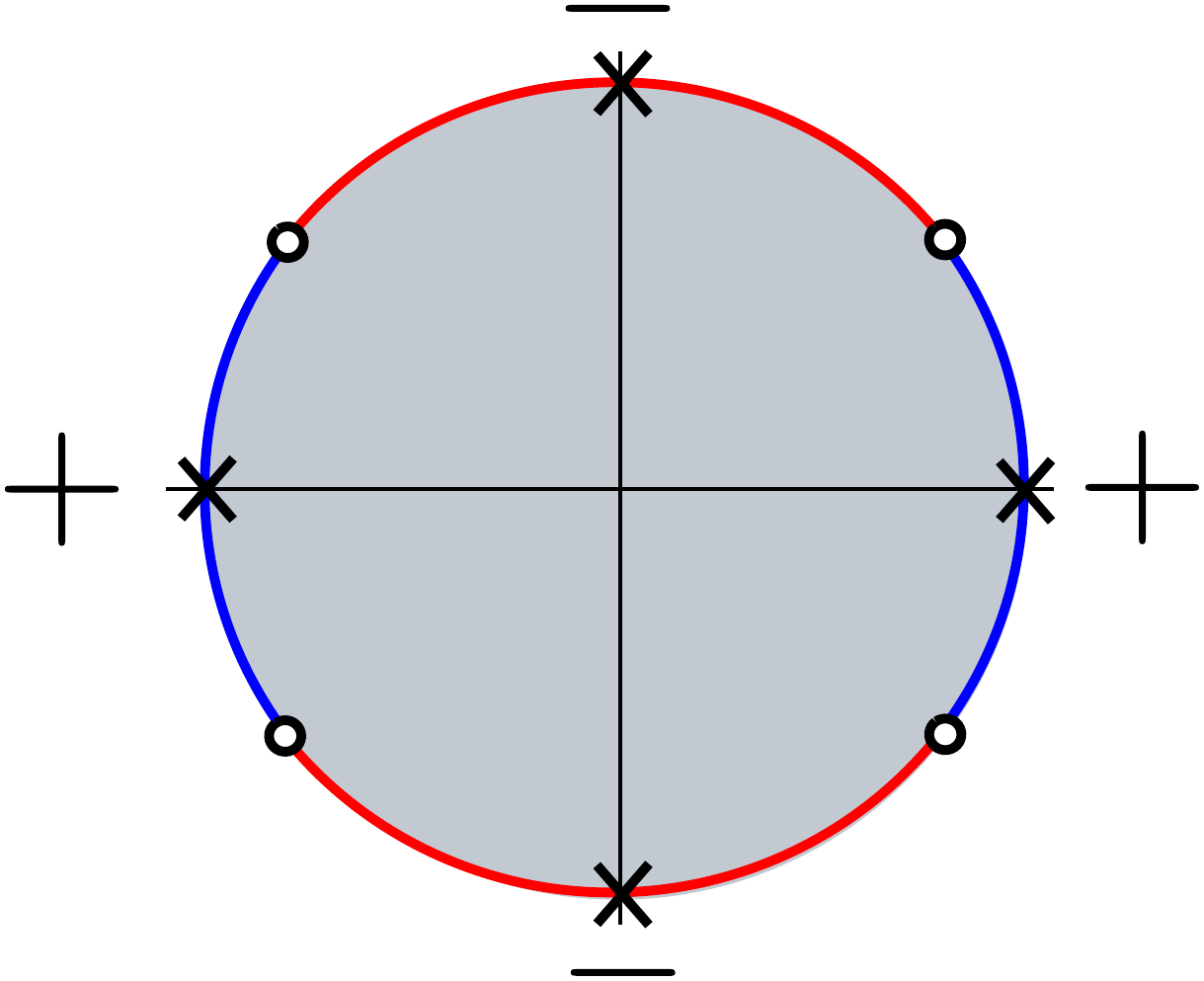}
    \caption{Illustration of zero-energy states (open circles) for the chiral-symmetric DMQI model on a circle geometry. Zero energy states are localized on domain walls for an edge mass term that is invariant under the two mirror symmetries, but has a different sign at the $x=0$ and $y=0$ lines, thus changing sign four times along the edge. These zero-energy states can only be removed by annihilating pairwise at the high symmetry points denoted by the x symbol. }
    \label{fig:DMQIcircle}
\end{figure}

\subsubsection{Relationship to higher-order topological insulators}

It is instructive to relate our understanding of the boundary obstruction in the DMQI in terms of surface domain walls to the corresponding understanding in higher-order topological insulators pioneered in Refs.~\cite{Fang17, Khalaf17, Khalaf18, Wieder18}. In these works, the surface states of higher-order topological insulators were understood in terms of symmetry-enforced surface domain walls where the transformation properties of the surface Dirac mass under spatial symmetry forced it to change sign. In the DMQI, the surface mass does not change sign under either mirror symmetry (see Appendix \ref{app:SSDMQI}). However, by requiring the edge spectrum to be gapped at HSSs, we can distinguish phases based on the sign of the mass term at these points which are associated with different patterns of surface domain walls. For instance, if the sign of the mass term at $x=0$ differs from that at $y=0$, then there are four domain walls as we go around the circle (cf.~Fig~\ref{fig:DMQIcircle}), which cannot be removed without changing the sign of one of these masses. Thus, we can understand the localized boundary modes of the DMQI as domain walls protected by the gap at HSSs. {This is verified by an explicit calculation of the surface states of the DMQI in Appendix \ref{app:SSDMQI}}. We note that the transformation properties of the surface mass term under the symmetries does not force it to change sign. Instead, two distinct, symmetry-allowed configurations of mass signs at HSSs are possible, but they cannot be deformed to each other (without closing a gap at an HSS). This is in contrast to higher-order topological insulators where the sign change of the mass is enforced by the symmetry.

It is worth noting that our topological distinctions are similar to a real-space version of those used in diagnosing semimetals whose spectrum is gapped at high-symmetry momenta \cite{Po17, SongSemimetal}. For instance, a Weyl semimetal with two inversion-related Weyl points with opposite chirality can be smoothly deformed to a trivial insulator by bringing the two Weyl points together. However, this process will necessarily involve a gap-closing at one of the inversion-invariant momenta \cite{hughes2011,turner2012}.

Another related concept is that of the so-called extrinsic higher-order topological insulators discussed in Refs.~\cite{Geier18, Trifunovic18}. These are insulators hosting ``higher-order" surface states with co-dimension $(d-D)$ ($D>1$) that can be removed by symmetrically adding a lower-dimensional SPT on the boundary. Our boundary-obstructed phases fall under this definition since their surface states (or filling anomalies) can be removed by adding some SPTs on the boundary while preserving the symmetry. For example, the corner modes (and corner charge) in the DMQI can be removed by symmetrically adding a pair of SSH chains at two symmetry-related edges as discussed in Sec.~\ref{sec:filling_anomaly}. However, for a BOTP such a process only redefines what we identify as the trivial phase, and \emph{does not} remove the distinction between different phases. In other words, the {\it relative} distinction between the phases is insensitive to what we add at the boundary as long as it is kept fixed when comparing the two phases, a  requirement that is already encoded in our definition.  Thus, we only consider boundary transitions driven by changes in the bulk, ruling out the 
cases in which a non-trivial SPT is glued to the surface of a trivial bulk. 

In general, relative topological distinctions are captured by a mathematical structure called a \emph{torsor} (rather than a group) \cite{Torsor}. A torsor can be thought of as a group without a clear notion of an identity element (trivial phase). We notice that this concept is not new as many topological distinctions for well-known bulk phases are also relative. This is the case, for example, for obstructed atomic phases where an arbitrary choice is made for the \emph{trivial} atomic limit \cite{bradlyn2017topological, Po17}. A simple example is provided by the 1D SSH chain which has two different values of the polarization that can only be distinguished relative to each other. Such a relative distinction can be made more absolute by taking into account the background of positive ions or fixing a convention for the unit cell, either of which can be used to distinguish the two polarization states in the SSH chain according to whether the charge centers lie on top of the positive ions (center of the unit cell) or not (edge of the unit cell) \footnote{ We note that there is still an integer ambiguity in defining polarization coming from the freedom to assign the electron position to any unit cell.}. It is, however, important to keep in mind that the topological distinctions in the electronic Hamiltonian (which does not contain information about the positive ions) in these cases are strictly relative. A similar concept applies to the topological distinction provided by BOTPs.

\subsection{Band representations of BOTPs}
\label{sec:bandreps}

In this section, we present an alternative formulation of BOTPs based on the formalism of band representations \cite{Bradlyn17, Bradlyn2017c} (see Appendix \ref{app:BRs} for a brief review). A band representation corresponds to a set of bands (over the usual momentum space Brillouin zone) generated by decorating the real space Wyckoff positions by atomic orbitals that transform in some representations of the local site symmetry group. In short, a band representation is generated by the Fourier transform of a real space representation of the space group $G$ generated by these local atomic orbitals. One of the recent developments of band theory \cite{Po17, Bradlyn17} was the classification of topological bands that identifies topological bands as any isolated set of bands that \emph{cannot} be written as a band representation, i.e., cannot be generated from the symmetry action on real space orbitals. In the following, we will show these ideas can be extended to describe BOTPs.

As discussed in Sec.~\ref{sec:GenDef}, we can restrict ourselves to BOTPs which admit a Wannier representation. This means that such BOTPs correspond to band representations of the space group $G$ generated from real space atomic orbitals in the periodic system. When we discussed the notion of boundary obstructions in the case of the DMQI in Sec.~\ref{sec:II_DMQI}, we found that there were two distinct notions of obstruction on a given boundary. First, there is a relative notion which distinguishes phases which are equivalent in the bulk,  but which are separated by edge-gap-closing transitions at the boundary. This notion does not specify which of the phases is trivial. Second, there is an absolute notion of obstruction identified by the presence of a filling anomaly which distinguishes the trivial phase (no filling anomaly) from the obstructed or non-trivial one (filling anomaly). In the following, we will use band representations to also define two notions of boundary obstructions.

Our main approach to define boundary obstructions is to identify band representations which do not describe atomic insulators in the open system. That is, a BOTP corresponds to an atomic insulator of the periodic system that cannot be written in terms of localized atomic orbitals which transform under the symmetry group $F$ of the open system. This defines an absolute notion for boundary obstruction (once a boundary is fixed) similar to the notion of filling anomalies discussed in Sec.~\ref{sec:filling_anomaly}. On the other hand, we can also define boundary obstructions in a relative way similar to Sec.~\ref{sec:edge_Wannier} such that two Hamiltonians are considered boundary obstructed if they belong to equivalent band representations of the space group in the bulk, but distinct representations of the point group in the open system. We will show below how these two notions -- the absolute and relative -- are related.

\subsubsection{Symmetry obstruction with open boundaries}

A band representation is specified by a set of Wyckoff positions and a set of irreps attached to each position. Any such band representation can be decomposed into a direct sum  of so-called  elementary band representations (EBRs). An EBR is a BR generated from a single site in a maximal Wyckoff position $\bq$ \cite{Cano2018a}, and will be the focus of what follows. Given a symmetry representation $\rho_\bq$ of the site symmetry group $G_\bq \subset G$ which leaves the site $\bq$ invariant, we can generate an EBR $\rho$ for the Wyckoff position invariant under the space group $G$ by a procedure called induction denoted by $\rho = \rho_\bq \uparrow G$. This procedure essentially extends the symmetry action from the site symmetries of the Wyckoff positions in the unit cell to the space group \cite{Bradlyn17, Bradlyn2017c} (see appendix \ref{app:BRs} for details). Note that band representations induced from different positions $\bq$ and $\bq'$ may turn out to be equivalent. For instance, we saw this to be the case in the DMQI model where the BRs generated by placing the two electrons in the $\bar E$ representation at each of the four maximal Wyckoff positions $1a,\dots,1d$ turned out to be the same. In general, two BRs $\rho$ and $\rho'$ are considered equivalent if there exists a symmetry preserving interpolation matrix $S(t)$ such that $S(0) = \rho$ and $S(1) = \rho'$ \cite{Bradlyn17, Bradlyn2017c}. The matrix function $S(t)$ parameterizes the adiabatic paths between phases. 

The crucial observation is that two representations $\rho$ and $\rho'$ which are equivalent in the periodic system may not be equivalent in the presence of a boundary. To see this, we note that the representation of the open system is obtained by restricting the representation of the space-group (i.e., the periodic system) to a representation of the point group $F \subset G$ (i.e., the open system) relative to a global origin $\O$ for the point group.\footnote{The explicit choice of boundary termination plays a role in the determination of the symmetry origin $\O.$  Hence the ambiguity in the definition of a BOTP due to the choice of boundary termination is implicitly embedded in the determination of the point-group relative to the origin $\O.$} Such a restriction, or ``subduction", which generates a representation $\pi$ of the open system is denoted by $\pi = \rho \downarrow F$. 
Thus, two open boundary representations $\pi$ and $\pi'$ which are generated from equivalent bulk representations may in fact not be the same depending on the original generating site $\bq$.

Let us now define the notion of boundary obstructions by employing the machinery of band representations reviewed in App. \ref{app:BRs}. 
First, we say that a site symmetry representation generated from site $\bq$ in the periodic system defines a (absolute) boundary obstructed phase in the open system with point group $F$ if $\pi = (\rho_\bq \uparrow G) \downarrow F$ is not a \emph{local} representation of $F$. Here, a local representation of $F$ means that it can be written as a sum of site symmetry representations of the Wyckoff positions of the \emph{open} system. Alternatively, we can define a \emph{relative} boundary obstruction between {two phases described by induced representations of $\rho_\bq$ and $\rho_{\bq'}$} on a given open boundary with point group $F$ if 
\beq
\rho=(\rho_{\bq}\uparrow G)~~\sim~~\rho'~~=~~(\rho_{\bq'}\uparrow G)
\eeq
and 
\begin{align}\pi=(\rho_{\bq}\uparrow G)\downarrow F~~\nsim~~\pi'=(\rho_{\bq'}\uparrow G)\downarrow F\end{align}\noindent where the symbols $\sim$ and $\nsim$ indicate equivalence or inequivalence of the representations as defined above.

The relative and absolute notions of boundary obstruction can be tied together by noting that a relative obstruction between $\rho_\bq$ and $\rho_{\bq'}$ implies that at least one of them is boundary obstructed in an absolute sense. In other words, there can never be a relative boundary obstruction between $\rho_\bq$ and $\rho_{\bq'}$ if both $\pi = (\rho_\bq \uparrow G )\downarrow F$ and $\pi'  = (\rho_{\bq'}\uparrow G) \downarrow F$ are local and $\rho = \rho_\bq \uparrow G$ and $\rho' = \rho_{\bq'} \uparrow G$ are equivalent. To see this, we note that, by definition, we can decompose a local representation as a sum of irreps of the point group $F$ as
\beq
\pi = \sum_l n_l(\pi) \pi_l
\eeq
where $l$ runs over the irreps of $F,$ and $n_l(\pi)$ denotes the multiplicity of 
$\pi_l$ in the expansion of $\pi$. Here, we assumed that we moved all the electrons in the open system symmetrically to the center $\O$ which can always be done if $\pi$ is local. The multiplicities can be obtained by employing the orthonormality of the irrep characters relative to the scalar product \cite{evarestov2007quantum}:
\beq
\langle \pi_l|\pi_m \rangle = \frac{1}{|F|} \sum_{g \in F} \tr \pi_l(g) \tr \pi^*_m(g) = \delta_{l,m}
\eeq
leading to
\beq
n_l(\pi) = \frac{1}{|F|} \sum_{g \in F} \tr \pi(g) \tr \pi^*_l(g) .
\eeq
We note now that for $g \in F$, $\tr \pi(g) = \tr \rho(g)$ which by virtue of the equivalence $\rho \sim \rho'$ is equal to $\tr \rho(g') = \tr \pi'(g)$ leading to $n_l(\pi) = n_l(\pi')$. That is, the two local representations $\pi$ and $\pi'$ are equivalent. This means that among the equivalence classes of representations  of the open system induced from equivalent bulk representations, there is only one class containing the local representations, and they are all equivalent. This class can then be labelled as the `trivial' phase. Any other phase with a boundary obstruction relative to it will also have an absolute obstruction.

\subsection{Diagnosing BOTPs using the Wannier spectrum}

{Since the BOTP classification is independent of the specific termination (as long as it is topologically trivial), we are going to choose a particular termination for which the calculation of the boundary spectrum simplifies. This corresponds to  choosing $\phi(\br)$ in Eq.~\ref{eq:Hopen} to be a linearly increasing potential such that the boundary spectrum is the same as the Wannier spectrum, as discussed in Sec.~\ref{sec:edge_Wannier} \cite{Fidkowski11}.} We can then apply band representation theory to the Wannier spectrum to diagnose BOTPs in  a periodic geometry without having to resort to an open system.

Let us begin by recalling some basic facts about the Wilson loop operator whose eigenvalues are the Wannier spectrum. {As we briefly introduced before, the} Wilson loop operator $\W^\bb$ is a unitary operator defined by parallel transporting the occupied-band projection operator in a closed path in the Brillouin zone along the reciprocal lattice vector $\bb$. Its eigenvalues have the form $\exp\{2\pi i \nu_\bb(\bk)\},$ and its eigenstates are the hybrid Wannier functions that are localized in the direction along the lattice vector $\ba$ which is dual to $\bb$, (i.e., $\ba \cdot \bb = 2\pi$) for a fixed momentum along the perpendicular directions $\bk_{\perp}$ \cite{Marzari97, Marzari12}. The dispersion of {the Wannier eigenvalues} $\nu_\bb(\bk_{\perp})$ with $\bk_{\perp}$ is known as the Wannier spectrum, and provides information about the $\bk$-resolved center of the charge along the $\ba$ direction, for $\bk_{\perp}$ momenta perpendicular to $\ba$, $\ba \cdot \bk_{\perp} = 0$. 
{The Wannier spectrum is periodic, reflecting the translation symmetry of a unit cell along $\ba$. The lattice constant is here set to 1}. 

 To see how boundary obstructions can be diagnosed using the Wannier spectrum, we note that since we only care about gaps at HSSs, we can restrict ourselves to boundary terminations consisting of intersections of $(d-1)$-dimensional high-symmetry hyperplanes to study a $d$-dimensional BOTP (a high-symmetry surface hyperplane is one which is left invariant by at least one non-trivial spatial symmetry). We can then study the surface spectrum at different surface hyperplanes, or at the intersection of any number of such hyperplanes. For example, in three dimensions, we can consider the spectrum on 2D surfaces or on 1D hinges lying at the intersection of two such surfaces. In the following, we focus on boundary obstructions accompanied by a gap-closing at a surface hyperplane (i.e., surface obstructions) rather than the intersection of surface hyperplanes (i.e., hinge obstructions). For these types of geometries, a boundary obstruction in our definition is equivalent to the statement that any trajectory connecting the two Hamiltonians must close the surface gap  for at least one high-symmetry surface hyperplane. 
 
 The key concept we apply is that the spectrum of a surface hyperplane perpendicular to a lattice vector $\ba$ for a linear potential is the same as the Wannier spectrum $\nu_\bb(\bk)$ \cite{Fidkowski11}.
This can be used to define a Wannier chemical potential $\mu_\bb$ (cf.~Sec.~\ref{sec:edge_Wannier}) at which the Wannier spectrum $\nu_\bb(\bk)$ should remain gapped if the surface energy gap is to remain open. Thus, we can identify actual surface gap-closing transitions in a surface hyperplane perpendicular to the vector $\ba$ with gap-closing transitions in the Wannier spectrum $\nu_\bb(\bk)$ at the WCP $\mu_\bb$. Unless otherwise stated, we choose a convention that the boundary termination is consistent with (our choice of) the unit cell. The boundaries of the unit cell are chosen to be at $1/2$ in units of the primitive lattice vectors, which corresponds to a Wannier chemical potential $\mu_\bb = 1/2$ for any primitive lattice vector $\ba$. 

The discussion above provides a recipe for diagnosing boundary obstructions by using Wannier spectra as follows. Given two bulk Hamiltonians $\H_1$ and $\H_2,$  a set of high symmetry surface hyperplanes specified by a set of reciprocal lattice vectors $\{ \bb \},$ and WCPs $\{ \mu_\bb \}$ describing the positions of these planes within the unit cell, then a boundary obstruction exists if every trajectory connecting $\H_1$ and $\H_2$ involves a gap closing in the Wannier spectra $\nu_\bb(\bk)$ for some direction $\bb$ at the Wannier chemical potential $\mu_\bb$.

\subsubsection{Wannier band representations}
 A gap closing of the Wannier spectrum indicates a change in the symmetry representations or topology of the Wannier bands. Therefore, it is convenient to relate the symmetry representations of the Wannier spectrum to the periodic boundary and open boundary representations discussed in Sec. \ref{sec:bandreps}.
The Wannier spectrum is obtained by the eigenstates of the Wilson loop operator $\mathcal W^\bb(\bk_\perp)$, whose eigenstates are hybrid Wannier functions $\ket{h^\bb_{\ell n}(\bk_\perp)}$ (HWF), localized in the $\ba$-direction at a layer index $\ell.$ The HWFs satisfy $\mathcal W^\bb(\bk_\perp)\ket{h^\bb_{\ell n}(\bk_\perp)}=\exp\{2\pi i\nu_{\bb n}(\bk_\perp)\}\ket{h^\bb_{\ell n}(\bk_\perp)}$, and $\ket{h^\bb_{\ell n}(\bk_\perp)}=(1/2\pi)\int_0^{2\pi}dk_\bb\exp{-i\ell k_\bb\cdot\ba}U_{nm}\ket{\psi_m(\bk)}$ where $U_{nm}$ is fixed by the Wilson loop operator and reflects the gauge freedom in the choice of these functions. 

Since the HWFs are just a linear combination of the occcupied states (i.e., a basis transformation of the occupied states), then the collection of the hybrid Wannier functions transforms under the same symmetry representation $\rho$ as the collection of occupied Bloch states $\ket{\psi_n(\bk)}.$ The translation symmetry along $\ba$ connects the HWFs in distinct layers $\ell.$ Thus we want to ask under what representation does the collection of HWFs transform at a given $\ell$. Here this representation is denoted $w^\bb$, and it is a representation of the subgroup $G_\bb\subset G$ that leaves $\ell$ invariant. Hence, $w^\bb$ is obtained by restricting $\rho$ to sites in the layer $\ell$, which can be expressed conveniently by $w^\bb=\rho\downarrow G_\bb$. This restriction amounts to disregarding all entries of $\rho$ acting on basis states outside of $\ell$ and group elements outside of $G_\bb$. Alternatively, for atomic bands we can also obtain $w^\bb$ by induction from the site symmetry representation $\rho_\bq$ of a site $\bq$ in this layer. These relations are symbolically written as
\begin{align}\rho=\rho_\bq\uparrow G,\quad w^\bb=\rho_\bq\uparrow G_\bb,\quad w^\bb=\rho\downarrow G_\bb.\end{align} 

Note that due to translation symmetry, all layers transform under the same band representation, and changing the layer corresponds to changing the origin of $G_\bb$. Since each $\ell$ preserves translations perpendicular to $\ba$, we can express the Wannier band representation (WBR)  as a function of $\bk_\perp$, and it is explicitly given by 
\[w^\bb_{\bk_\perp}(h)=\sum_{ij}\exp{i(h_i-h_j)\bq\cdot \bk_\perp}\rho_\bq(h_j\inv h h_i)\]
where $h\in G_\bb,$ and the summation is taken over the coset representatives $h_i\in G_\bb/G_\bq$ that generate all sites $\bq_i=h_i\bq$ in the layer. This formula is analogous to the induction formula for the bulk representation
\[\rho_{\bk}(g)=\sum_{ij}\exp{i(g_i-g_j)\bq\cdot \bk}\rho_\bq(g_j\inv gg_i),\]
with $g_j\inv gg_i\in G_\bq$, $g\in G,$ and $g_i\in G/G_\bq$
with a derivation presented in App. \ref{app:BRs} to be self-contained.

With this structure in place, it is now crucial to notice that the Wannier band representation $w^{\bb}$ may correspond to disconnected Wannier bands,  \begin{align}w^{\bb}=\bigoplus_a w^{\bb}_{a},\end{align}
labelled by an index $a$. In this case, the representations $w^{\bb}_a$ may have their own topological characterization, and they are not required to be local even when the full bulk representation $\rho$ is local. Indeed each $w_a^\bb$ either corresponds to an elementary WBR, or to a topological band when its characters at high symmetry momenta are not compatible with a local atomic description. Reiterating, while $w^{\bb}$ in total forms an EBR, each individual component describing separate Wannier spectrum bands, $w^{\bb}_a$, does not need to be a an EBR. An example of topological Wannier bands is shown in Sec.~\ref{sec:3DSS}. Such phases  with topological Wannier bands result in anomalous boundary modes rather than the boundary charges that would appear if the Wannier bands were in obstructed atomic limit representations.

The consistency relations above allow us to infer the bulk representation $\rho$ and the open boundary representation $\pi$ from the different Wannier representations $w^\bb$. Closing the Wannier spectrum implies a change in the Wannier representations $w_a^\bb$. 
The explicit calculation of the Wannier representations together with the bulk and open system representations are shown in Tables. \ref{tab:WannierBandsWyckoffPositions}, \ref{tab:WannierBandsWyckoffPositionsCn} and  \ref{tab:WannierBandsWyckoffPositionsChern} for the models discussed in this text.



\section{BOTP and filling anomalies in other 2D systems}
\label{sec:2DFA}
{Now that we have laid out the general framework for BOTPs we would now like to discuss the possibility of finding BOTP phases in other 2D systems besides the DMQI model. We are going to focus on BOTPs which can be identified through a filling anomaly in the open system similar to the DMQI (Sec.~\ref{sec:filling_anomaly}).}

~

{\subsection{Generalities on filling anomalies}}
{Let us begin by distinguishing the spatial symmetry groups in a few different settings. First, we can consider an infinite translationally symmetric system with a space group $\G$. Among the possible wallpaper groups \cite{WiederScience}, we restrict ourselves to symmorphic ones which can be written as a direct product of a point group $G$ relative to a point $\O$ and the group of translations in the infinite system. We can instead consider a \emph{finite} translationally symmetric system with periodic boundary conditions such that for translation along a given lattice vector $T_\ba$, there is an integer such that $T^n_\ba=1,$ i.e., the group of translations is a product of cyclic groups. We can then consider the group of spatial symmetries of such a periodic system even if we allow translation symmetry to be broken. We will denote such a group by $G$ which denotes the set of spatial symmetries on a torus without any translations. Finally, there is the symmetry group of the open system which is given by $F,$ which can be non-trivial if the boundary is chosen symmetrically. 

For example, in the SSH chain, the group of symmetries of the infinite system is generated by a reflection around a given point and translation by one unit cell. This includes mirror reflections around all the points $n$ and $n + 1/2$ for any integer $n$. For a finite periodic system with length $N$, the symmetry group is obtained from the one for the periodic system by the condition $T_x^N = 1$. For a system with periodic boundary conditions but not translations, the symmetry group is generated by two mirrors about the origin $x=0$ and the point $x=N/2$ (which sends $x$ to $N-x$ which is equivalent under periodic boundary conditions to $-x$). Finally, for the open system, the symmetry group is generated by a single mirror relative to a single fixed origin $\O$ which is determined by the choice of boundary termination.}

In order to define filling anomalies for BOTP systems, we need a way to go back and forth between the open system and the periodic system. For this reason, we start by considering a special class of boundaries which can be folded back to a torus.  
We will call such boundaries foldable and define them as terminations which are obtained by cutting a torus into a $d$-dimensional region topologically equivalent to the disk such the the following conditions are satisfied: (i) the open system is invariant under the point group symmetry $F$ relative to $\O$, (ii) the $d$-dimensional torus can be obtained by identifying opposite edges/faces of the open region, and (iii) it includes an integer number of unit cells of the translationally symmetric system. This definition means that the open system has the shape of a (large) symmetric unit cell for the periodic system. This special class of boundaries enables us to compare the filling in the open and periodic systems and make the idea of ``resolving" the filling anomaly by identifying boundaries more precise. Once we have established the existence of a filling anomaly in the open system, we will show afterwards how we can define it for more general boundaries, hence establishing in the process that it captures a stable topological invariant of the open system which does not change under any boundary or bulk deformation which preserves the gap at high symmetry surfaces (as well as the bulk gap).

We now define a filling anomaly as follows. Given a periodic Hamiltonian with filling of $\nu$ per unit cell, we can define a corresponding open Hamiltonian (or projector) on a foldable boundary whose filling is $\nu N$ with $N$ denoting the number of unit cells. \emph{A Hamiltonian (or a projector) is characterized by a filling anomaly at $\nu N$ if a gapped, symmetric, Wannierizable and charge-neutral ground state is possible in the periodic system, but impossible on the foldable open boundary.} For a system with a filling anomaly we can find a gapped, symmetric and Wannierizable ground state in the open system if we remove the condition of charge-neutrality. Hence, the filling anomaly can be quantified by an integer $\nu^{\rm fa}$ denoting the number of electrons we need to add to obtain a gapped, symmetric and Wannierizable ground state in the open foldable system. 

Now let us try to generalize this invariant to more generic types of boundaries. Our definition for boundary obstructions identifies Hamiltonians in an open system which are related by symmetric deformations which do not close the gap in the bulk or at any HSS on the boundary. This does allow for gap-closing transitions away from HSSs at the boundary which can change the total filling by pushing some eigenstates across the chemical potential. Such gap-closings have to occur at $|F|$ symmetry related points/surfaces at the boundary which means that the total change in the filling induced by such a process has to be a multiple of $|F|$. As a result, the integer $\nu^{\rm fa}$ can change by a multiple of $|F|$ but is invariant modulo $|F|$. This defines a topological invariant protected by the gap at HSS which is additive modulo $|F|$. As a result, the topological phases distinguished by filling anomaly invariants form a subgroup of the integers $\Z_{|F|}$. We can now lift the restriction of foldable boundaries by considering more general symmetric boundary terminations which can be deformed to a foldable boundary without closing the gap at a HSS. Since $\nu^{\rm fa}\, {\rm{mod}}\, |F|$ is protected by the gap at HSSs, it will remain well-defined for such boundaries and can be used to define filling anomalies in a more general context.

{To understand how a filling anomaly arises when we open the boundary,  we notice that the little group of a point $\bq$, i.e. the subset of symmetries which leave $\bq$ invariant, is generally larger on the torus compared to the open system. More specifically, the little group of $\bq$ is defined in the open system and the torus as:}
\begin{gather}
F_\bq = \{g \in F, g \bq = \bq\}, \nonumber \\
G_\bq = \{g \in G , g \bq = \bq \}.
\end{gather}
{It is clear from the definition that $F_\bq \subseteq G_\bq$ since $F \subseteq G$. As a result, the size of the orbit of $\bq$ under the action of the point group $F$ is $|F|/|F_\bq|$ ($|F|/|G_\bq|$) in the open (periodic) system. For example, in the DMQI a general point $(x,y)$ whose little group contains only the identity element has an orbit with 4 elements in total given by $(x,y)$, $(-x,y)$, $(x, -y)$ and $(-x,-y)$ under the action of $M_x$ and $M_y$. On the other hand, a point lying on one of the mirror lines, let's say $(x,0)$ has an orbit with only two points: $(x,0)$ and $(-x,0)$. When we open the boundary, a point $\bq$ in the periodic system maps to $n_\bq$ points in the open system where $n_\bq$ is an integer larger than or equal one given by
\beq
n_\bq = |G_\bq|/|F_\bq| .
\eeq
Thus, a filling anomaly occurs if (and only if) there is a point $\bq$ whose filling in the periodic system is not divisible by $n_\bq$. The anomaly invariant $\nu^{\rm fa}$ can be expressed in terms of $n_\bq$, the size of the full group $|G|,$ and the filling of $\bq$ in the periodic system $\nu^{\rm pbc}_\bq$ as
\beq
\nu^{\rm fa} = \frac{\nu^{\rm pbc}_\bq|F|}{n_\bq} \mod |F| = \frac{\nu^{\rm pbc}_\bq|F| |F_\bq|}{|G_\bq|} \mod |F|,
\label{eq:nq}
\eeq
which is always an integer since $|F|$ is divisible by $|G_\bq|$ (which follows from $G_\bq \subseteq F$).
Now we can state in particular that \emph{a filling anomaly describes a BOTP if there are two gapped bulk Hamiltonians which are symmetrically deformable in the periodic systems, but which correspond to distinct anomaly invariants $\nu^{\rm fa}$ in the open system.} 

~

\subsection{Filling anomalies in 2D systems}
\label{sec:2DProof}

To investigate possible filling anomalies for BOTPs in general 2D systems, we consider the possible point group symmetries for which the anomaly condition derived above is satisfied. Crystallographic point groups in 2D belong to 2 categories: $n$-fold rotation groups $C_n$, $n=1, 2,3,4,6$ or the dihedral groups $D_n$ generated by $n$-fold rotation and an in-plane mirror reflection with $n=1, 2,3,4,6$. We denote the filling of a point $\bq$ in the open and periodic system by $\nu_\bq$ and $\nu^{\rm pbc}_\bq$, respectively. We now note that a point $\bq$ whose little group $G_\bq= C_n$ will be characterized by $n_\bq = n/|G_\bq|$. As a result, a filling anomaly is only possible if $\nu^{\rm pbc}_\bq$ is not divisible by $n$. On the other hand, the value of $\nu^{\rm pbc}_\bq$ modulo $n$ cannot be symmetrically changed in the periodic system since the only possible symmetric operations involve the addition or removal of $n$ electrons to $\bq$ which will change its filling by multiples of $n$. As a result, all filling anomalies for points $\bq$ with point group $C_n$ are associated with bulk phases not BOTPs.

Next, consider points $\bq$ whose point group is $G_\bq = D_n$ for which $|G_\bq| = 2n$ such that a filling anomaly is realized whenever $\nu^{\rm pbc}_\bq$ is not divisible by $2n$. In contrast to the $C_n$ case, there are symmetric deformations which can change the value of $\nu^{\rm pbc}_\bq \mod 2n$. Namely, by symmetrically moving $n$ electrons in or out of $\bq$ along the mirror invariant lines. As a result, a filling anomaly is associated with a bulk phase whenever $\nu^{\rm pbc}_\bq \neq n \mod 2n,$ and it is associated with a BOTP when $\nu^{\rm pbc}_\bq = n \mod 2n$. For BOTPs this is equivalent to the conditions
\beq
\nu^{\rm pbc}_\bq = n, \qquad |G_\bq| = 2n, \qquad |F_\bq| = 1,
\label{eq:BOTPFA}
\eeq
which gives a necessary (but not sufficient) condition for a BOTP filling anomaly. One possibility of realising (\ref{eq:BOTPFA}) is at the corner of a rectangular sample which does not have any symmetry in the open system ($|F_\bq| = 1$) but lies at the intersection of two mirrors in the periodic system such that $|G_\bq| = 4$. Thus, the filling $\nu^{\rm pbc}_\bq = 2$ in the periodic system leads to a filling anomaly not associated with a bulk phase. This is what happens in the DMQI model. Other cases involve a point $\bq$ lying at an edge in the open system with $|F_\bq|=1$ which becomes a mirror invariant line in the periodic system with $G_\bq = 2$. If the periodic filling $\nu^{\rm pbc}_\bq = 1,$ and this position can be symmetrically deformed to the origin, then it satisfies all the conditions of a boundary filling anomaly. An example is illustrated in Fig.~\ref{fig:D4FA} with the group $D_4$ with the points $\bq$ lying at the edges of the square (with $|G_\bq|=2$) with $\nu^{\rm pbc}_\bq = 1$. Such filling anomalies can be observed on generic $D_4$ symmetric boundaries likes the ones shown in Fig.~\ref{fig:D4FA}b,c. However, on a simple square boundary with approximate translation symmetry along the edges, the localized charges (or alternatively the zero modes if we assume particle-hole or chiral symmetry) cannot exist at any arbitrary point along the edge and will be pushed towards the corners where they annihilate. Hence, for simple boundary terminations, e.g., a square or hexagon, we are not able to identify any interesting physical signatures of BOTPs other than what was already known for the DMQI.  For the rest of this work, we will instead focus on BOTPs in 3D.

\begin{figure}[t]
\center
\includegraphics[width=0.85\columnwidth]{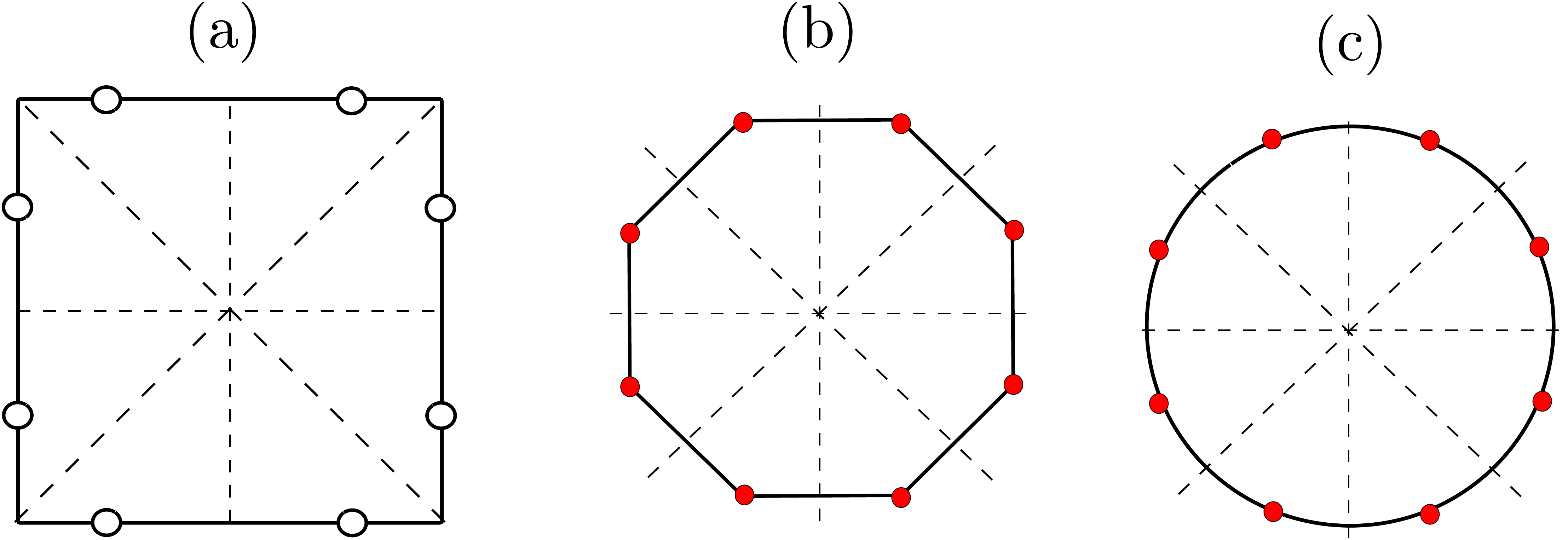}
\caption{(a) schematic illustration of a BOTP filling anomaly in the 2D point group $D_4$ associated with edge charges on a square geometry. (b) and (c) more complicated edge terminations which allows the localization of the zero modes so that they can only be removed by closing the gap at a high-symmetry edge.}
\label{fig:D4FA}
\end{figure}

\section{3D models with boundary obstructions}
\label{sec:3D}

\subsection{Recipe for constructing 3D BOTPs}
\label{sec:Recipe}

We now switch our attention to BOTPs in 3D.
In the following we present a simple recipe that can be used to construct 3D models with different types of boundary obstructions. The recipe works in analogy to the construction of the 2D DMQI model. For that model, we took a pair of inversion symmetric SSH chains in each unit cell parallel to say the $x$-direction, and then coupled them with dimerized couplings, i.e., intra- and inter-cell couplings $\gamma_y$ and $\lambda_y$ along the $y$-direction (i.e., the stacking direction). Additionally, to generate the necessary symmetry structure, we threaded a $\pi$-flux per unit cell (Fig.~\ref{fig:quad_lattice}). In the limit $\gamma_y = 0,\lambda_y \neq 0$, there is an isolated/unpaired SSH chain on each of the edges parallel to the $x$-axis that can be in the trivial ($|\gamma_x|>|\lambda_x|$) or obstructed ($|\gamma_x|<|\lambda_x|$) atomic limit, and thus may have an associated  boundary obstruction. 

This construction suggests a route to obtaining 3D BOTPs: we start with any 2D Hamiltonian $\H_{\rm 2D}(\bk)$ associated with boundary signatures such as corner/edge charges or 1D edge states. We then consider a pair of these Hamiltonians such that their sum is topologically trivial, and hence all the associated boundary signatures can be removed when they are coupled. Next, we couple these pairs via dimerized SSH-like couplings $\gamma_z, \lambda_z$ along the (stacking) $z$-direction. This arrangement guarantees that in the limit $\gamma_z = 0$, $\lambda_z \neq 0$, there is a single isolated copy of $\pm \H_{\rm 2D}(\bk)$ at the upper/lower surface when the system is considered with open boundaries. Additionally, we need to to ensure that the bulk gap is open if either $\H_{\rm 2D}(\bk)$ is gapped or $|\gamma_z| \neq |\lambda_z|$. This can be done by choosing the 2D Hamiltonian to have alternating sign between layers. In most cases, this is realized by inserting $\pi$ fluxes between the stacked layers (i.e., in the $xz$ and $yz$ plaquette types) {possibly with the addition of some other modifications to the Hamiltonian parameters between layers as we will show later}. The resulting 3D Hamiltonian is given by
\begin{multline}
\H_{\rm 3D}(\vec k) = \H_{\rm 2D}(\bk) \tau_3 + (\lambda_z \sin k_z )\tau_2  + (\gamma_z + \lambda_z \cos k_z) \tau_1,
\label{H3D}
\end{multline}
where $\tau_{1,2,3}$ denote the Pauli matrices in the layer subspace along $z$. Due to the $\pi$ fluxes, the different terms of the Hamiltonian anticommute which implies that it can only be gapless if each of the three terms separately vanish, i.e., only if $\H_{\rm 2D}(\bk)$ is gapless \emph{and} $|\lambda_z| = |\gamma_z|$. {As in the DMQI case, the existence of $\pi$ fluxes means that spatial symmetries are generally combined with gauge transformations and transform projectively. In particular, since the $\pi$-fluxes here are only considered in the vertical $xz$ and $yz$ planes but not the horizontal $xy$ plane, the resulting projective representation cannot be recast as a spinful representation as we will see later.} 

It is instructive to highlight the relationship between our construction and the layer construction used to build higher-order and topological crystalline phases from lower dimensional topological phases \cite{Huang17, song2017topological, Song2018, Kim2019}. In the conventional layer construction, one starts from a lower dimensional topological phase placed at a symmetry-invariant region and then symmetrically adds layers or copies of this lower dimensional system to realize a higher-dimensional system. A system constructed this way will always be topologically non-trivial in the bulk as long as the protecting spatial symmetry is unbroken. In contrast, we start with a lower dimensional system $\H_{\rm 2D}$ which is topologically trivial but consists of two components which are individually non-trivial. Upon repeating this lower-dimensional unit using the layering procedure, we can spatially separate these two non-trivial components such that they appear at opposite edges/surface of the system. The resulting system is topologically trivial in the bulk with lower dimensional non-trivial systems living at its boundary.

\begin{figure}
    \centering
    \includegraphics[width=.6\columnwidth]{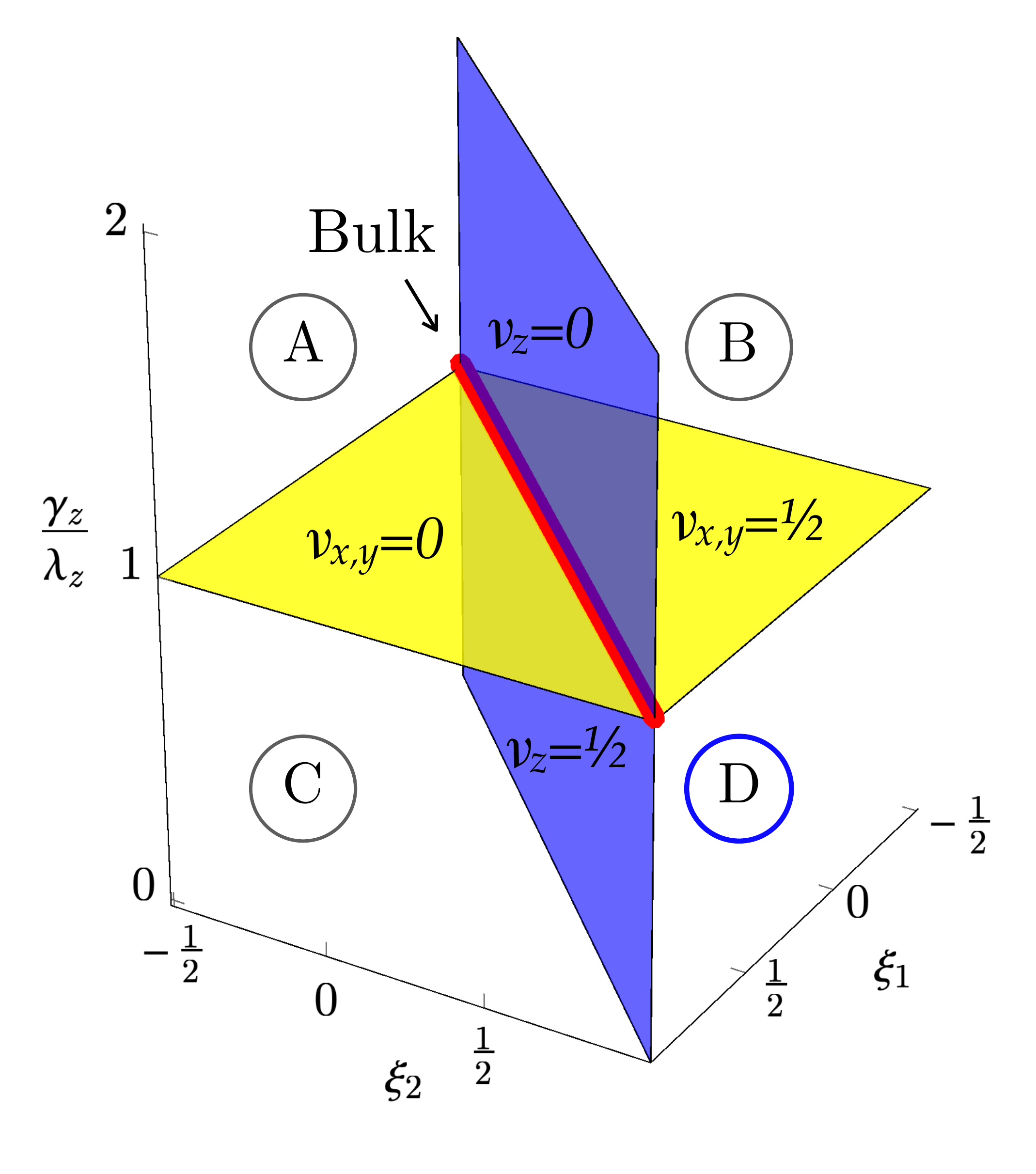}
    \caption{Generic phase diagram for $\H_{\rm 3D}(\bk)$ obtained by the three dimensional construction in Sec. \ref{sec:Recipe}, where the two dimensional Hamiltonian $\mathcal{H}_{\rm 2D}({\bf{k}})$ is parameterized by $\{\xi_1,\xi_2\}.$ As an illustration we choose the 2D model to exhibit two topologically distinct phases depending on which of $\xi_1$ or $\xi_2$ is the largest. Hence, $\mathcal{H}_{\rm 2D}({\bf{k}})$ has a bulk gap closing transition at $\xi_1=\xi_2$ that will subsequently lead to a Wannier gap closing in $\mathcal{H}_{\rm 3D}({\bf{k}})$ in the $\W^z$ Wannier spectrum (blue plane), which can happen at $\nu_z=0$ or $\nu_z=1/2$ as labeled in the figure. On the other hand, the gap in the $\W^x$ and $\W^y$ Wannier spectra closes when the stacks of $\mathcal{H}_{\rm 2D}({\bf{k}})$ are uniformly coupled in the stacking direction, i.e., when $\lambda_z=\gamma_z$ (yellow plane),  again either at $\nu_{x,y}=0$ or $\nu_{x,y}=1/2$ as labeled in the figure. The intersection between the two planes corresponds to a line where the bulk gap is closed in the phase diagram. This construction yields four phases $A,~B,~C$ and $D$ separated by Wannier gap closings. A choice of boundary may select one of the four to be topologically distinct from the other three (in the sense of a boundary obstruction) when the Wannier gap closing happens at the Wannier chemical potential. Here we highlighted phase D as a non-trivial boundary obstructed topological phase for a choice of boundary termination that coincides with the unit cell structure, $\mu_{i}=1/2$. }
    \label{fig:genphasediag}
\end{figure}

The type of boundary obstruction in the 3D Hamiltonian (\ref{H3D}) depends on the type of obstruction for $\H_{\rm 2D}(\bk),$ as well as the dimension of its surface states. For the top/bottom surfaces, $\H_{\rm 3D}(\bk)$ will have a surface (hinge) obstruction if $\H_{\rm 2D}(\bk)$ has a bulk (edge) obstruction. For side surfaces, $\H_{\rm 3D}(\bk)$ will have a surface (hinge) obstruction if $\H_{\rm 2D}(\bk)$ has edge states/charges (corner charges). The latter statement can be understood by investigating the dimension of the SPT needed to add to the side surfaces to cancel the surface states/charges (remember all BOTPs are also extrinsic higher-order topological insulators whose surface states can be removed by adding a lower dimensional SPT on the surface). If the 2D Hamiltonian has corner charges, then it is enough to add 1D SSH chains to the vertical hinges to cancel them. On the other hand, a 2D Hamiltonian with edge states or charges requires the addition of a 2D SPT on the side surfaces to cancel these states/charges. It follows from the previous discussion that $\H_{\rm 3D}(\bk)$ has a surface (hinge) obstruction if and only if $\H_{\rm 2D}(\bk)$ has a bulk (edge) obstruction with 1D (0D) edge states/charges. We note that it is also possible to have mixed surface-hinge obstructions where connecting two BOTPs involves either a surface gap-closing on the upper/lower surfaces or a hinge gap-closing on the side surfaces. This would be the case if, for example, we take $\H_{\rm 2D}(\bk)$ to be a bulk-obstructed atomic insulator with corner charge, e.g., the $C_{4z}$ quadrupole model. It is worth noting that the distinction between surface and hinge types of obstructions is only possible for boundaries consisting of several intersecting 2D planes. On a more smooth boundary like a sphere, such a distinction is ill-defined. 

 Now let us more explicitly describe the properties of our 3D parent model. A representative phase diagram of the above construction can be found in Fig.~\ref{fig:genphasediag}. There we consider a two dimensional Hamiltonian parametrized by generic parameters $\xi_1$ and $\xi_2.$ For illustration we assume the 2D Hamiltonian admits a topologically nontrivial phase with boundary modes if $\xi_1>\xi_2$, a trivial phase when $\xi_1<\xi_2$ and a critical point when $\xi_1=\xi_2$. When stacking into a three dimensional system, the bulk 3D Hamiltonian will be gapped even when $\xi_1=\xi_2$ as long as there is a dimerization along $z$, $\lambda_z\neq\gamma_z$. As a function of $\xi_1, \xi_2,$ and $\gamma_z/\lambda_z$ there is a line in the phase diagram (red line) where the bulk is gapless. By construction, this model allows for four adiabatically connected bulk phases, separated by Wannier gap transitions where the Wannier spectra changes its topology. A choice of a symmetric boundary termination will single out one of the phases, where the Wannier gap closing happens at the WCP, (for our choice this is at $\nu_{x,y}=1/2$ and $\nu_z=1/2)$ (c.f. Fig.~\ref{fig:genphasediag}). For this boundary termination, which coincides with the unit cell structure, phase $D$ is a non-trivial boundary obstructed topological phase since $\lambda_z>\gamma_z$ leaving a dangling, unpaired $\H_{\rm 2D}(\bk)$ on both the top and bottom surfaces that is in its topologically nontrivial phase.

In the following two subsections, we will focus on models with surface obstructions while briefly discussing an example of a hinge obstruction. We will separately consider the cases where the surface obstruction is associated with a filling anomaly or gapless surface states.

\subsection{3D BOTPs with filling anomalies}
\label{sec:3DFA}
In this subsection, we will discuss 3D BOTPs associated with filling anomalies. We will propose a class of models with $C_{2nh}$ symmetry $n=1,2,3$ exhibiting surface obstructions associated with a {filling anomaly}. After introducing the models, we analyze their properties using a real-space approach as well as through the Wannier spectrum. At the end, we will briefly discuss an example of a hinge-obstructed model with corner charge.

\FloatBarrier

\subsubsection{A new class of $C_{2nh}$ Hamiltonians}\label{sec:c2nhmodels}

Now, let us introduce a new class of 3D models defined for {the space groups 10, 83, and 175 which are symmorphic space groups whose} point group $F$ is $C_{2nh}$ with $n=1,2,3$, {respectively}.  These symmetry groups are characterized by a $2n$-fold rotation $C_{2nz}$ accompanied by a mirror reflection perpendicular to the rotation axis $M_z$. 

We build the 3D Hamiltonians in each case following the recipe of Sec. \ref{sec:Recipe} using 2D models that realize a $C_{2nz}$ insulator at filling $n.$ The full 3D model will be a 3D BOTP at filling of $2n$ which has a \emph{surface} obstruction associated with {a filling anomaly}.
The 2D tightbinding Hamiltonians we use are shown in Fig.~\ref{fig:2Dmodel_246}. Each of these 2D models has several distinct atomic limit phases separated by a bulk transition: one atomic limit where all the $2n$ electrons are at the center of the unit cell (with site symmetry group $C_{2nz}$)
and others where the $2n$ electrons are distributed in pairs to $n$ symmetry related edges of the unit cell (with site symmetry $C_{2z}$).
The non-trivial obstructed atomic limit of these models is characterized by edge charges and bulk polarization so they can be understood as weak 2D phases if translation symmetry in the 2D plane is preserved.

\begin{figure}[t]
\center
\includegraphics[width=.9\columnwidth]{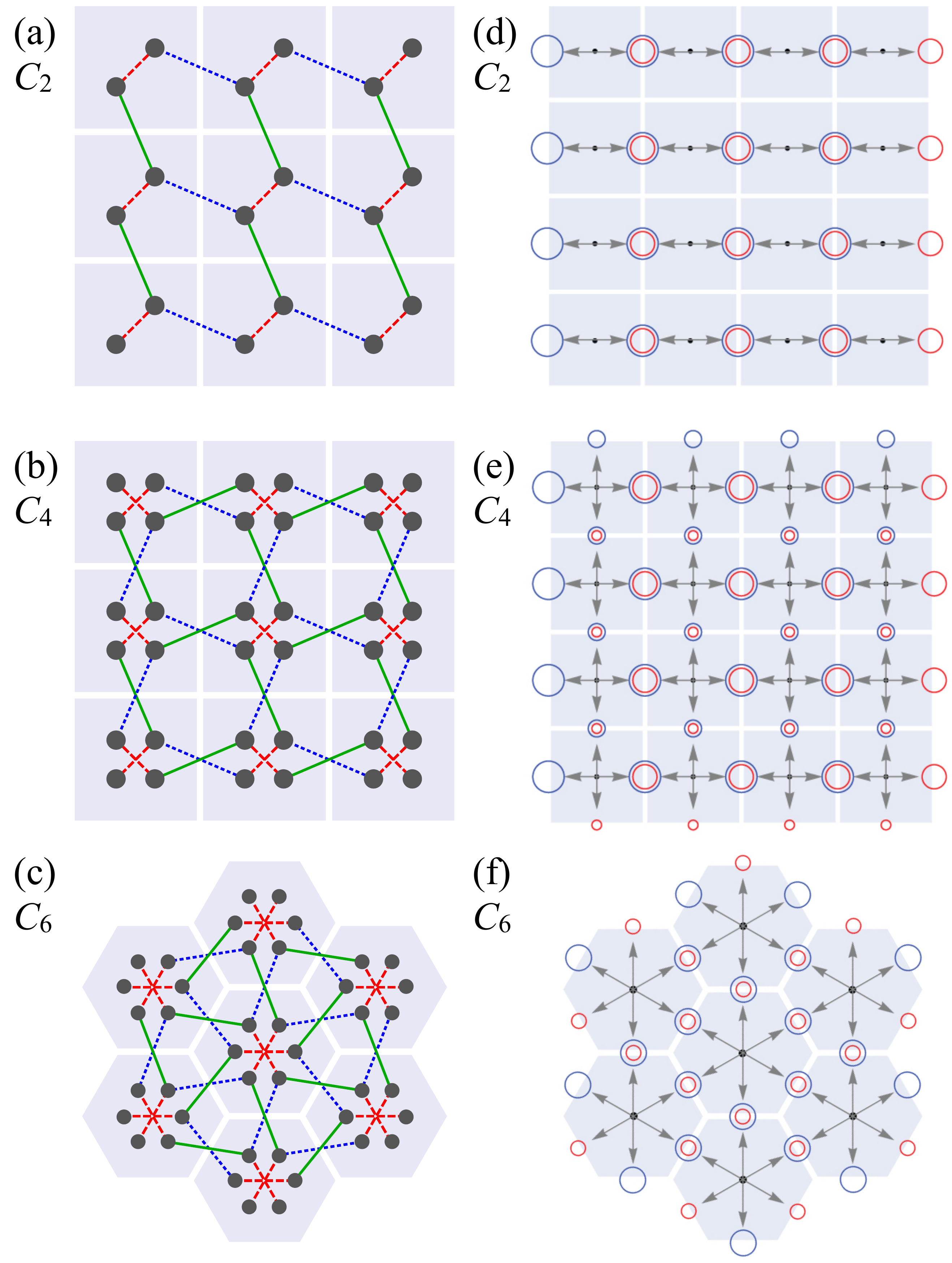}
\caption{(a,b,c) Schematic illustration of the 2D hopping Hamiltonians used to build the 3D surface obstructed models with (a) $C_{2z}$, (b) $C_{4z}$, and (c) $C_{6z}$ symmetries. The hopping parameters $\lambda_0$, $\lambda_x$ and $\lambda_y$ are denoted by red, green, and blue lines respectively. (d,e,f) Top view illustration of the deformation process explained in the text between position $1a$ and position $nc$ for (d) $C_{2z}$, (e) $C_{4z}$, and (f) $C_{6z}$ symmetries.}
\label{fig:2Dmodel_246}
\end{figure}

 Explicitly, the $C_{2z}$ model is given by
\begin{multline}
 \H^{C_{2z}}_{\rm 2D}(\vec k) = (\lambda_x \sin k_x + \lambda_y \sin k_y) \sigma_2 \\+ (\lambda_0 + \lambda_x \cos k_x + \lambda_y \cos k_y) \sigma_1,
 \label{H2D}
\end{multline}
where $\sigma_{1,2,3}$ denote the Pauli matrices distinguishing the two atomic orbitals inside the unit cell. The hopping parameters $\lambda_{0,x,y}$ are represented by the black, green, and blue links in Fig.~\ref{fig:2Dmodel_246}a, respectively.
The spectrum of Eq. \eqref{H2D} is given by 
\beq
\varepsilon_{\rm 2D}^2=\lambda^2+2\lambda_0(\lambda_x\cos k_x+\lambda_y\cos k_y)+2\lambda_x\lambda_y\cos (k_x-k_y),
\eeq
with the shorthand $\lambda^2=\lambda_0^2+\lambda_x^2+\lambda_y^2$. The 2D model is gapped whenever the largest of $\lambda_{0,x,y}$ exceeds the sum of the other two. In addition, the gapped phases at half filling correspond to distinct atomic insulators with a single electron localized either at the $1a$, $1b,$ or $1c$ positions
whenever the largest $\lambda$ is $\lambda_0$, $\lambda_x,$ or $\lambda_y$, respectively.

The 2D Hamiltonians for $C_{4z}$ and $C_{6z}$ symmetric cases can be constructed in a straightforward way. We simply stack two or three copies of $\H^{C_{2z}}_{\rm 2D}(\bk)$ related by $C_{4z}$ or $C_{6z}$ rotations, respectively. They are shown schematically in Fig.~\ref{fig:2Dmodel_246}b,c. Let us here explicitly show the resulting $C_{4z}$ model, by adding to the $C_{2z}$ model a $\pi/2$ rotated copy of itself, 
\beq
\H^{C_{4z}}_{\rm 2D}(\vec k) = \left(\begin{array}{cc}
\H^{C_{2z}}_{\rm 2D}(\vec k) & 0 \\
0 & e^{i \frac{\pi}{4} \sigma_1}\H^{C_{2z}}_{\rm 2D}( C_{4z}\vec k) e^{-i \frac{\pi}{4} \sigma_1}
\end{array}\right),
\eeq
where $C_{4z}\vec k=(k_y, -k_x)$ is the momentum space action of $C_{4z}$. This Hamiltonian describes a 2D four-band model. 

The 3D models can be constructed following the recipe of Sec.~\ref{sec:Recipe} stacking pairs of 2D models with dimerized couplings $\gamma_z$ and  $\lambda_z$ along the z-direction. Additionally, we thread a $\pi$-flux per unit cell in the $xz$ and $yz$ plaquettes but not in the $xy$ plaquette. This flux threading modifies the symmetry representation for the spatial symmetries to a projective representation by combining spatial symmetries with gauge transformations such that the gauge invariant condition
\beq
C_{2z} M_z C_{2z}^{-1} M_z^{-1} = -1
\label{eq:C2Mz}
\eeq
is satisfied. This condition captures the $\pi$ flux in the vertical ($xz$ and $yz$) plaquettes. Notice that this condition implies that we \emph{cannot} write this projective representation as a non-projective spinless or spinful representation: the former has $M_z^2 = C_{2z}^2 = +1$ whereas the latter has $M_z^2 = C_{2z}^2 = -1$ with both having $[M_z, C_{2z} ] = 0$ \cite{WiederKane}. As a result, the inversion operator which is given by $I = C_{2z} M_z$ squares to $+1$ in \emph{any} non-projective spinful or spinless representation whereas it squares to $-1$ when (\ref{eq:C2Mz}) is satisfied meaning that it cannot be consistent with any non-projective spinless or spinful representation. This arises from the special property of the models we consider where $\pi$-fluxes are attached only to the vertical plaquettes but not the horizontal ones.

For the 3D Hamiltonian defined by (\ref{H3D}) with $\H_{\rm 2D}$ given in (\ref{H2D}), the spatial symmetries $C_{2z}$ and $M_z$ are implemented as
\begin{align}
M_z \H(\vec k) M_z^\dag &=\H( m_z\vec k),  \quad &M_z = \sigma_3 \tau_1 \nonumber \\
C_{2z} \H(\vec k) C_{2z}^\dagger &= \H(C_{2z}\vec k), \quad &C_{2z} = \sigma_1\tau_0,
\end{align}
where $m_z\vec k=(k_x,k_y,-k_z)$ and $C_{2z}\vec k=(-k_x,-k_y,k_z)$.
The eight-band $C_{4h}$ model has the symmetries
\beq
C_{4z} = e^{-i \frac{\pi}{4}} \left(\begin{array}{cc}
0 & e^{i \frac{\pi}{4} \sigma_1} \\
e^{i \frac{\pi}{4} \sigma_1} & 0
\end{array}\right)\tau_0, \quad M_z =  \left(\begin{array}{cc}
\sigma_3 & 0 \\
0 & \sigma_3
\end{array}\right)\tau_1.
\eeq
Here $\sigma_{1,2,3}$ act in the space of orbitals within the same 2D layer, whereas $\tau_{1,2,3}$ act in the space of orbitals in different layers. The $C_{6z}$ model has a similar structure.

 \begin{figure}
    \centering
    \includegraphics[width=\columnwidth]{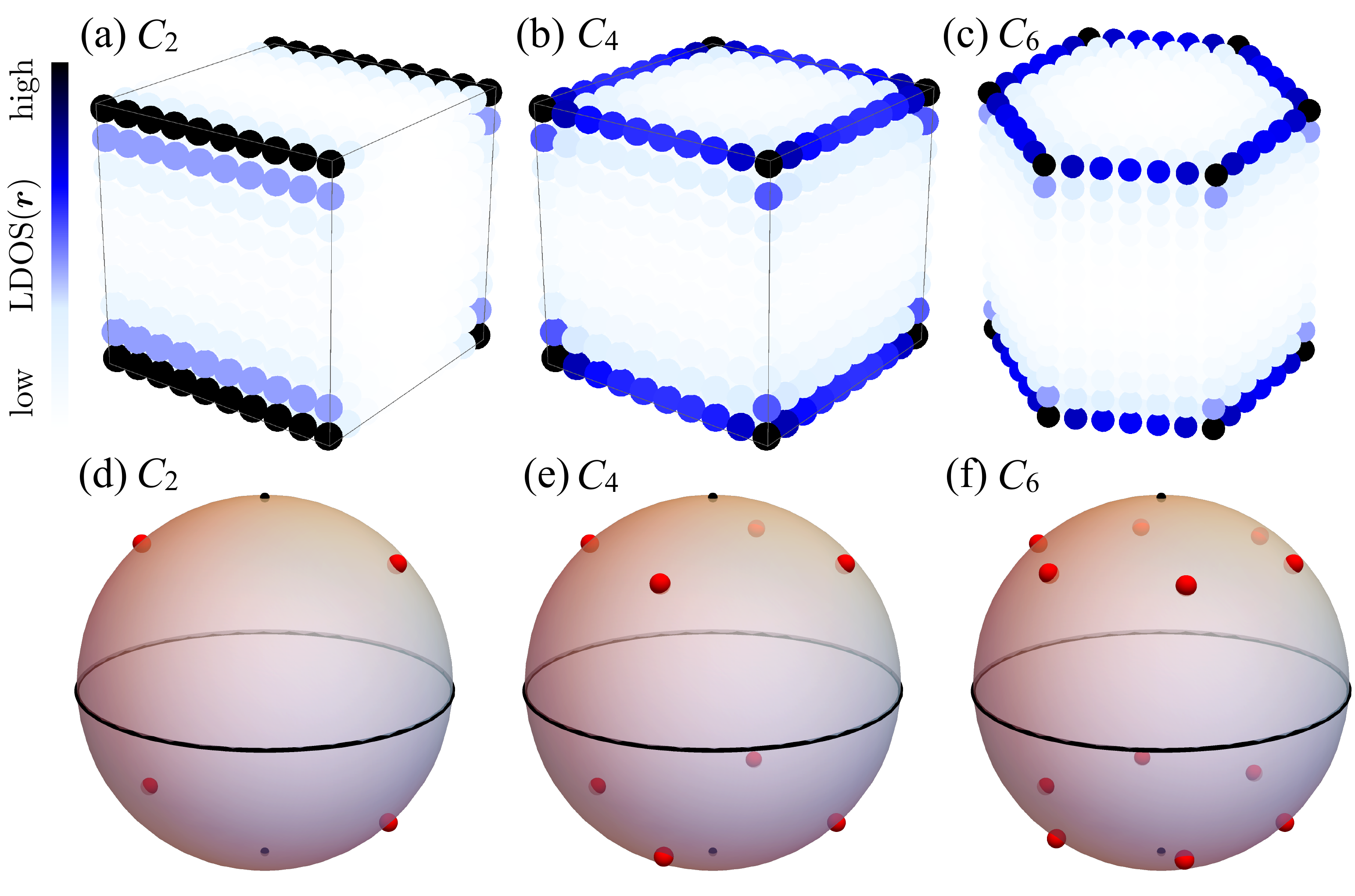}
    \caption{(a,b,c) local density of states at zero energy for the surface obstructed insulators constructed by stacking the tight binding models from Fig. \ref{fig:2Dmodel_246} with (a) $C_{2z}$, (b) $C_{4z},$ or (c) $C_{6z}$ symmetry. We can clearly notice the existence of hinge-localized states associated with hinge charges if we require the system to be symmetric and gapped, or alternatively a partially filled hinge band if we require the system to be symmetric and neutral; both phenomena are a result of the filling anomaly. (d,e,f): illustration of the 0D surface states associated with the filling anomaly for a spherical surface termination. Such states can only be annihilated at the high symmetry regions at the equator or north/south poles leading to a gap-closing at these points.}
    \label{fig:3dnumerics}
\end{figure}

\subsubsection{Real space picture and filling anomaly}

We now explain the boundary obstruction in the models described in the previous subsection.
We can start with a similar picture to that of Sec.~\ref{sec:2DReal}. In this picture, we consider two configurations of Wyckoff positions which are smoothly deformable in the bulk, but not in the presence of a surface. The precise meaning of this procedure follows from the machinery of Sec.~\ref{sec:GenDef} wherein the deformation trajectory in the periodic system induces a corresponding trajectory in the open system via the boundary map $\sigma$. By tracing the movement of the charge centers, we can then show that such a trajectory cannot be both symmetric and smooth. We consider the real-space deformation between two distinct configurations of the electrons: one where all electrons are at the center of the unit cell $1a$ (c.f. Fig. \ref{fig:phasediag3dmodels}, red $\Q$) in the $z=0$ plane, and the other with two electrons at the edge of the unit cell  $1c$, $2c$ or $3c$ ($nc$ for shorthand) in the three different models (c.f. Fig. \ref{fig:phasediag3dmodels}, purple $\Q$) in the $z = 1/2$ plane. This can be done while preserving the symmetry by first moving the $2n$ electrons at $1a$ horizontally in the $z=0$ plane into $nc$ as shown in Fig.~\ref{fig:2Dmodel_246}; and then moving the $2$ electrons at this site vertically in opposite directions. We could also start by first moving $n$ electrons vertically in each direction, then moving them horizontally to the position $nc$. 

Now we can see that the first trajectory (which involves moving horizontally then vertically) will be obstructed in the presence of side surfaces at the unit cell edge. The reason is that at such side surfaces, the filling of the $nc$ position is 1, and thus we cannot further move the electrons vertically while preserving the mirror symmetry in the $z$ direction. Similarly, the second trajectory involving vertical then horizontal movement is not possible in the presence of the top/bottom surfaces at the $z = 1/2$ plane. The reason is that the filling of the position $1a$ at this surface is $n$ (rather than $2n$ which is its bulk filling). This does not allow us to symmetrically move the electrons in the plane to position $nc$ without breaking the symmetry. This discussion parallels the one of Sec.~\ref{sec:2DReal} for the DMQI, and it shows that in the presence of a surface termination consistent with the unit cell shown in Fig.~\ref{fig:2Dmodel_246}, it is not possible to connect the atomic configuration where the $2n$ electrons are at $1a$ in the $z=0$ plane to the one where $2$ electrons are at each position $nc$ at the $z=1/2$ plane, despite the fact that these two atomic configurations are symmetrically deformable to each other in the bulk (with periodic boundaries). We also note that, similar to the DMQI case, the possibility of interpolating the two atomic configurations in the bulk imposes some constraints on the commutation properties of the different symmetries, i.e.,
\begin{equation}
    \{C_{2z},M_z\}=0.\label{C2M},
\end{equation}
which is shown to be a necessary condition for the smooth deformation between these different bulk trajectories in Appendix \ref{app:mirrors}. 

Similar to the DMQI, the 3D models introduced in this section are chatacterized by filling anomalies which are not associated with bulk topology. They can be understood by employing a similar argument to that of Sec.~\ref{sec:filling_anomaly}. Let us start with the $n=1$ model corresponding to 3D space group 10 whose point group is $C_{2h}$. Taken on an open symmetric boundary, let us consider the filling of the low-dimensional symmetry-invariant region $\Gamma$ given by the union of the $xy$ plane and the $z$-axis. This region consists of the disjoint union of three regions: (i) the origin $\O$, (ii) the $z$-axis excluding the origin (described by the line $x = y = 0$), and (iii) the $xy$-plane excluding the origin. The filling of (ii) and (iii) is necessarily even due to the action of $M_z$ and $C_{2z}$, respectively. The filling of (i) is even due to the anticommutation of $M_z$ and $C_{2z}$ which enforces the (projective) symmetry representation to be two-dimensional. This two dimensional representation is neither spinless or spinfull, and we will refer to it in Table \ref{tab:WannierBandsWyckoffPositions} as $\bar E$ is analogy to the DMQI model, however this is slight abuse of notation since this projective representation is not tabulated or equivalent to a purely spinless or spinfull representation \cite{aroyo2006bilbao}. As a result, the total filling of $\Gamma$ is even. On the other hand, adding or removing electrons from outside this region can only change $\nu_\Gamma$ by 4 corresponding to the four symmetry-related locations $(x,y,z)$, $(-x,-y,z)$, $(x,y,-z)$, and $(-x,-y,-z)$ (that is to say that the general Wyckoff position in the open system has multiplicity 4). Thus, the parity of $\nu_\Gamma/2$ describes a $\Z_2$ invariant which cannot be changed without changing the filling of the high-symmetry region $\Gamma$. This is only possible by closing the gap at a high symmetry surface (or in the bulk). As in the DMQI case, the discrepancy between this number and the total filling indicates a filling anomaly described by the $\Z_2$ invariant.

Such an anomaly can be resolved by identifying opposite side faces of the sample. This introduces new $C_{2z}$-invariant two-fold rotation axes whose filling is even, but can take a value of 0 or 2 modulo 4, which makes it possible for it to compensate for the discrepancy between the total filling and $\nu_\Gamma$ modulo 4. The orbitals in these new $C_{2z}$-invariant lines can be symmetrically brought to $\Gamma$ without closing the bulk gap or breaking the symmetry, thereby changing the value of $\nu_\Gamma$ and removing the bulk signature of the filling anomaly observed in the open system. This indicates that such filling anomaly is associated with a BOTP rather than a bulk phase. Similar analysis can be carried for the models with point groups $C_{4h}$ and $C_{6h}$.

Another approach to filling anomalies is the one introduced in Sec.~\ref{sec:2DFA}. In this case, we choose a foldable boundary termination as shown in Fig.~\ref{fig:3dnumerics}a,b,c which resembles a symmetric large cell for the periodic system. Recall the criterion derived in Sec.~\ref{sec:2DFA} where a filling anomaly is realized whenever the filling of a position $\bq$ on the periodic system $\nu^{\rm pbc}_\bq$ is not divisible by $n_\bq$, defined as the ratio of the order of its site symmetry group in the periodic and the open system (Eq.~\ref{eq:nq}). For all the $C_{2nh}$ models discussed above, the bulk filling for the $nc$ position at $z=1/2$ in the obstructed ``phase" is $\nu^{\rm pbc}_\bq = 2$, and its symmetry group on the torus is $G_\bq = C_{2h}$ (generated by $C_{2z}$ and $M_z$) which has four elements, $|G_\bq| = 4$. For the surface termination where the vertical surfaces are at the edges of the unit cell (shown in Fig.~\ref{fig:2Dmodel_246}a,b,c), and the horizontal surfaces are at $z=1/2$, the sites $\bq$ lying at the horizontal hinges (see Fig.~\ref{fig:3dnumerics}a,b,c) have no symmetry left. Thus $|F_\bq| = 1$ for these points which implies that the condition of Eq.~\ref{eq:BOTPFA} is satisfied.

Finally, let us discuss the observable consequences of the filling anomaly. In the DMQI, the filling anomaly manifests itself in the existence of fractional corner charge. Here, we similarly expect fractional charges placed at $4n$ symmetry related points at the corners of a generic boundary termination. These corner charges can be more easily undersood by first imposing chiral symmetry or particle-hole symmetry and noting that we can associate the anomaly with $2n$ 0D surface states at generic symmetry related points on an arbitrary symmetric surface as shown in Fig.~\ref{fig:3dnumerics}(d,e,f). Such 0D states can only be annihilated at the equator or the north/south poles which are HSSs. Following the discussion of Sec.~\ref{sec:filling_anomaly}, we can show that the filling anomaly will survive even after breaking chiral symmetry. In this case, it will appear as excess charge fractionally distributed among these symmetry related points for the symmetric gapped insulator. We notice that on highly symmetric foldable terminations such as the ones shown in Fig.~\ref{fig:3dnumerics}, the edges have an approximate translation symmetry. As a result, the filling anomaly for a symmetric insulator implies the existence of fractional charges per unit cell along the hinge for a symmetric gapped insulator, i.e., a hinge charge. Alternatively, we can impose the condition of charge neutrality (and unbroken symmetry) to deduce that the filling anomaly gives rise to partially filled bands corresponding to gapless hinges in these models.

\FloatBarrier

\subsubsection{Wannier spectrum}

We now turn our attention to the evolution of the Wannier bands and how they encode the boundary obstructions described in the previous subsection. We diagnose the boundary obstruction by analyzing the Wannier spectrum and choosing a lattice termination that coincides with the edges of the unit cell. That is, the WCP is given by $\mu=(1/2, 1/2, 1/2)$, which means that only gap closings of the Wannier spectrum at these values correspond to actual boundary gap closings. A phase diagram of the models is shown for reference in Fig. \ref{fig:phasediag3dmodels}.

\emph{$C_{2h}$ model}---  A gapped Wannier spectrum along $\hat z$ is possible provided $\H^{C_{2z}}_{\rm 2D}$ is in a gapped phase. This only happens when $2\lambda_M > \lambda_S$ (blue regions in  Fig. \ref{fig:phasediag3dmodels}), where we defined the parameters $\lambda_S = \lambda_x + \lambda_y + \lambda_0$ and $\lambda_M = \max(\lambda_x, \lambda_y, \lambda_0)$. On the other hand, the Wannier spectrum along $\hat x$ or $\hat y$ is gapped as long as $|\gamma_z| \neq |\lambda_z|$  (yellow plane in  Fig. \ref{fig:phasediag3dmodels}). The bulk Hamiltonian is gapped if either condition is satisfied, and is therefore only gapless in the intersections of the blue region and yellow  plane in the phase diagram. The region of intersection is marked in red. It is evident that the Wannier gap may close while preserving the bulk gap, a necessary condition for a BOTP.
We show the Wannier spectra $\nu_z(k_x, k_y)$ and $\nu_x(k_z, k_y)$ in Figs.~\ref{fig:Wz}a and \ref{fig:Wx}a. In these figures, we can see  six distinct regions with gapped Wannier spectra separated by Wannier gap closings, as will be explained in detail below. 

Similar to the DMQI case, the different gapped phases of the $C_{2h}$ model are characterized by distinct Wannier band representations, which can be reflected in distinct values of the nested Wannier band polarizations. 
Let us first notice that the model has four quantized Wannier band polarizations: $p_{z,x}$ and $p_{z,y}$ are quantized due to $C_{2z}$ symmetry, and $p_{x,z}$ and $p_{y,z}$ are quantized by $M_z$ symmetry. In our model, $p_{z,x}$ and $p_{z,y}$ cannot be simultanuously non-zero, and $p_{x,z}$ and $p_{y,z}$ are equal, leading to six distinct gapped phases (see Fig. \ref{fig:phasediag3dmodels}). Let us label the different phases by the maximal Wyckoff position $\Q$ around which the Wannier centers are located, as shown in Fig. \ref{fig:phasediag3dmodels}b.  While the electrons may be symmetrically displaced from this maximal Wyckoff position in some cases, they remain centered around $\Q$, and both their band representations, and Wannier band representations, are topologically equivalent to those of the phase where the electrons lie exactly at $\Q$. Therefore it is useful to label the phases by such maximal Wyckoff positions.  
We can use the notation $(p_{z,x},p_{z,y},p_{x,z},p_{y,z})_\Q$, to label the six distinct phases, which are given by: $(0,0,0,0)_{1a}$, $(0,1/2,0,0)_{1b}$, $(1/2,0,0,0)_{1d}$, $(0,0,1/2,1/2)_{1c}$, $(0,1/2,1/2,1/2)_{1g}$, $(1/2,0,1/2,1/2)_{1f}$. In our model we cannot tune parameters such that the Wannier centers lie at the $1e$ or $1h$ Wyckoff positions. 

The Wannier transitions between the different phases can be studied numerically leading to the phase diagram in Fig.~\ref{fig:phasediag3dmodels}a. As we can see, all of the pairwise Wannier transitions between the phases $1a$, $1b,$ and $1d$ are associated with a gap-closing of the Wannier spectrum in the $z$-direction at $\nu_z = 0$. On the other hand, the pairwise Wannier transitions between the phases $1c$, $1g,$ and $1f$ are associated with a gap-closing for the Wannier spectrum in the $z$-direction but at $\nu_z = 1/2$. This reflects the fact that transitions between $1a$, $1b,$ and $1d$ take place in the $z = 0$ plane, whereas transitions between $1c$, $1f,$ and $1g$ take place in the $z = 1/2$ plane. The Wannier transitions $1a \leftrightarrow 1c$, $1b \leftrightarrow 1f$ and $1g \leftrightarrow 1d$ are associated with a gap-closing in both Wannier spectra along the $x$ and $y$ directions occuring at $(\nu_x,\nu_y) = (0,0)$, $(0,1/2)$, and $(1/2,0)$, respectively.

Following the previous discussion, we can see that the phases $1f$ and $1g$ have a surface obstruction, which can be diagnosed by $p_{x,z} = p_{y,z} = 1/2$ and $(p_{z,x},p_{z,y}) = (0,1/2)$ or $(1/2,0)$, respectively. They are characterized by a quantized hinge charge of $1/2$ per unit cell, at the hinges along $ y$ or $ x$ respectively. A real space calculation of the low energy states is found in Fig.~\ref{fig:3dnumerics}(a,b,c). 
As an example, in Figs.~\ref{fig:Wz} and \ref{fig:Wx} we show how the two gapped phases, $1a$ and $1g$, which correspond to the parameters  $(\lambda_0, \lambda_x, \lambda_y, \gamma_z, \lambda_z) = (0.5, 0, 0.25, 0.5, 1)$ and $(\lambda_0, \lambda_x, \lambda_y, \gamma_z, \lambda_z) = (0.5, 1, 0.25, 1.5, 1)$ respectively, are separated by a Wannier gap closing transition at the WCP, and consequently a surface gap closing. We follow two distinct paths, either by first increasing $\lambda_x: 0 \rightarrow 1$, and  then increasing $\gamma_z: 0.5 \rightarrow 1.5,$ or vice versa. When increasing $\lambda_x$ first, the gap at the $z=1/2$ surface closes in the range of $0.25 \leq \lambda_x \leq 0.75$, (c.f. gapless $\nu_z(k_x,k_y)$ in Fig.~\ref{fig:Wz} a2 and a3).
Instead, if we increase $\gamma_z$ first we find the gap closing at the WCP at $\gamma_z = 1$, (c.f. gapless $\nu_x(k_y,k_z)$ in Fig.\ref{fig:Wx} b). 
Thus, a surface gap-closing at the $z = 1/2$ or $x=1/2$ surface is unavoidable whenever we connect the $1a$ and $1g$ phases.

\begin{figure}
    \centering
    \includegraphics[width=\columnwidth]{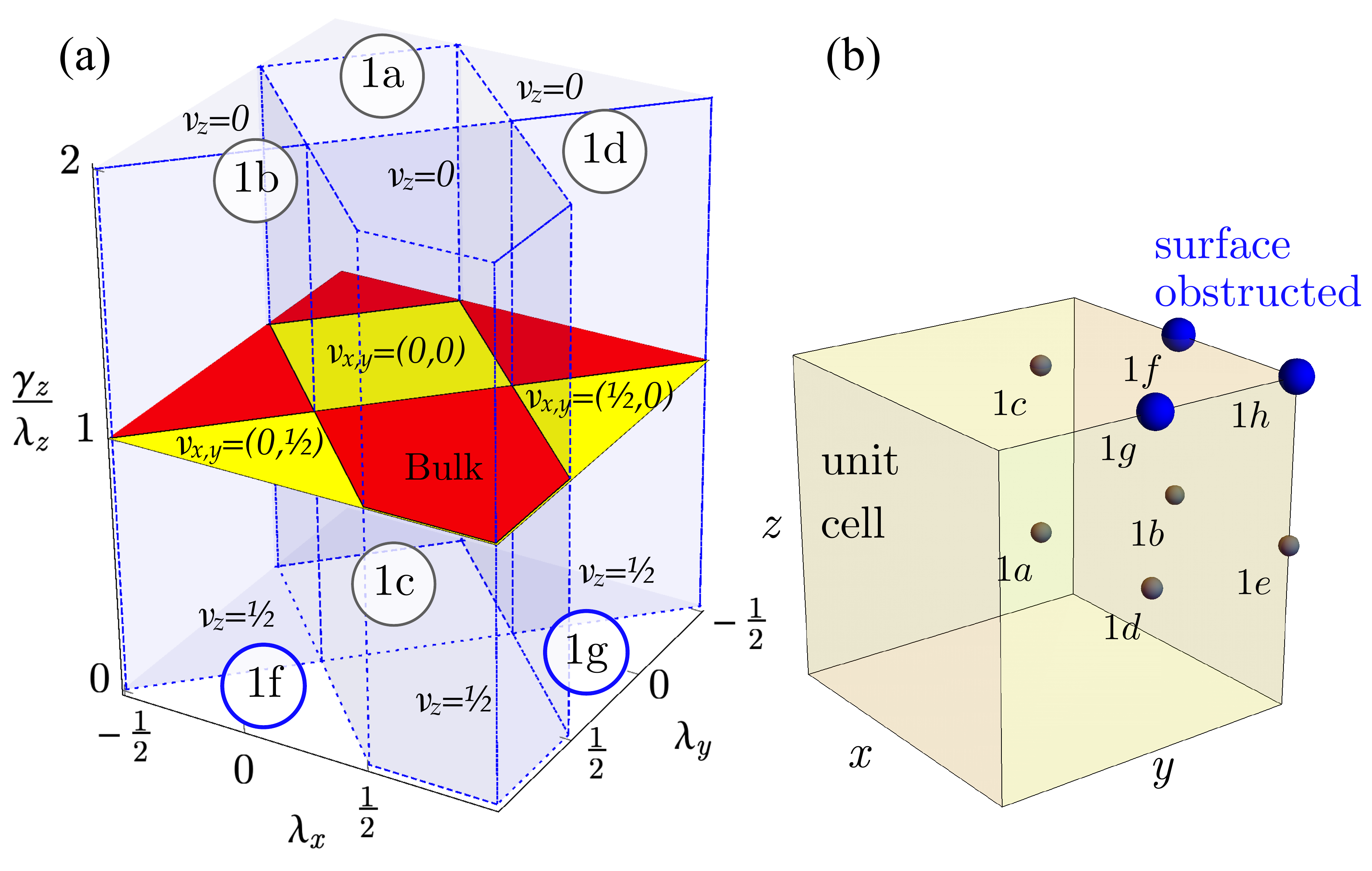}
    \caption{(a) Phase diagram for the $C_{2h}$ model introduced in Sec. \ref{sec:c2nhmodels} with a fixed value of $\lambda_0=1/2$. The Wannier spectrum $\W^z$ closes at the blue
    regions which span the $\gamma_z/\lambda_z$ axis in the areas marked with $\W^z$; while the Wannier spectrum $\W^x$ closes at the yellow  plane at $\gamma_z/\lambda_z=1$. Due to the constraints in the model, a gap closing in $\W^x$ implies a gap closing in $\W^y$. In the intersection between the two, the red regions, the bulk is gapless
    The gapped phases with gapped Wannier spectra are marked by the representative maximal $\Q$ from which the Wannier band representations may be induced. (b) Spatial location of the maximal $\Q$'s compatible with the $C_{2h}$ point group. With the conventional choice of Wannier chemical potential $\mu=(1/2, 1/2, 1/2)$, phases induced from $1f,~1g$, and $1h$ may be surface obstructed; from $1h$ they could also be hinge obstructed. In our model we realize $1f$ and $1g$ surface obstructions. 
    }
    \label{fig:phasediag3dmodels}
\end{figure}

 \begin{table}[h!]
\centering
\begin{tabular}{c|cccc|cc|c}
$\bk$ & $\bf \Gamma$ & $\bf X$ & $\bf Y$ & $\bf M$ & $\bf \Gamma$ & $\bf Z$  \\
\hline
$\rho$ & $w^{z}_+(C_{2z})$ & $w^{z}_+$ & $w^{z}_+$ & $w^{z}_+$ & $w^{x}_+(M_z)$ & $w^{x}_+$ &$\nu_\Gamma$  \\
\hline
$(1a,\bar E)$ &$+$ & $+$ & $+$ & $+$ & $+$ & $+$ &$2$\\
$(1b,\bar E)$ &$+$ & $+$ & $-$ & $-$ & $+$ & $+$&$2$\\
$(1c,\bar E)$ &$+$ & $+$ & $+$ & $+$ & $+$ & $-$&$2$\\
$(1d,\bar E)$ &$+$ & $-$ & $+$ & $-$ & $+$ & $+$&$2$\\
$(\vec{1f},\bar E)$ &$\vec +$ & $\vec+$ & $\vec-$ & $\vec-$ & $\vec+$ & $\vec-$&$\vec 0$\\
$(\vec{1g},\bar E)$ &$\vec+$ & $\vec-$ & $\vec+$ & $\vec-$ & $\vec+$ & $\vec-$& $\vec 0$\\
\hline
$(1e,\bar E)$ &$+$ & $-$ & $-$ & $+$ & $+$ & $+$&$2$ \\
$(\vec{1h},\bar E)$ &$\vec+$ & $\vec-$ & $\vec-$ & $\vec+$ & $\vec+$ & $\vec-$&$\vec 0$
\end{tabular}
\caption{ Upper section of rows: Diagnosis of boundary obstructions through Wannier band representations in the $C_{2h}$ model introduced in Sec. \ref{sec:c2nhmodels}.  The left column indicates the maximal Wyckoff position $\Q$ that labels the phase with periodic boundaries. In the columns to the right we show the symmetry representations at high symmetry momenta of one of the two WBRs. They are evaluated for $C_{2z}$ and $M_z$ according to which symmetry is preserved in the respective direction. 
Lower section of rows: Other possible WBR configurations compatible with the $C_{2h}$ point group, but not realized by our model. Throughout the table, the bold lines correspond to the phases with a boundary obstruction when the conventional Wannier chemical potential is chosen, i.e., $\mu=(1/2,1/2,1/2)$ and the center of the sample is located at the position $1a$. The obstruction in these cases is of the filling anomaly type, where the multiplicity of the sites $1g$, $1f$ and $1h$ is larger in the open boundary system compared to the periodic system,  see Fig.~\ref{fig:phasediag3dmodels}b. We see that for this model the obstructed phases are uniquely indicated by the symmetry eigenvalues of the Wannier bands.  Here we use $\bar E$ in analogy to the DMQI model, however this projective representation is not tabulated in the usual symmetry representation of the three-dimensional point group $C_{2h}$. The definition of this representation is discussed in the main text. }
\label{tab:WannierBandsWyckoffPositionsCn}
\end{table}

\begin{figure*}
    \centering
    \includegraphics[width=2\columnwidth]{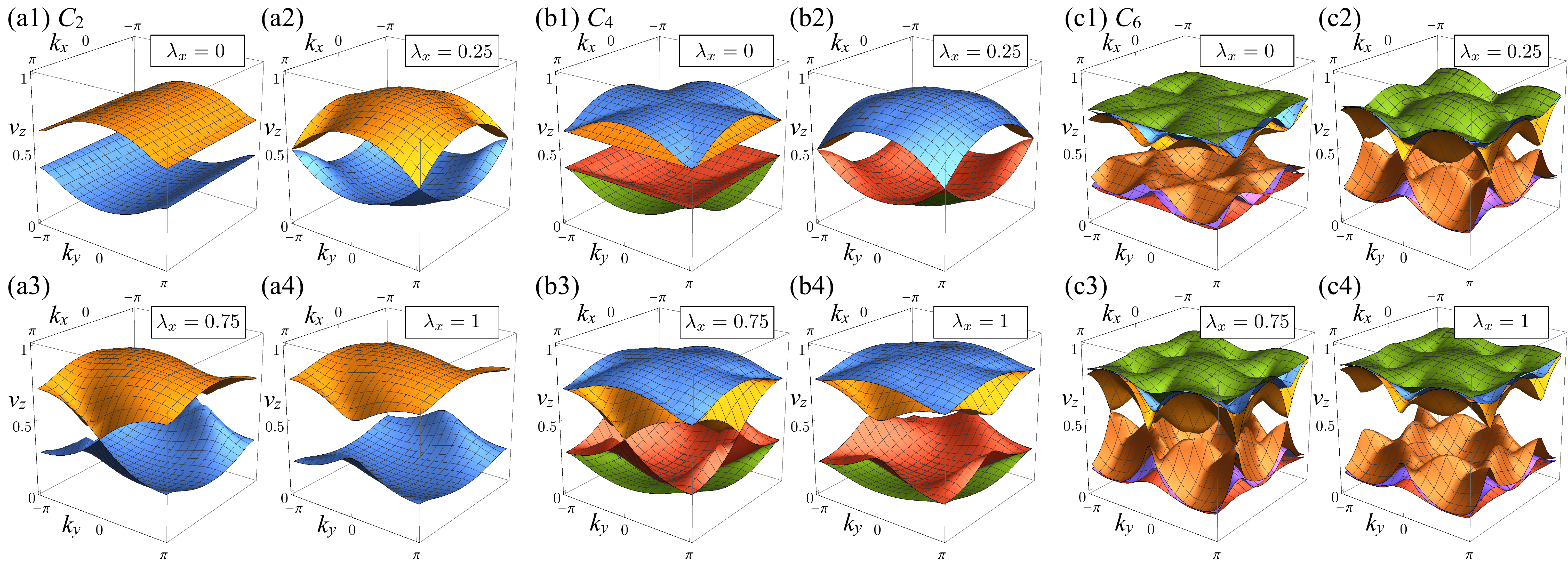}
    \caption{Evolution of the Wannier spectrum $\W^z$ with eigenvalues $\nu_z(k_x,k_y)$ for different values of $\lambda_x$ in the three-dimensional boundary obstructed models with filling anomalies and $C_{2h}$, $C_{4h}$ and $C_{6h}$ symmetries. The other parameters are kept fixed at the values of $\lambda_0=.5$ and $\lambda_y=.25$ with $\lambda_z=1.5$ and $\gamma_z=1$.  With gapped $\nu_x$ (see text), we find that varying $\lambda_x$ allows for two gapped phases bordered by an extended region where $\nu_z$ is gapless. This happens when $.25<\lambda_x<.75$ where the condition $2\max(\lambda_0,\lambda_x,\lambda_y)<(\lambda_0+\lambda_x+\lambda_y)$ is satisfied.  Panels (a1) and (a4) correspond to the gapped phases $1f$ and $1g$ in the phase diagram of Fig.\ref{fig:phasediag3dmodels}, respectively.}
    \label{fig:Wz}
\end{figure*}

Finally, let us comment on the Wannier band representations. The Wannier spectrum of $\W^z$, corresponds to the upper and lower surfaces (which preserve $C_{2}$), while the Wannier spectrum of $\W^{x/y}$ correspond to the side surfaces (which preserve $M_z$). The WBRs corresponding to $w^{z}$ and $w^{x/y}$ are split into two disconnected WBRs when $2\lambda_M > \lambda_S$ or $|\gamma_z| \neq |\lambda_z|$, respectively. The values that the WBRs take at high symmetry momenta may be used to determine the topological phases {since in this case they can be uniquely associated with a representation $\pi$ of the open boundary system. The symmetry eigenvalues for the WBRs are shown in Table.~\ref{tab:WannierBandsWyckoffPositions}}. To determine these symmetry indicators we induce the bulk BR's from the maximal $\Q$'s and subduce them to the subgroup maintained by  the respective Wannier spectrum, as described in Sec. \ref{sec:bandreps}. {The subduction of the bulk BRs to the point group $F$ for a fully open system (assuming the sample center to be at $\O=1a$) are also computed to directly associate the WBRs to the obstructed or trivial phases under this choice of boundary. The table shows exhaustively the WBRs compatible with this band representation, therefore we can observe that in this case the obstructed phases are symmetry indicated from the symmetry eigenvalues of the Wannier bands.}  As anticipated earlier, we can see that the $C_{2z}$ eigenvalues for $1a/1c$, $1b/1f,$ and $1d/1g$ correspond to a 2D inversion symmetric insulator with a single electron localized at $(0,0)$, $(0,1/2),$ and $(1/2,0)$, respectively. Similarly, The $M_z$ eigenvalues correspond to a 2D mirror-symmetric system where the electron is localized at the $0$ or $1/2$ mirror lines for $1a/1b/1d$ or $1c/1f/1g$, respectively.

\begin{figure}
    \centering
    \includegraphics[width=\columnwidth]{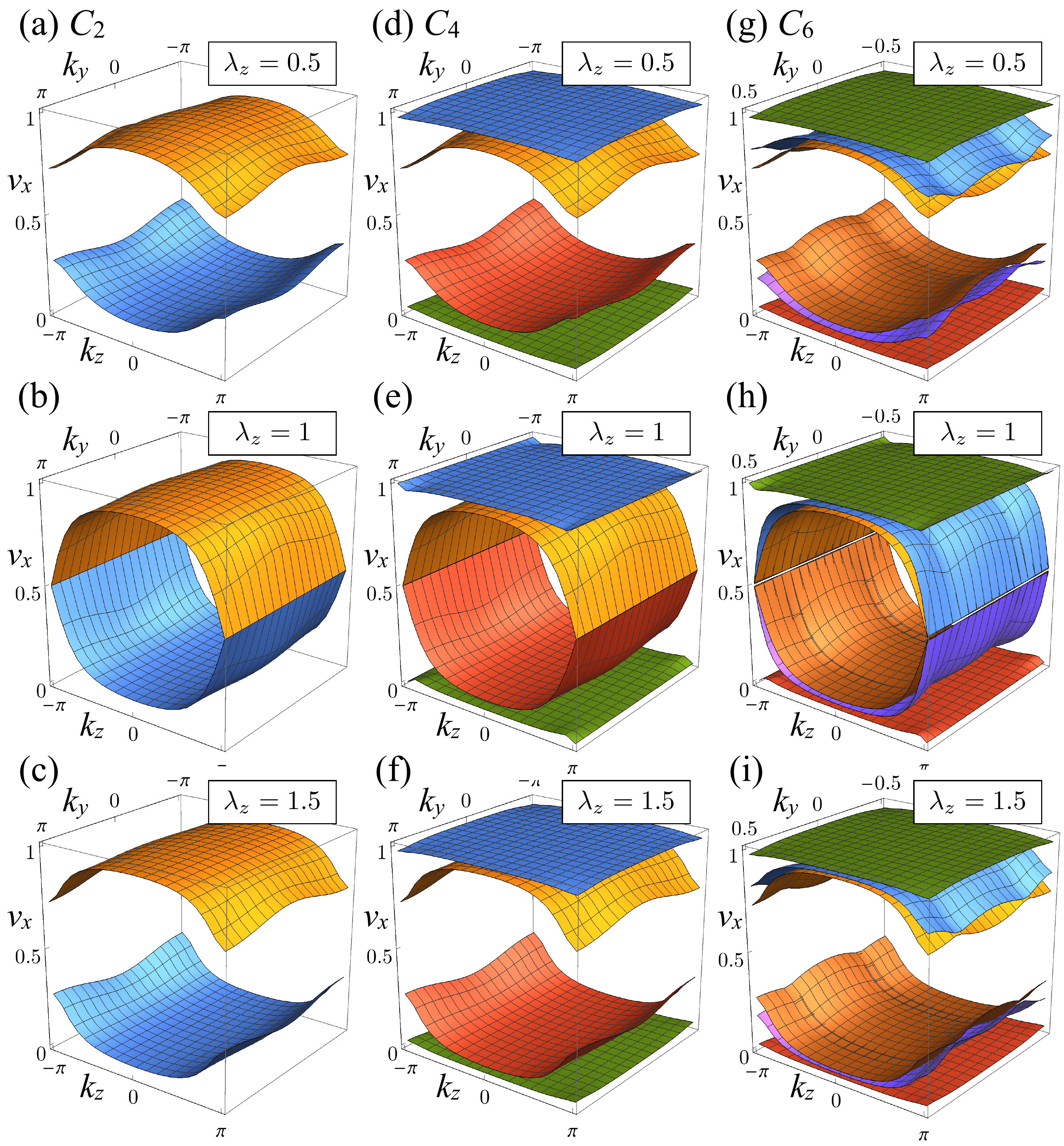}
    \caption{Evolution of the Wannier spectrum $\W^x$ with eigenvalues $\nu_x(k_y,k_z)$ for different values of $\lambda_z$ and fixed $\lambda_0=.5$, $\lambda_x=1$,  $\lambda_y=.25$, and $\lambda_z=1$. Across these that $\nu_z$ is gapped, we find the gappless transition of $\nu_x$ to happen when $\lambda_z=1$.  Panels (a) and (c) correspond to the gapped phases $1g$ and $1d$ in the phase diagram of Fig. \ref{fig:phasediag3dmodels}, respectively.}
    \label{fig:Wx}
\end{figure}

\emph{$C_{4h}$ model}--- We now look at the 8 band model with $C_{4h}$ symmetry. The model differs from the $C_{2h}$ model in two main aspects. First, $C_{4z}$ symmetry dictates that the Wannier spectra $\W^x$ and $\W^y$ are identical, hence the nested polarizations $p_{z,x}$ and $p_{z,y}$ are equal. This means the model has only \emph{one} non-trivial phase, and this phase has hinge charge along both the $x$ and $y$ directed hinges. In this phase, the occupied electrons are distributed around the $1f$ and $1g$ positions in Fig. \ref{fig:phasediag3dmodels}b. Second, with four occupied electrons per unit cell, the Wannier bands allow for more complicated gap-closing patterns, as we will see below.

The analysis of the $C_{4h}$ model parallels the $C_{2h}$ model above. The model has four distinct regions of gapped Wannier spectra characterized by the values of the two distinct nested polarizations $(p_{z,x}, p_{x,z})$ equal to  $(0,0)_{1a1a}$, $(0,1/2)_{1c1c}$, $(1/2, 0)_{1b1d},$ and $(1/2, 1/2)_{1f1g},$ and separated by Wannier gap-closings. {Here the sequence of two Wyckoff position labels indicates one orbital in one position and another orbital in the second position}. The transitions $1a1a \leftrightarrow 1c1c$ 
and $1b1d \leftrightarrow 1f1g$ are characterized by Wannier gap-closings at $\nu_x = \nu_y =0$ and $\nu_x = \nu_y = 1/2$, respectively, whereas the transitions  $1a1a \leftrightarrow 1b1d$ and $1c1c \leftrightarrow 1f1g$ are characterized by Wannier gap-closings at $\nu_z = 0$ and $\nu_z = 1/2$, respectively. As a result, the $1b1d$ phase is a BOTP that cannot be reached from any of the three other phases without closing a gap at the boundary for our chosen boundary termination (corresponding to a WCP $\mu_{x,y,z} = 1/2$).

This can be seen more directly by studying the evolution of the Wannier bands. The Wannier bands in the $z$-direction calculated via $\W^z$ are shown in Fig.~\ref{fig:Wz}b. The four bands are split into two groups of two bands. Akin to the $C_{2h}$ model, when $\lambda_M = \lambda_0 > \lambda_S/2$, the two filled bands are located around $1a$ with $p_{z,x} = p_{z,y} = 0$. On the other hand, when $\lambda_M = \lambda_{x,y} > \lambda_S/2$, the two filled bands describe one electron at $1f$ and one at $1g$, each having $p_{z,x} = p_{z,y} = 1/2$. In both cases, the Wannier bands form elementary bands and cannot be split further. In particular, this implies that the Wannier gap closing when $\lambda_M < \lambda_S/2$ involves all four bands. 

Focusing on $\W^x$, we can similarly investigate the transition between the two distinct phases by tuning $\gamma_z$. Here, we find a gap closing transition when $|\gamma_z| = |\lambda_z|$ (c.f., Fig.~\ref{fig:Wx}). Unlike $\W^z$, the bands are split into four separate bands (four EBRs) since the symmetries of $\W^x$ do not relate them. This follows from the fact that, with only mirror symmetry, all non-general Wyckoff positions have multiplicity 1. This  implies that all EBRs that are not derived from the general position are one-dimensional. As a result, the gap closing at the WCP takes place only between two of the four Wannier bands. This implies that we can identify the different phases by looking at the (nested) Wannier band polarization of a single band. The phase with hinge charge is characterized by $p_{x,z}=1/2$. 

\emph{$C_{6h}$ model}---We can define a model with $C_{6h}$ symmetry in a very similar fashion. The 2D Hamiltonian used to construct this model (via Eq. (\ref{H3D})) is schematically illustrated in the third panel of Fig.~\ref{fig:2Dmodel_246}. The behavior of the model is qualitatively similar to the models above. Namely, there are two phases distinguished by a surface obstruction as shown in Figs.~\ref{fig:Wz}c and ~\ref{fig:Wx}c. The non-trivial phase is characterized by a fractional hinge charge per unit cell (c.f. Fig.~\ref{fig:3dnumerics}c).

\subsubsection{A hinge-obstructed 3D BOTP}

One of the simplest 3D BOTPs we can construct is a trivial dimensional extension of the DMQI. Following the recipe of Sec.~\ref{sec:Recipe}, we can create a 3D BOTP by stacking pairs DMQI's {with $\pi$ fluxes at each plaquette along the stacking direction}. Since the DMQI is edge-obstructed and has corner charges, the resulting 3D BOTP has a hinge obstruction associated with a filling anomaly which translates to a corner charge of $e/2$ at each of the 8 corners.

The filling anomaly can be understood by noting the 3D model has mirror symmetries about three perpendicular planes, i.e., $M_x, M_y,$ and $M_z$. When considered on a torus, i.e., with periodic boundary conditions, then there are 8 points $\bq$ with maximal symmetry $|G_\bq|=8$, where the three perpendicular mirror planes intersect. These points correspond to the 8 maximal Wyckoff positions of multiplicity 1. Whenever the filling of such points is a multiple of two, we can symmetrically move the electrons along the intersection of any two of the three mirror planes to reach any of the other symmetric positions. On the other hand, with open boundaries, the site $\bq$ that lies on a corner of the open system has its site symmetry group reduced such that $|F_{q}| = 1$. At a bulk filling of $\nu_\bq^{\rm pbc} = 4$ (which is natural for a pair of DMQI's), the sites at the corners will have a filling of $1/2$ signaling the existence of a BOTP filling anomaly. Such a filling anomaly is associated with a hinge rather than surface obstruction. Indeed, this model is the octupole insulator of Refs.~\onlinecite{Benalcazar17, Benalcazar17b}, which was the first example of a 3D BOTP {when protected by mirror symmetries (additionally, bulk-obstructions exist in this model when additional crystalline symmetries are considered)}. Since this model is hinge-obstructed, its boundary obstruction is encoded in the band representations of the \emph{nested} Wannier bands rather than the WBR themselves.

\subsection{3D BOTPs with gapless surface states}
\label{sec:3DSS}

In the previous section, we discussed models with boundary obstructions associated with filling anomalies. Here, we construct two models with gapless hinge states due to a surface obstruction: the dimerized weak Chern insulator and the dimerized weak quantum spin Hall (QSH) insulator. Both of these models are constructed using the recipe of Sec.~\ref{sec:Recipe}.

\subsubsection{Dimerized weak Chern insulator}

 \begin{table}[h!]
\centering
\begin{tabular}{c|cccc|c}
$\bk$ & $\bf \Gamma$ & $\bf X$ & $\bf Y$ & $\bf M$ &  \\
\hline
$\rho$ & $w^{z}_+(C_{2z})$ & $w^{z}_+$ & $w^{z}_+$ & $w^{z}_+$ & $nu_\Gamma$  \\
\hline
$(1a,\{{}^1\bar E,{}^2\bar E\})$ &$+$ & $+$ & $+$ & $-$  &$2$\\
$(\vec{1c},\{{}^1\bar E,{}^2\bar E\})$ &$\vec +$ & $\vec +$ & $\vec +$ & $\vec -$ & $\vec 0$\\
\hline
$(1a,\{{}^1\bar E,{}^2\bar E\})$ &$+$ & $+$ & $+$ & $+$  &$2$\\
$(1c,\{{}^1\bar E,{}^2\bar E\})$ &$+$ & $+$ & $+$ & $+$ &$2$\\
\end{tabular}
\caption{ Example of Wannier symmetry representations in the magnetic space group $P2/m'$ generated by the irreducible representation of the $2/m'$ point group $\{{}^1\bar E,{}^2\bar E\}$ at different maximal Wyckoff positions allowed through different parameter choices in the model of Eq.~\eqref{eq:dimchern}. In this case $M_z$ is not a symmetry and does not label the phase. Top rows: Phases with  These correspond to boundary obstructed phases with anomalous surface states, where the Wannier bands in $\W^z$ are characterized by a Chern number. The bold line corresponds to the phase with a boundary obstruction when the conventional Wannier chemical potential is chosen, i.e., $\mu=(1/2,1/2,1/2)$. Neither 
the positions $1a$ or $1c$ admit a filling anomaly in this space group (See Fig. \ref{fig:phasediag3dmodels}) however the bulk representation $\rho$ admits a restriction the Wannier bands which are \emph{topological} or in other words nonlocal. Such configuration makes the position $1c$ obstructed. The Wannier configuration on the top rows are characterized by a Chern number and the obstructed cases will host anomalous chiral modes at the boundary. Note that this phase, unlike the models with filling anomalies presented above, is not uniquely indicated by the symmetry eigenvalues of the Wannier bands, but it is still diagnosed by the representation of the open boundary system.
}
\label{tab:WannierBandsWyckoffPositionsChern}
\end{table}

\begin{figure}
    \centering
    \includegraphics[width=\columnwidth]{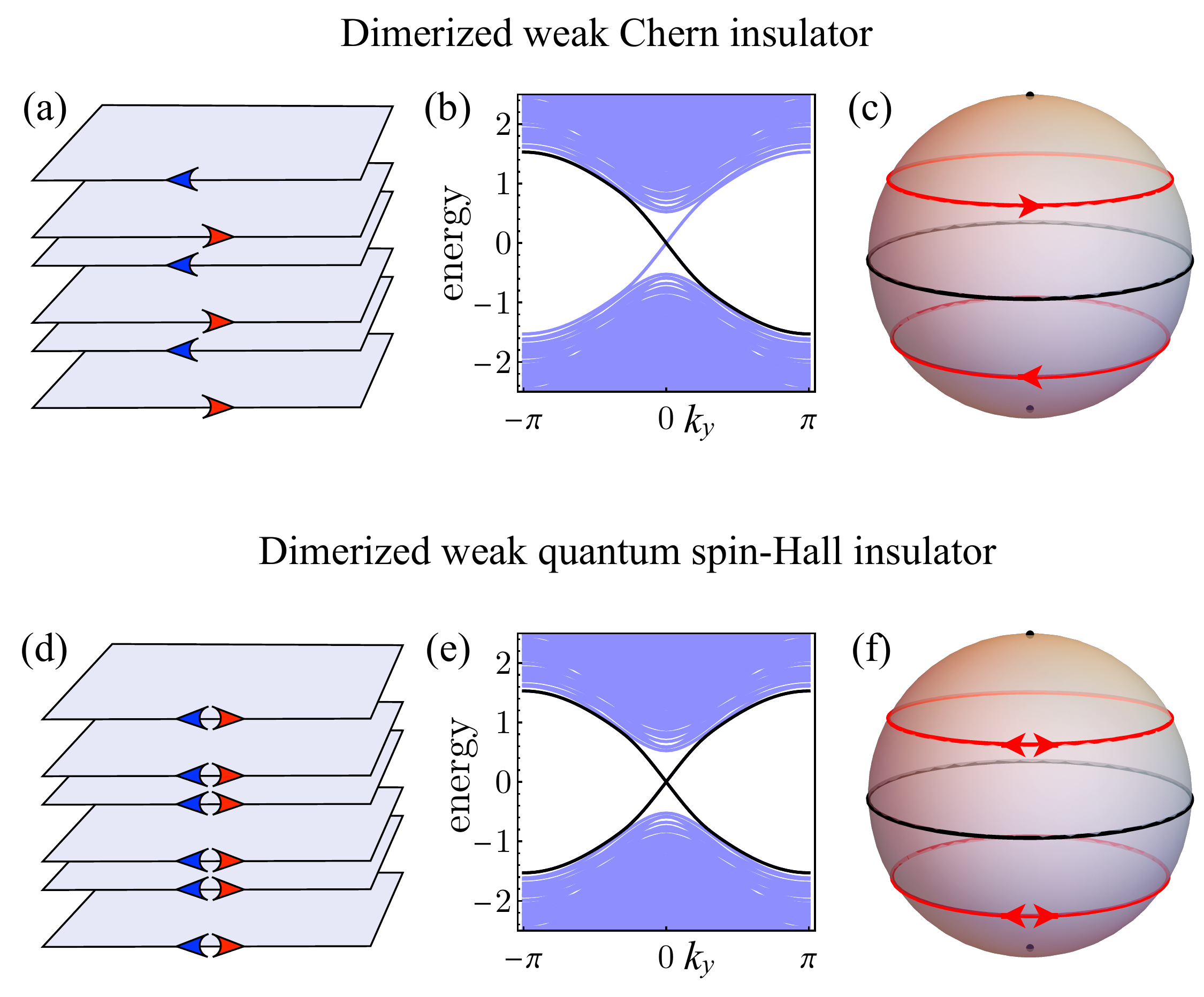}
    \caption{Three dimensional boundary obstructed phases without a Wannier representation: dimerized weak Chern insulator protected by $M_z\T$ and $C_{2z}$ (top panel), and the dimerized weak quantum spin Hall insulator  protected by $M_z$, $\T$, and $C_{2z}$ (bottom panel). Left: Scheme of the model where the arrows indicate chiral one-dimensional modes that can be gapped by a coupling along $z$. In the two nontrivial cases, either a 2D Chern insulator or a 2D quantum spin Hall insulator is left unpaired at the surface. Center: Energy spectrum of a system with translation symmetry along $y$ but open along both $x$ and $z$. The ingap states are located at the top and bottom hinges. We have marked in black the band that is exponentially localized at one hinge. Right: illustration of the 1D surface states on a spherical surface termination. The equator and north/south pole are invariant under $M_z$ and $C_{2z}$, respectively. The 1D surface states can only be symmetrically removed by closing the gaps at one of these high symmetry regions.
    }
    \label{fig:dimerized3dmodels}
\end{figure}

We start by considering a 2D Chern insulator whose Hamiltonian is given by 
 \begin{multline}
\H_{\rm CI,\pm}(\bk) = \pm[\sin k_x \sigma_1 + \sin k_y \sigma_2 \\ + (2 + m - \cos k_x - \cos k_y) \sigma_3].
\label{HCI}
\end{multline}
Let $m$ be restricted to the interval $-2 < m < 2$. In this case, the Hamiltonian $\H_{\rm CI,\pm}$ has Chern number $C = \mp 1$ for $m < 0$ (with the sign determined by the sign in Eq. \ref{HCI}), Chern number 0 for $m>0,$ and it is gapless for $m = 0$. 
The 3D dimerized Chern insulator is built by stacking copies of the 2D Chern insulator above with alternating sign which results in two layers per unit cell with alternating Chern number. Then we couple the layers with a dimerized coupling that alternates between two values $\lambda_z$ and $\gamma_z$. Physically, the terms multiplying $\sin k_{x,y}$ and $\cos k_{x,y}$ correspond to hopping between unit cells, so flipping their sign is equivalent to threading a  $\pi$-fluxes through all the vertical plaquettes. On the other hand, the term $2 + m$ denotes an onsite potential whose sign should be chosen to alternate between layers. The resulting 3D Hamiltonian is given in Eq.~\ref{H3D}. This Hamiltonian is invariant under the product $M_z \T = \sigma_1 \tau_1 \K,$ and $C_{2z} = i\sigma_3$. This set of symmetries places our model in the magnetic space group $P2/m'$(10.45), which has the point group $2/m'$ that admits a single spinful complex representation $\{{}^1\bar E,{}^2\bar E\}$ \cite{doublemagneticrep}.


The energy spectrum can be understood by noting that the 3D Hamiltonian has the form of a Dirac Hamiltonian
\begin{multline}
\H =  \sin k_x \Gamma_1 + \sin k_y \Gamma_2 + (2 + m - \cos k_x - \cos k_y) \Gamma_3 \\ +  \lambda_z \sin k_z \Gamma_4 + (\gamma_z + \lambda_z \cos k_z) \Gamma_5 \label{eq:dimchern}
\end{multline}
with $\Gamma_{1,2,3} = \sigma_{1,2,3} \tau_3,$ and $\Gamma_{4,5} = \tau_{2,1}$ such that $C_{2z} = \Gamma_1 \Gamma_2,$ and $M_z \T = \Gamma_1 \Gamma_4 \K$. It is then obvious that for this Hamiltonian to be gapped all 5 terms have to vanish simultaneously which is only possible for $m = 0$ \emph{and} $|\gamma_z/\lambda_z|=1$. This set of points where the Hamiltonian is gapped forms a line in the three dimensional parameter space which means that it is always possible to connect any set of parameters $(m, \gamma_z, \lambda_z)$ for which the Hamiltonian is gapped without going through a bulk gap closing. That is, the bulk phase is always trivial. We note that we can also understand the triviality of the model from the fact that none of the symmetries $C_{2z}$ and $M_z \T$ quantizes the $\theta$ angle in the axion term $\theta {\vec E} \cdot {\vec B}$ \cite{hughes2011, Wieder18, Venderbilt}.

In order to study this transition, we compute the Wannier spectra in Fig.~\ref{fig:WilsonDC}. We can see in the top panel that changing the sign of $m$ induces a gap-closing in the Wannier spectrum in the $z$-direction indicating gap-closing in the top/bottom surfaces. On the other hand, keeping the value of $m$ fixed and negative, we find a gap-closing transition in the Wannier spectra along the $x$ or $y$-directions (bottom panel). This means that for the chosen termination, the model exhibits a boundary obstruction separating the gapped phase with parameters $(|\gamma_z|>|\lambda_z|, m<0)$ from $(|\gamma_z|<|\lambda_z|, m \lessgtr 0)$ and $(|\gamma_z|\lessgtr|\lambda_z|, m>0)$.

An interesting aspect of this model is that the Wannier obstruction for $\W^z$ has a different origin from $\W^x$ and $\W^y$. In the Wannier spectrum of $\W^z$, the obstruction is associated with a non-zero Chern number, i.e., the Wannier bands do not separately form WBRs, as illustrated  by the non-trivial spectral flow of the \emph{nested} Wannier spectrum shown in Fig.~\ref{fig:W2}a and e. On the other hand, the Wannier obstruction in $\W^x$ or $\W^y$ is associated with changes in the nested Wannier band polarization between values of $0$ and $1/2$ as shown in Fig.~\ref{fig:W2}c, d, g, and h. {Unlike the filling anomaly scenario, the obstruction in this model is not due to the change in multiplicity of a Wyckoff position when going between $G$ and $F,$ but rather the fact that generating the open boundary representation from the site $1c$ is not compatible with a local representation, i.e., it cannot be represented as the symmetry action on localized atomic orbitals of the open system.  . This is shown in Table \ref{tab:WannierBandsWyckoffPositionsChern}. When the Chern layers are in the topological phase (upper rows), the two Wannier bands of the Wannier spectrum $\W^z$ form an irreducible representation of $G_z=\{C_{2z},T_{x},T_y\}$ when grouped together. However, the separate Wannier bands are topological and their eigenvalues are not compatible with an atomic insulator. If the Wannier bands are both located around the center of the unit cell ($1a$), then this configuration is not obstructed, since for each unit cell the two Wannier bands of opposite Chern number combine to form a band representation whose local support lies on a site with site symmetry group $F$ admitting the two dimensional irreducible representation. However, if the bulk band representation is generated from the position $1c$, with site symmetry group $C_2$, which admits only one dimensional representations, it is necessarily obstructed for a boundary where $1c$ lies on the boundary. Intuitively, the Chern bands of opposite Chern number centered around the edge of the unit cell are spatially separated leaving behind chiral modes exponentially localized to the hinges, which can either be removed by closing the gap of the Chern bands and inverting their mass (closing the gap on the poles, see Fig. \ref{fig:dimerized3dmodels}), or changing the coset decomposition of $G$ such that it can be generated from a Wyckoff position in the $z=0$ plane, resulting in the surface gap closing at the equator.}

 It is important to highlight the role played by the $C_{2z}$ symmetry here. Although it is not needed as a traditional protecting symmetry for the Wannier topology, it is crucial to single out the upper and lower surfaces as high symmetry planes where Wannier gap-closings imply a surface phase transition. Similar to the discussion of Sec.~\ref{sec:GenDef}, this can be illustrated by considering a spherical geometry as shown in Fig.~\ref{fig:dimerized3dmodels}c. 
Under $M_z \T$ alone, only the equator is a high symmetry region, and we are generally allowed to smoothly deform the chiral hinge modes by shrinking them to a point without touching the equator. In the presence of $C_{2z}$, however, the north and south poles become HSSs where surface gap-closings indicate a surface phase transition per our definition. Note that allowing for distinctions captured by gap-closings at generic, non-HSSs will generate the same spurious, uncountable boundary distinctions as those discussed in Sec.~\ref{sec:GenDef}. 

\begin{figure}
    \centering
    \includegraphics[width=\columnwidth]{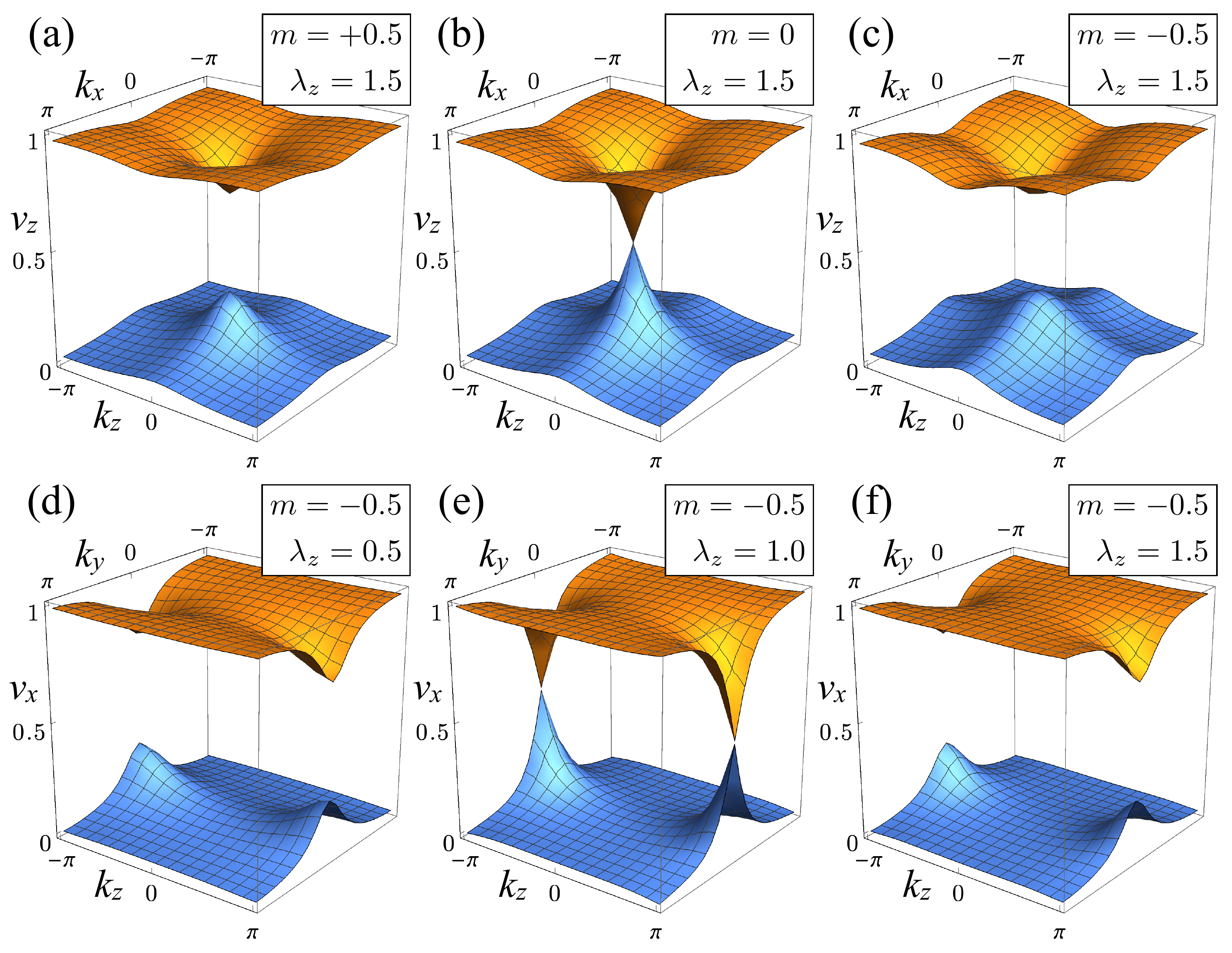}
    \caption{Evolution of the $\W^z$ (a-c) and $\W^x$ (d-f) Wannier spectra of the dimerized weak Chern insulator model for different values of $m$ and $\lambda_z$ with fixed $\gamma_z=1$. Entering the phase with $m=-0.5$ and $\lambda_z=1.5$ requires a Wannier gap closing at $\nu=1/2$ both when tuning $m$ or $\lambda_z$. This phase is therefore separated by a boundary phase transition for a boundary with the conventional Wannier chemical potential $\mu_{x,y,z}=1/2$.} 
    \label{fig:WilsonDC}
\end{figure}

\begin{figure}
    \centering
    \includegraphics[width=\columnwidth]{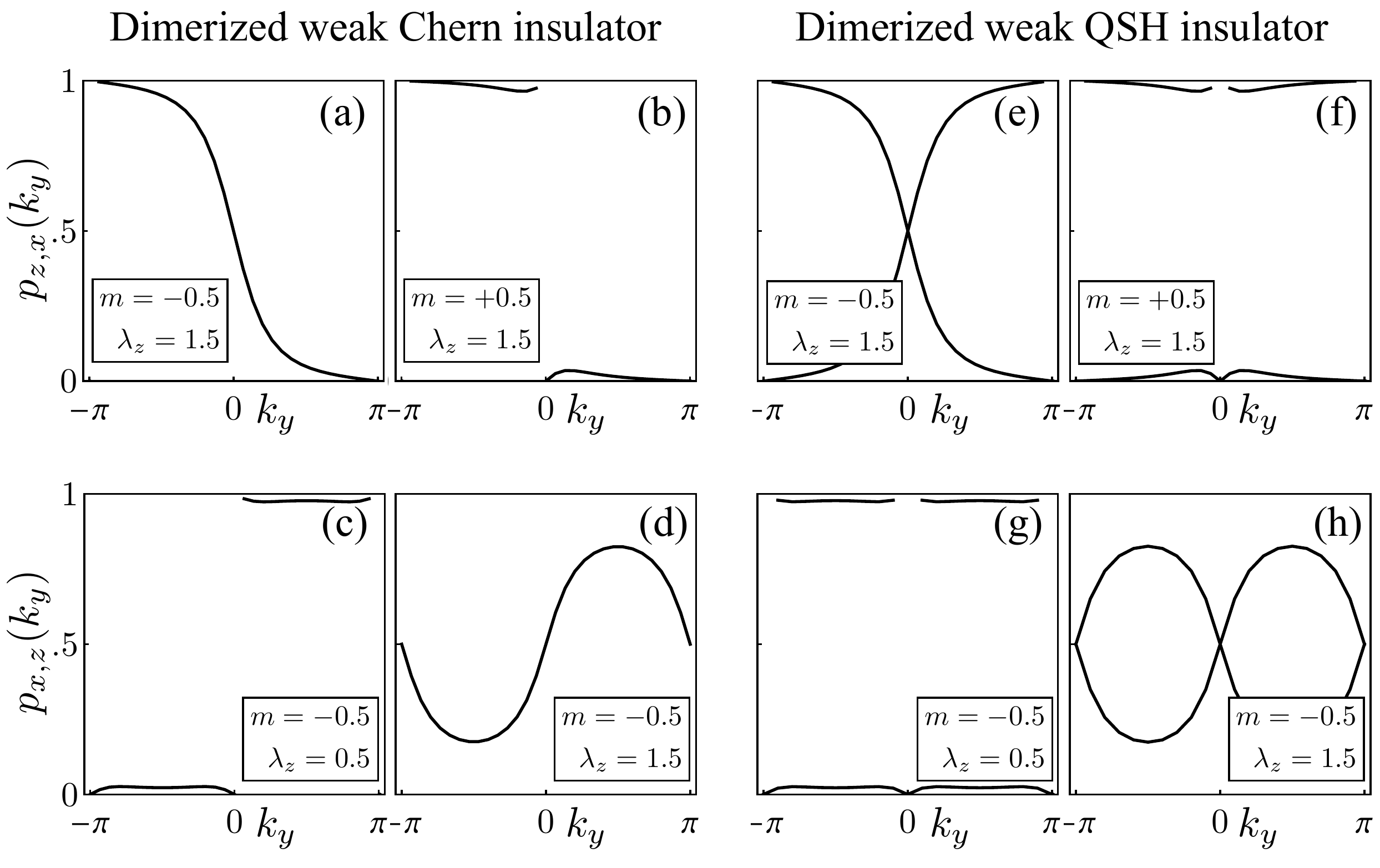}
    \caption{Nested Wannier polarization $p_{z,x}(k_y)$ and $p_{x,z}(k_y)$ for different values of $m$ and $\lambda_z$ with fixed $\gamma_z=1$ in the dimerized weak Chern insulator (a-d) and weak quantum spin-Hall insulator (e-h).  The phase with $m=-0.5$ and $\lambda_z=1.5$ (a,e) is characterized by a winding of the nested polarization which indicates that the Wannier bands in $\W^z$ do not form band representations and are associated with hinge modes. On the other hand, $p_{x,z}(k_y)$ is gapped with a different average value in the two phases, either $0$ (c,g) or $1/2$ (d,h), reflecting the fact that $\W^x$ decomposes into distinct Wannier band representations.}  
    \label{fig:W2}
\end{figure}

\subsubsection{Dimerized weak quantum spin Hall insulator}\label{sec:weakqsh}

Similar to the dimerized Chern insulator, we can define a 3D BOTP by stacking layers of the 2D quantum spin Hall (QSH) topological insulators with dimerized coupling. We start with the 2D Bernevig-Hughes-Zhang (BHZ) model \cite{Bernevig06} whose Hamiltonian is given by
\begin{multline}
\H_{\rm BHZ}(\bk) = \sin k_x \sigma_1 s_3 + \sin k_y \sigma_2 \\+ [2 + m - \cos k_x - \cos k_y] \sigma_3
\label{eq:HQSH}
\end{multline}
where $s_{1,2,3}$ denote the Pauli matrices in  spin space. We also take $|m| < 2$ such that the Hamiltonian describes a topological insulator for $m < 0$ and a trivial phase for $m > 0$. The Hamiltonian is invariant under spinful time-reversal symmetry $\T = i s_2 \K$. It also has $\rm U(1)$ $S_z$ rotation symmetry, but that is not required for its topological protection \cite{Kane05a, Kane05b}. The Hamiltonian is also invariant under 2D inversion ($\I_{\rm 2D} = \sigma_3$) which, when combined with $\pi$ $S_z$ rotation, yields the spinful two-fold rotation $C_{2z} = i \sigma_3 s_3$.

We can then build a 3D BOTP by stacking layers of the 2D Hamiltonian (\ref{eq:HQSH}) with \emph{alternating signs}. As explained earlier, this amounts to threading $\pi$ fluxes threaded in the vertical plaquettes and choosing the sign of the parameter $2 + m$ to alternate between layers. The resulting 3D Hamiltonian has the form (\ref{H3D}) with $\H_{\rm 2D} = \H_{\rm QSH}$. Due to the $\Z_2$ topology of the QSH, switching the sign of the Hamiltonian between layers does not alter the topological index, but it is needed to ensure that the bulk Hamiltonian describes a BOTP rather than a weak TI. This distinction will be encoded in the symmetry action as we will see now.  In addition to time-reversal and $C_{2z}$, the 3D Hamiltonian is also invariant under $M_z = i\tau_1 \sigma_1 s_1$. Similar to Secs.~\ref{sec:II_DMQI} and \ref{sec:3DFA}, we can think of the symmetry representations considered here as a projective representation which combines spatial symmetry transformations with internal and gauge transformations. We note that the 2D inversion symmetry $\I_{\rm 2D}$ anticommutes with $M_z$ as required by the presence of the $\pi$ fluxes in the vertical plaquettes, whereas the spinful $C_{2z}$ rotation commutes with $M_z$. Thus, the projective representation of the symmetry group generated by $C_{2z}$ and $M_z$ (together with translations) can be thought of as a non-projective spinful (double) representation of the space group 10 (P2/m) (if we allow for perturbations which break $S_z$ spin rotation but not $C_{2z}$ or $M_z$).

As in the dimerized weak Chern insulator, the Hamiltonian has the form of a Dirac Hamiltonian
\begin{multline}
\H =  \sin k_x \Gamma_1 + \sin k_y \Gamma_2 + (2 + m - \cos k_x - \cos k_y) \Gamma_3 \\ +  \lambda_z \sin k_z \Gamma_4 + (\gamma_z + \lambda_z \cos k_z) \Gamma_5 \label{eq:dimchern}
\end{multline}
with $\Gamma_1 = \sigma_1 s_3 \tau_3$, $\Gamma_{2,3} = \sigma_{2,3} \tau_3$, $\Gamma_{4,5} = \tau_{2,1},$ and $\Gamma_{6,7} = \tau_3 \sigma_1 s_{2,1}$ with $C_{2z} = \Gamma_1 \Gamma_2$, $\T = \Gamma_1 \Gamma_7 \K,$ and $M_z = \Gamma_4 \Gamma_7$. Similar to the dimerized weak Chern insulator, the Hamiltonian above is only gapless for $m=0$ and $|\lambda_z/\gamma_z| = 1$ which means that all the parameters for which the Hamiltonian is gapped can be deformed to one another without closing the gap. One consistency check for the triviality of this Hamiltonian is to note that the Hamiltonian has the inversion symmetry $\I = C_{2z} M_z = \tau_1 \sigma_2 s_2$ which squares to $+1$. Thus, one can check its triviality by computing the symmetry indicators \cite{Po17, Khalaf17} obtained from the inversion eigenvalues at the time-reversal invariant momenta (TRIM). These correspond to a $\Z_4$ index labelling the possible bulk phases (strong TI or second-order TI), and three weak $\Z_2$ invariants for the $x$, $y$, and $z$ directions \cite{Po17, Khalaf17}. There are four filled bands which come in pairs of Kramers' doublets at the TRIMs. The symmetry indicators are obtained by computing the product of inversion eigenvalues at each TRIM (counting only one state out of each Kramers' doublet) \cite{FuKane, Po17, Khalaf17}. By an explicit calculation, we find that the product of inversion eigenvalues at every TRIM is equal to $-1$. As a result, all symmetry indicators are trivial.

It is important to highlight the role of $\pi$-fluxes through the vertical plaquettes. Unlike the models considered in Sec.~\ref{sec:3DFA}, it is possible here to choose a gauge such that $M_z$ and $C_{2z}$ commute. This arises from the fact that the total flux per plaquette in the vertical direction (including both spins) vanishes, although the there is a $\pi$-flux per spin. Thus, the the $\pi$-fluxes do not alter the commutation relations between $M_z$ and $C_2$

 {We can investigate the boundary obstruction numerically by computing the Wannier bands. Here again we consider a slightly modified model with the same topological properties but on a cubic lattice with the Hamiltonian given by}
\begin{multline}
\H = (2 + m - \cos k_x - \cos k_y) \Gamma_1 + \sin k_x \Gamma_2 + \sin k_y \Gamma_7 \\ +  \lambda_z \sin k_z \Gamma_4 + (\gamma_z + \lambda_z \cos k_z) \Gamma_3. 
\end{multline}} 
The Wannier spectra of the model are very similar to the weak Chern insulator since this model is just two time-reversed copies of the Chern insulator model. Thus, we see a Wannier transition at the WCP in $\W^z$ when $|\lambda_z|>|\gamma_z|$, and $m=0$; as well as a Wannier transition at the WCP in $\W^x$ and $\W^y$ when $m<0$ but $|\gamma_z|=|\lambda_z|$. However, in contrast to the dimerized weak Chern insulator, this model has four filled bands. Each of the Wannier bands is two-fold degenerate at time-reversal invariant momenta, and the additional $C_{2z}$ symmetry causes this degeneracy to appear across the entire surface Brillouin zone.

The nested Wannier spectra of this model are shown in Fig.~\ref{fig:W2}. We note that $\nu_{z,x}(k_y)$ resembles the spectral flow in a quantum spin Hall insulator. Two Wannier bands flow in opposite directions, crossing at $k_y = 0$ and $\pi$  where the degeneracy cannot be lifted due to Kramers' theorem. This nested Wannier winding captures the dangling QSH at the upper and lower surfaces. On the other hand, $\nu_{x,z}(k_y)$ in the nontrivial phase shows two bands related by time-reversal symmetry, each with average polarization of $1/2$. Although the total polarization vanishes, the polarization for each component of the Kramers' pair does not, and it is such a quantized (time-reversal) polarization \cite{fu2006time} that captures the Wannier transition in $\nu_x(k_y,k_z)$. In general one could use a description of the time-reversal polarization in terms of a Pfaffian invariant\cite{fu2006time}, but we will not pursue that further here. In this case the Wannier transition resembles the transition in a spinful SSH chain with time-reversal and inversion symmetry where the edge anomaly is associated with Kramers' pairs of electrons rather than an individual electron.

 In summary, both the dimerized weak Chern model and the dimerized weak QSH model exhibit one-dimensional propagating hinge states that can be removed without closing the bulk gap. However, removing the hinge states requires closing the energy gap on at least one high-symmetry surface. These systems only admit a Wannier representation for periodic boundaries, but not for open boundaries. This serves to illustrate that boundary obstructions are not restricted to Wannier representable systems.

\section{Discussion}
\label{sec:Discussion}
In this work, we introduced the notion of boundary topological obstructions that captures obstructions which do not exist on periodic boundaries, but exist on geometries with symmetric open boundaries. In more precise terms, we call two Hamiltonians $\H_1$ and $\H_2$ boundary-obstructed {on a given symmetric open boundary} if they can be symmetrically deformed to each other without closing the gap with periodic boundary conditions, but not for the open boundary. Another way to formulate this is by saying that any symmetric deformation of $\H_1$ to $\H_2$ involves a gap-closing at a high-symmetry surface rather than in the \emph{bulk}, as in the case of conventional SPTs. Although the possibility of an obstruction between a given pair of Hamiltonians is termination dependent, the group of equivalence classes distinguished by boundary obstructions is termination-independent and is characterized by physical boundary signatures such as filling anomalies and gapless states. 
 
It is worth stressing that boundary signatures associated with BOTPs are not anomalous as they 
can be removed by adding a lower-dimensional SPT on the boundary. As a result, we need to be careful when defining the boundary termination to distinguish surface features which arise from the bulk from those due to the termination. This is done by first considering a simple reference class of terminations achieved by ``interpolating" the bulk Hamiltonian to the vacuum Hamiltonian by means of a single real scalar function. This is always possible by changing the value of the chemical potential from the inside to the outside of the sample. In a more general boundary termination, we can always separate the contribution coming from the termination by comparing to this reference boundary (cf.~Eq.~\ref{eq:HBoundary}). This enables us to isolate the boundary contribution to the open Hamiltonian making it possible to compare two open Hamiltonians \emph{with the same boundary}.

We also offered an alternative definition of boundary obstructed phases using the theory of band representations. In this definition, BOTPs can be identified with equivalent band representations of the space group in the periodic system which become inequivalent upon restricting to a point group of the open system. This definition does not only bridge the idea of boundary obstruction to the bulk band representation, it also enables us to define an \textit{absolute} notion of boundary obstruction, related to the existence of filling anomalies and gapless surface states, from the knowledge of a \textit{relative} obstruction as defined above. This is achieved by observing that whenever two band representations which are equivalent in the periodic system are inequivalent in the open system, at least one of them has to be non-local and exhibit a filling anomaly or gapless surface states. This enables us to identify a trivial BOTP phase (the one corresponding to a local representation in the open system) such that all phases with a boundary obstruction relative to it are non-trivial.


The motivation for introducing the concept of BOTP is two-fold. First, there is the advent of new ways to implement 2D topological systems, e.g., in synthetic materials \cite{noh2018, peterson2018,serragarcia2019,imhof2018,xie2018,khanikaev2018,xue2018,kempkes2019,xue2019b,mittal2019,bao2019,xue2019,khanikaev2019,peterson2020} and cold atoms \cite{aidelsburger2011,li2013,atala2013,atala2014,aidelsburger2015,mancini2015,li2016,leder2016}, which can be used to realize the DMQI and similar models not associated with a bulk topological invariant. The observation and robustness of corner modes in these systems motivate developing a concept that goes beyond standard SPTs to understand the features of these models responsible for these observables' stability. We note that these systems are characterized by a large degree of control over the boundary termination, which is usually done by sharply terminating the lattice or through a confining potential.  Even for real crystals where there is less control regarding the termination, we expect our concept to be relevant to understanding surface states which may arise through tuning the bulk parameters by applying pressure, strain, magnetic field, etc. or by applying a spatially dependent gate voltage inside the sample to implement an edge termination as described in Eq.~\ref{eq:HBoundary}.

The second motivation is conceptual. Since the introduction of the concept of higher-order topological phases \cite{Schindler17, Benalcazar17, Benalcazar17b}, there has been a large body of works that discussed models with corner or surface states (see Ref.~\cite{HOTIReview} for a recent review). In many cases, the existence of surface states was established by numerically analyzing the Hamiltonian on a simple boundary without making the distinction whether these states arise from a bulk invariant or not. To distinguish these, the concept of intrinsic vs. extrinsic HOTIs was introduced in Ref.~\cite{Geier18, Trifunovic18} with the former corresponding to phases whose surface states are anomalous and cannot be removed with symmetric boundary addition, whereas the latter corresponding to systems whose surface states are removable through symmetric boundary additions. However, the concept of extrinsic HOTIs, while useful, is too broad since it lumps together models like the DMQI, which has robust boundary features when considered on most simple boundaries and which can be related to bulk quantities such as those described in this text, with models consisting of a completely trivial bulk attached to a lower dimensional SPT on the surface. The concept of BOTP helps make this distinction and explains why some models exhibit some topologically robust surface features without being associated with a bulk invariant. 

In Sec.~\ref{sec:II_DMQI}, we used the double-mirror quadrupole insulator (DMQI) of Ref.~\cite{Benalcazar17, Benalcazar17b} as a prototypical example to illustrate the notion of BOTPs. Although this model has been intensively studied as one of the first examples of a higher-order topological insulator, one crucial aspect of it had so far been overlooked. Namely, the fact that in the absence of $C_{4z}$ symmetry, the two `phases' of the model are not SPTs in the standard sense since they can be deformed to each without closing the bulk gap in periodic boundary conditions. Using the notion of BOTP, we provide a resolution to this problem by showing that such `phases' are related by a boundary rather than bulk obstruction. This was investigated in detail from several different perspectives including real space orbital deformations, Wannier spectra, and symmetry representations. The latter was particularly useful in establishing the connection between the boundary obstruction in the model and the existence of a filling anomaly. Indeed the existence of a filling anomaly allows for a topological boundary signature that does not rely on the existence of zero-energy corner states (which, for insulators, typically require fine-tuned symmetries such as particle-hole or chiral symmetry). Instead, it reflects the fact that the model, when filled with electrons to neutrality, cannot be simultanuously gapped, symmetric and charge-neutral. Thus, a symmetric, gapped DMQI at half-filling will necessarily have excess/deficit charge of $2e$ equally distibuted among the four corners yielding a fractional corner charge of $e/2$.

In Sec.~\ref{sec:3D}, we introduced several 3D models for BOTPs. The first family of such BOTPs have similar phenomenology to the DMQI and exhibit surface obstructions. The non-trivial BOTPs in this family are associated with filling anomalies similar to the one in the DMQI, and they manifest in fractional hinge charge (per unit cell) when the system is symmetric and gapped at half-filling. In addition, we introduced a second family of BOTPs characterized with chiral/helical propagating hinge modes. In both cases, the existence of a boundary obstruction is established explicitly by studying the Wannier spectra and the symmetry indicators of the corresponding WBRs.

Before closing, let us discuss the possible extension of the concepts discussed here beyond free fermions. We note that the existence of chiral/helical hinge states or filling anomalies are robust features that are expected to survive in the presence of interactions. Thus, we expect the boundary topological distinctions we found here to still be relevant for interacting systems. On the other hand, defining boundary obstructions is likely more tricky for interacting phases since it requires the implementation of a surface termination in a way which isolates the surface and bulk contributions. It is unclear whether this is generally possible which makes it difficult to decide what it means to ``keep the boundary fixed," or to compare two phases with the ``same boundary." We leave the investigation of such questions to future works.

\acknowledgements{We thank Barry Bradlyn, Jennifer Cano, Dominic Else, Adrian Po, Ryan Thorngren, and Ashvin Vishwanath for stimulating discussions. E. K. was supported by a Simons Investigator Fellowship, by NSF-DMR 1411343, and by the German National Academy of Sciences Leopoldina through grant LPDS 2018-02 Leopoldina fellowship. W.~A.~B. thanks the support of the Eberly Postdoctoral Fellowship at the Pennsylvania State University. T. L. H. thanks US National Science Foundation (NSF) grants EFMA-1627184 (EFRI), and DMR-1351895 (CAREER) for support.  R. Q. was funded by the Deutsche Forschungsgemeinschaft (DFG, German Research Foundation) – Projektnummer 277101999 – TRR 183 (project B03), the Israel Science Foundation, and the European Research Council (Project LEGOTOP).}

\emph{Author Contributions:}
E.~K. developed the general definition for boundary obstructions, the general correspondence between Wannier and boundary topology, proposed 3D model examples, aided the development of the real-space, band representation, and filling anomaly approaches to BOTPs, and was the primary manuscript writer. W.~A.~B. developed the connection between real-space Wannier center deformation and boundary obstructions, helped develop the Wannier band representation, real-space, and filling anomaly approaches to BOTPs, and helped in the writing of the manuscript. T.~L.~H. contributed to the refinement of all of the conceptual advances of the paper and helped in the writing of the manuscript. R.~Q. developed the formalism of band representation theory of BOTPs, helped develop the real-space and Wannier spectrum approaches to BOTPs, developed the filling anomaly arguments in BOTPs, helped in the writing of the manuscript, and led the initiation of the collaboration.  
\appendix

\section{Character table of $D_2^\pi$}
\label{app:charactertable}
\FloatBarrier

\begin{table}[h!]
\centering
\begin{tabular}{l|rrrrr}
Rep. / class & $\{1\}$ & $\{C_{2z}^2\}$ & $\{C_{2z}\}$ & $\{M_y\}$ & $\{ M_x\}$\\
\hline
$A_1$ & 1 & 1 & 1 & 1 & 1\\
$A_2$ & 1 & 1 & 1 & -1 & -1\\
$B_1$ & 1 & 1 & -1 & 1 & -1\\
$B_2$ & 1 & 1 & -1 & -1 & 1\\
$\bar E$ & 2 & -2 & 0 & 0 & 0
\end{tabular}
\caption{Character table of the point group $D_2^\pi$ defined in the main text. The point group is isomorphic to $D_4$ in the absence of a $\pi$ flux.}  
\label{tab:CharacterQuaternion}
\end{table}
\FloatBarrier

\section{Real Space Proof of Anti-commuting Symmetries}\label{app:mirrors}
\subsection{Anticommuting Mirrors in the 2D DMQI}
We now show that in order to carry out the deformation between the configurations at $1a$ and $1d$ in the DMQI, it is necessary that the two mirror operators anticommute. This is suggested by the fact that the two configurations actually correspond to \emph{different} symmetry representations for commuting mirrors, but the same symmetry representation for anticommuting mirrors ($\bar E$). We will see that the anticommuting mirrors can be inferred from the real space picture in Fig.~\ref{fig:edge_obstruction}. Let us start with the orbitals at position $1a$ and then move them horizontally using position $2e$ into position $1c.$ The two orbitals in position $2e$ are related by mirror symmetry $M_x$ and have the form $\ket{(\pm x,0)}$,  where $\ket{\vec r}$ denotes an orbital localized at point $\vec r$. In this basis, $M_x$ is off-diagonal and acts generally as $M_x \ket{(\pm x,0)} = e^{\pm i \phi} \ket{(\mp x,0)}$ (with some arbitrary phase $\phi$), leading to the eigenvectors $\ket{(+x,0)} \pm e^{i \phi} \ket{(-x,0)}$
whose eigenvalues are $\pm 1$ respectively. Since these orbitals lie on the $y = 0$ mirror line, they are eigenvectors of $M_y$.
Next, to bring these orbitals to position $1d$, we move them vertically using position $2h$. In this position, the action of $M_y$ is off-diagonal, thus, following the same argument as for $M_x$, we can deduce that it has two distinct eigenvalues $+1$ and $-1$. Since the symmetry eigenvalues cannot change under smooth, symmetry-preserving deformations, we deduce that $M_y = \pm \sigma_3$ in the $2e$ basis. In the same basis, we showed that $M_x = \sigma_1 \exp\{i \phi\sigma_3\},$ which in turn implies $\{M_x, M_y\}=0$. Since anticommutation is basis-independent, we deduce that connecting the $1a$ and $1d$ positions at a filling of two electrons per cell requires $M_x$ and $M_y$ to anticommute. We note that the connection of mirror anticommutation in the DMQI model to important quadrupole properties was proposed in Refs.~\cite{Benalcazar17, Benalcazar17b} based on a study of the Wannier spectrum.
\subsection{Anticommuting $C_{2z}$ and $M_z$ for 3D $C_{2nh}$ Models}
 To show that $C_{2z}$ and $M_z$ anticommute for the $C_{2nh}$ models that we discuss in Section \ref{sec:c2nhmodels}, we note that in order to occupy a movable Wyckoff postion that interpolates between $1a$ and $nc$ in the $M_z$ invariant planes, the Wannier centers permute into each other under the action of $C_{2nz}$. This implies the eigenvalues of the $C_{2nz}$ operator must span all the $2n$ roots of unity $e^{\pi i l/n}$ with $l=0,\dots,2n-1$. Similarly, to occupy a vertically movable Wyckoff position at $C_{2z}$ invariant lines, the eigenvalues of $M_z$ should be $\pm 1$ since its action is a permutation of the two sites in this position. Noting that the two electrons at position $nc$ originate from electrons which have moved away from the center $1a$ in diametrically opposite directions, and are thus related by $C_{2z}$, we find that the requirement of deformability in the vertical direction forces any pair of $C_{2z}$-related orbitals to have opposite mirror eigenvalues. Thus, in a basis where the mirror symmetry $M_z$ is diagonal, $C_{2z}$ is purely off-diagonal and it switches the positive and negative mirror sectors which implies the anticommutation condition 
\beq
C_{2z} M_z C_{2z}^\dagger = - M_z.
\eeq
This discussion is very similar to the one for the DMQI above.

\section{Band representations of BOTPs}
\label{app:BRs}
\subsection{Induction of a band representation from a local representation}

Restricting ourselves to atomic bands, it follows from the work of Zak \cite{zak80} that one can write a representation $\rho$ of a space group $G$  from two ingredients. The first ingredient is the Wyckoff position $\mathcal Q$, comprised of a site $\bq$ with a site symmetry group $G_\bq$ and the sites $\bq_i$ obtained by the coset expansion of $G$ with respect to $G_\bq$.  Explicitly, the full space group can be decomposed by $G=\sum_i g_i G_\bq$ with $g_i\in G/G_\bq$. The lattice sites $\bq_i$ in the crystal are obtained by the left action of the coset representatives $\bq_i=g_i \bq$, with $i=1,...,N_{\mathcal Q}$ with $N_{\mathcal Q}$ the multiplicity of $\mathcal Q$, the number of times the site $\bq$ is repeated over the lattice.  The second ingredient needed to define $\rho$ is the representation $\rho_\bq$ of $G_\bq$ under which the state $|\bq\rangle$ transforms. That is, for any $h\in G_\bq$, $h|\bq\rangle=\rho_\bq(h)|\bq\rangle$. Finally, the collection of the representations $\rho_{\bq_i}$ and the associate basis functions $|{\bq_i}\rangle$, form a complete basis for the electronic states in the system, and the complete representation $\rho$ of the space group $G$, acts on this space. Naturally, $\rho$ is not necessarily diagonal in this basis since it mixes orbitals at different lattice sites $\bq_i$. Making use of translation symmetry in $G$, and the fact that the crystal momentum $\bk$ diagonalizes lattice translations $T$, it is possible to express $\rho$ in such basis, where the diagonal entries $\rho_\bk$ represent Bloch states labelled by $\bk$ with little group $G_\bk$, at a Brillouin zone point $\bk$. The collection of $\rho_\bk$ at high symmetry momenta determine the band representation \cite{zak80,zak1982band,Bradlyn2017c,Cano2018a}.

The construction of the band representation is straight forward. Let us start with the representation $\rho_\bq$ of $G_\bq$
for any element $h\in G_\bq$, $h|\bq\rangle=\rho_\bq(h)|\bq\rangle$, here $|\bq \rangle$ denotes a basis state localized at site $\bq$.
Other basis states $\ket{\bq_i}$ are obtained by the action coset elements $g_i$ as $|g_i\bq\rangle$. Any group element $g$ can be expressed as a product of an element of $G_\bq$ and elements of the coset as $g=g_j h g_i^{-1}$ or alternatively $g_jh=gg_i$ with $h$ and $g_j$ uniquely determined by $g$ and $g_i$. Then, it follows that the action of any element $g\in G$ on any basis state is given by \begin{align}g|g_i\bq\rangle=g_jh|\bq\rangle=\rho_\bq(h)|g_j\bq\rangle.\end{align} 
It follows that a symmetric state in the full Hilbert space 
transforms into itself under all elements of $G$ under a (reducible) induced representation $\rho$ of $G$, which in the real space basis $\ket{\bq_i}$ is  given by 
\begin{align}[\rho(g)]_{ij}=\rho_\bq(g_j^{-1}gg_i) ~~~\text{if}~~~ g_j^{-1}gg_i=h\in G_\bq,\end{align} and zero otherwise. This representation is nothing but the local representation $\rho_\bq$ induced to the entire Wyckoff position $\mathcal Q$, and it is usually abbreviated as $\rho=\rho_\bq\uparrow G$. The induced representation is unique up to the choice ambiguity of the coset representatives $g_j$. 

While the statement above is completely general, we can take a shortcut in symmorphic space groups by inducing the representation from a site $\bq$ with the point group $G_\bq=F$ as its site symmetry group. In this case, all the degrees of freedom in the unit cell collapse into the dimension of the local representation at $\bq$. The action of translations on the center of the unit cell $\mathcal O$ (fixed to coincide with the center of the lattice with open boundaries) defines the Bravais lattice which may or may not coincide with the lattice $\mathcal Q$ obtained by the action of translations on $\bq$. Given that $G=F\times T$, the different lattices correspond to distinct coset decompositions $G=\sum_it_iF$. For each space group $G$ there are usually more than one site $\bq$ of maximal symmetry, where its site symmetry group is isomorphic but not the same. They can be distinguished by the symmetry action on the basis states and cannot be connected under conjugation by any element in $G$. Explicitly in the DMQI model, these points generate the Wyckoff positions $1a$ to $1d$. The space group $p2mm$ admits four distinct coset decompositions up to conjugation by an element in $G$, with four isomorphic factor groups $F=G/T$. 

 Directly applying the induction procedure we find that an element of $G$ acts on the maximal lattice as $g|g_i\bq\rangle=[\rho(g)]_{ij}|g_j\bq\rangle$ with $[\rho(g)]_{ij}=\rho_\bq(g_j^{-1}gg_i)$ if $g_j^{-1}gg_i\in G_\bq,$ and zero otherwise. In order to describe the representation of Bloch states in the Brillouin zone it is convenient to change the basis $|\bq_i\rangle$ to the crystal momentum basis $|\bk\rangle$ related by the basis transformation $\langle \bq_i|\bk\rangle=\exp(-i\bk\cdot \bq_i)$. After this Fourier transform we have $g|\bk\rangle =[\rho(g)]_{\bk\bk'}|\bk'\rangle$ with diagonal terms given explicitly by \begin{align}\rho_\bk(g)=\sum_{ij}\rho_\bq(g_j^{-1}gg_i)e^{i\bk\cdot(g_j-g_i)\bq}.\end{align} When $\rho$ corresponds to an elementary band representation it is either connected in momentum space or disconnected into topological subsets, which separately are not compatible with any tabulated band representation \cite{Bradlyn17}.

\subsection{Movable Wyckoff positions}

Two representations $\rho$ and $\rho'$ are equivalent if and only if there is a unitary and smooth matrix-valued function $S(t,g)$ respecting $S(0,g)=\rho(g)$ and $S(1,g)=\rho'(g)$ for all symmetry group elements $g$.
In a crystalline system, the unitary matrix that generates these changes $S(t,g)=U(t-t',g)^\dag S(t',g)U(t-t',g)$, continuously implements a basis transformation of $\rho$. As discussed in Ref.~\cite{Cano2018a}, when $\rho$ is an induced representation from a point $\bq$ into $G,$ and $\rho'$ is an induced representation from a different site $\bq'$, the $S$ matrices can be explicitly constructed by inducing a family of representations from intermediate sites $\vec p$ that continuously interpolate between $\bq$ and $\bq'$. A Wyckoff position $\mathcal P$ with continuously tunable sites $\vec p$ is called \emph{movable} or \emph{non-maximal} when it can continuously preserve $G_{{\vec p}}$. 

This structure is the mathematical background for the pictures presented in the main text: $S(t,g)$ is the induced representation from the states located at $\vec p$ in a path that connects $\bq$ and $\bq'$, and guarantees there exists a gapped, symmetry-preserving path between the different Wannier configurations located at $\bq$ and $\bq'$. 
We can try to directly find the unitary transformation $U(1,g)$ that transforms $\rho(g)$ into $\rho'(g)$. If $\rho$ and $\rho'$ are band representations, labelled by $\bk$, this implies finding a unitary matrix that is periodic in $\bk$ and satisfies
\begin{align}\rho_\bk(g)=U_{g\bk}^\dag(1,g) \rho'_\bk(g) U_\bk(1,g)\end{align}
for all $g$ and $\bk$. 
While we have argued that all \emph{representations} along a movable path are equivalent, the Wannier states $\ket{\bR;q}$ and $\ket{\bR;q'}$ are generally different. Additionally, the Hamiltonian itself is not generally left invariant by these transformations, and the path is implemented by varying the Hamiltonian parameters. However, in some cases two different points along the path may simply correspond to a \emph{basis transformation}. In these cases $U_\bk$ leaves the Hamiltonian invariant.
This is relevant for the DMQI example, where the mirror operators anticommute, and the Wannier states can be either at the $2e$ or the $2g$ depending on which $M_x$ or $M_y$ we choose to act diagonally. Thus, whether we draw the Wannier functions in $2e$ or $2g$ is a choice of basis (at least when periodic boundary conditions are chosen). 

\section{Surface states of the DMQI}
\label{app:SSDMQI}
{In this appendix, we analyze the edge spectrum for the DMQI on a generic boundary termination. To do this, we start by noting that the DMQI can be obtained from the $C_{4}$-symmetric quadrupole model by breaking the $C_{4}$ symmetry. Since the $C_{4}$-symmetric model is a proper Higher-order topological phase, we can employ the same techniques used to analyze HOTI surface states as in Refs.~\cite{Khalaf17, Khalaf18, Liu18, AhnJung}. We note that recently, an analysis of the surface states of the $C_{4}$ quadrupole model was employed in Ref.~\cite{Wieder2020} by adding a symmetry-breaking mass term to the surface of a 2D TI. Our analysis differs in the fact that we do not need to explicitly employ spinful representation of the symmetries and we can work directly in terms of the projective representations defined in the main text.}

{The $C_4$-symmetric model is obtained from Eq.~\ref{eq:QuadHamiltonian} by setting $\gamma_x = \gamma_y = \gamma$ and $\lambda_x = \lambda_y = \lambda$. In this case, the Hamiltonian has the extra $C_4$ symmetry implemented as
\begin{gather}
    C_{4z} \H(\bk) C_{4z}^\dagger = \H(O_4 \bk), \quad O_4(k_x,k_y) = (k_y,-k_x) \nonumber \\ C_{4z} = \Gamma_2 e^{\frac{\pi}{4}(\Gamma_2 \Gamma_4 - \Gamma_1 \Gamma_3)} = -\sigma_2 \tau_2 e^{i\frac{\pi}{2} \frac{\sigma_0 - \sigma_3}{2} \tau_2}
\end{gather}
We will find it useful to define the modified $\Gamma$ matrices $\Gamma_\pm = \frac{1}{\sqrt{2}}(\Gamma_2 \pm \Gamma_4)$ such that
\beq
C_{4z} \Gamma_\pm C_{4z}^\dagger = \pm \Gamma_\pm
\label{C4Gpm}
\eeq
The Hamiltonian for the $C_4$ symmetric model can then be written as

\begin{gather}
    \H_{C_4} = \lambda [\sin k_x \Gamma_3 + \sin k_y \Gamma_1] + m_+(\bk) \Gamma_+ + m_-(\bk) \Gamma_- \\
    m_+(\bk) = \frac{1}{\sqrt{2}}[2\gamma + \lambda (\cos k_x + \cos k_y)], \\
    m_-(\bk) = \frac{1}{\sqrt{2}}(\cos k_y - \cos k_x) 
\end{gather}
where $m_\pm(O_4 \bk) = \pm m_\pm(\bk)$.
}

\begin{figure}
    \centering
    \includegraphics[width = 0.3 \textwidth]{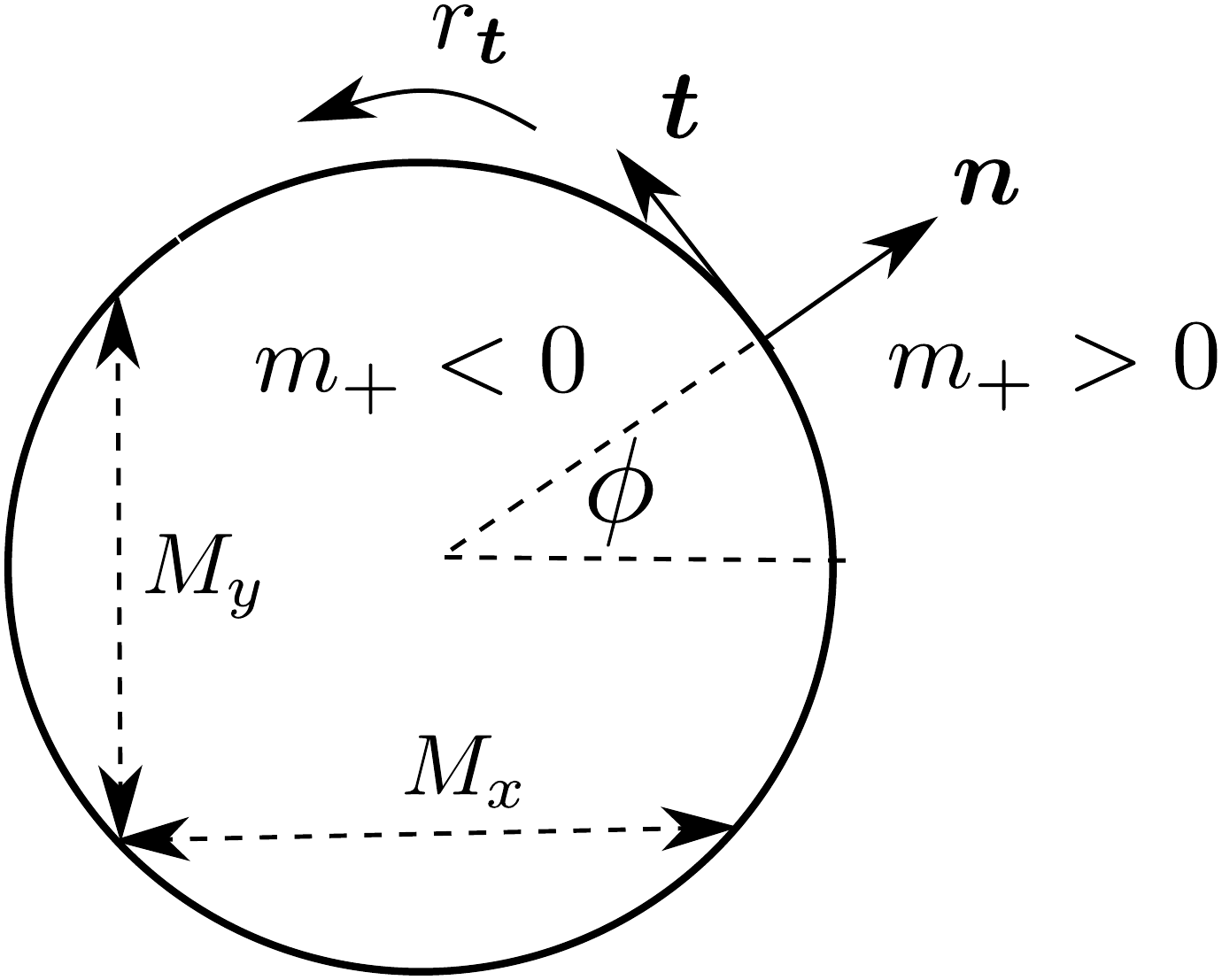}
    \caption{Illustration of the edge theory of the DMQI on a genetic surface. Starting from the $C_4$-symmetric model, we can derive the edge theory by taking the mass term $m_+$ to change sign from the inside to the outside of the sample. The edge theory at any point can be expressed in terms of the normal and tangent vectors $\bn$ and $\bt$. The variable $r_\bt$ parametrizes the 1D edge which can be alternatively expressed in terms of an angular variable $\phi$.}
    \label{fig:DMQIEdge}
\end{figure}

{The $C_4$ symmetric model is characterized by two topologically distinct bulk phases depending on whether $|\lambda/\gamma|$ is smaller or larger than 1. These can be distinguished by the eigenvalues of $C_{4z}$ symmetry at the $\Gamma$ and $M$ points as follows. For $|\lambda| < |\gamma|$, the $C_{4z}$ eigenvalues are given by $\{e^{3i \pi/4}, e^{-3i \pi/4}\}$ at both the $\Gamma$ and $M$ whereas for $|\lambda| > |\gamma|$, the $C_{4z}$ eigenvalues are given by $\{e^{3i \pi/4}, e^{-3i \pi/4}\}$ at $\Gamma$ and by $\{e^{i \pi/4}, e^{-i \pi/4}\}$ at $M$. Thus, the two phases of the $C_{4z}$ symmetric quadrupole model are distinguished by a bulk phase transition at $M$. Close to such transition, we can write a continuum model by expanding the Hamiltonian close to the $M$ point and going to real space as
\beq
\H_{C_4} = i\lambda (\Gamma_3 \partial_x + \Gamma_1 \partial_y) + m_+(\br) \Gamma_+ + m_-(\br) \Gamma_-
\eeq
The action of the spatial symmetries on $m_\pm(\br)$ can be deduced from the conditions (\ref{C4Gpm}) and $M_{x,y} \Gamma_{\pm} M_{x,y}^\dagger = \Gamma_{\pm}$ leading to
\begin{gather}
    m_\pm(O_4 \br) = \pm m_\pm(O_4 \br), 
    \label{C4mm}\\
    m_\pm(-x,y) = m_\pm(x,y), \qquad m_\pm(x,-y) = m_\pm(x,y)
    \label{Mxymm}
\end{gather}
The edge is implemented by choosing $m_+(\br)$ to change sign from being negative inside the sample (corresponding to the non-trivial phase) to being positive outside (corresponding to the trivial phase). Let us denote the normal to the edge at point $\br$ by $\bn(\br) = (\cos \phi(\br), \sin \phi(\br))$ with the corresponding tangent $\bt(\br) = (-\sin \phi(\br), \cos \phi(\br))$ (cf.~Fig.~\ref{fig:DMQIEdge}). We note that, in general, the normal vector depends on the position on the edge. In the following, we will drop this $\br$ dependence to simplify the notation. We also introduce
\begin{gather}
    \Gamma_n = \cos \phi \Gamma_3 + \sin \phi \Gamma_1, \quad \Gamma_t = -\sin \phi \Gamma_3 + \cos \phi \Gamma_1 \\
    r_{\bn} = \br \cdot \bn, \quad r_\bt = \br \cdot \bt
\end{gather}
which enables us to write
\beq
\H_{C_4} = i\lambda (\Gamma_n \partial_{r_\bn} + \Gamma_t \partial_{r_\bt}) + m_+(\br) \Gamma_+ + m_-(\br) \Gamma_-
\eeq
To obtain the edge theory, we will assume that the mass $m_-(\br)$ is much smaller than the value of $m_+(\br)$ in the bulk. If we also consider wavefunctions which vary slowly along the tangent direction to the edge, then we find that the low energy states are spanned by eigenfunctions of the form
\beq
\psi(\br) = \frac{1}{2}(1 - i \Gamma_n \Gamma_+) e^{-\frac{1}{\lambda} \int^\br dr_\bn' m_+(r_\bn',r_\bt)} \chi(r_\bt) 
\eeq
which are annihilated by $i\lambda \Gamma_n \partial_{r_n} + m_+(\br) \Gamma_+$ for any function $\chi(r_\bt)$. Thus, the edge Hamiltonian is given by
\beq
\H_{\rm edge} = i\lambda \gamma_t \partial_{r_\bt} + m_-(r_t) \gamma_-
\eeq
where $\gamma_{t}, \gamma_-$ are obtained by projecting $\Gamma_{t}, \Gamma_-$ onto the space spanned by $\psi(\br)$. The edge Hamiltonian is a one-dimensional Dirac Hamiltonian defined in terms of the variable $r_\bt$ along the direction parallel to the edge. Changing the variable from $r_\bt$ to the angular variable $\phi$ (cf.~Fig.~\ref{fig:DMQIEdge}), and using (\ref{C4mm}), we find that
\begin{gather}
    m_-(\phi + \pi/2) = -m_-(\phi) 
    \label{C4phi}\\
    m_-(-\phi) = m_-(\phi), \quad m_-(\pi - \phi) = m_-(\phi)
    \label{Mxyphi}
\end{gather}
where the first condition follows from the transformation properties under $C_4$ rotation (\ref{C4mm}) and the second follows from $M_x$ and $M_y$ (\ref{Mxymm}). Eq.~\ref{C4phi} implies that the edge Dirac Hamiltonian hosts four domain walls where the mass term $m_-$ changes sign.}

{Notice now that if we break $C_4$ symmetry, these four domain walls can be freely moved to annihilate. However, Eq.~\ref{Mxyphi} would dictate that they can only annihilate at $\phi = 0$ or $\pi$, i.e. at a high symmetry surface. If we enforce the condition that $m_-(0), m_-(\pi) \neq 0$, we can never get rid of these domain walls. We notice also that Eq.~\ref{Mxyphi} is also satisfied by the trivial phase where $m_-(\phi) = \rm const.$ leading to a completely gapped edge. Thus, For the $C_4$-broken model, the symmetries allow for two possibilities with or without zero energy states at the edge. In contrast, for the $C_4$ symmetric model which is a HOTI, the symmetry forces the mass term to change sign enforcing the existence of four symmetry-related zero energy states. 
}

\bibliography{refs}

\begin{thebibliography}{96}%
\makeatletter
\providecommand \@ifxundefined [1]{%
 \@ifx{#1\undefined}
}%
\providecommand \@ifnum [1]{%
 \ifnum #1\expandafter \@firstoftwo
 \else \expandafter \@secondoftwo
 \fi
}%
\providecommand \@ifx [1]{%
 \ifx #1\expandafter \@firstoftwo
 \else \expandafter \@secondoftwo
 \fi
}%
\providecommand \natexlab [1]{#1}%
\providecommand \enquote  [1]{``#1''}%
\providecommand \bibnamefont  [1]{#1}%
\providecommand \bibfnamefont [1]{#1}%
\providecommand \citenamefont [1]{#1}%
\providecommand \href@noop [0]{\@secondoftwo}%
\providecommand \href [0]{\begingroup \@sanitize@url \@href}%
\providecommand \@href[1]{\@@startlink{#1}\@@href}%
\providecommand \@@href[1]{\endgroup#1\@@endlink}%
\providecommand \@sanitize@url [0]{\catcode `\\12\catcode `\$12\catcode
  `\&12\catcode `\#12\catcode `\^12\catcode `\_12\catcode `\%12\relax}%
\providecommand \@@startlink[1]{}%
\providecommand \@@endlink[0]{}%
\providecommand \url  [0]{\begingroup\@sanitize@url \@url }%
\providecommand \@url [1]{\endgroup\@href {#1}{\urlprefix }}%
\providecommand \urlprefix  [0]{URL }%
\providecommand \Eprint [0]{\href }%
\providecommand \doibase [0]{http://dx.doi.org/}%
\providecommand \selectlanguage [0]{\@gobble}%
\providecommand \bibinfo  [0]{\@secondoftwo}%
\providecommand \bibfield  [0]{\@secondoftwo}%
\providecommand \translation [1]{[#1]}%
\providecommand \BibitemOpen [0]{}%
\providecommand \bibitemStop [0]{}%
\providecommand \bibitemNoStop [0]{.\EOS\space}%
\providecommand \EOS [0]{\spacefactor3000\relax}%
\providecommand \BibitemShut  [1]{\csname bibitem#1\endcsname}%
\let\auto@bib@innerbib\@empty
\bibitem [{\citenamefont {Kane}\ and\ \citenamefont
  {Mele}(2005{\natexlab{a}})}]{Kane05a}%
  \BibitemOpen
  \bibfield  {author} {\bibinfo {author} {\bibfnamefont {Charles~L}\
  \bibnamefont {Kane}}\ and\ \bibinfo {author} {\bibfnamefont {Eugene~J}\
  \bibnamefont {Mele}},\ }\bibfield  {title} {\enquote {\bibinfo {title}
  {Quantum spin {Hall} effect in graphene},}\ }\href@noop {} {\bibfield
  {journal} {\bibinfo  {journal} {\prl}\ }\textbf {\bibinfo {volume} {95}},\
  \bibinfo {pages} {226801} (\bibinfo {year} {2005}{\natexlab{a}})}\BibitemShut
  {NoStop}%
\bibitem [{\citenamefont {Kane}\ and\ \citenamefont
  {Mele}(2005{\natexlab{b}})}]{Kane05b}%
  \BibitemOpen
  \bibfield  {author} {\bibinfo {author} {\bibfnamefont {Charles~L}\
  \bibnamefont {Kane}}\ and\ \bibinfo {author} {\bibfnamefont {Eugene~J}\
  \bibnamefont {Mele}},\ }\bibfield  {title} {\enquote {\bibinfo {title}
  {{$Z_2$} topological order and the quantum spin {Hall} effect},}\ }\href@noop
  {} {\bibfield  {journal} {\bibinfo  {journal} {\prl}\ }\textbf {\bibinfo
  {volume} {95}},\ \bibinfo {pages} {146802} (\bibinfo {year}
  {2005}{\natexlab{b}})}\BibitemShut {NoStop}%
\bibitem [{\citenamefont {Bernevig}\ \emph {et~al.}(2006)\citenamefont
  {Bernevig}, \citenamefont {Hughes},\ and\ \citenamefont
  {Zhang}}]{Bernevig06}%
  \BibitemOpen
  \bibfield  {author} {\bibinfo {author} {\bibfnamefont {B.~Andrei}\
  \bibnamefont {Bernevig}}, \bibinfo {author} {\bibfnamefont {Taylor~L.}\
  \bibnamefont {Hughes}}, \ and\ \bibinfo {author} {\bibfnamefont {Shou-Cheng}\
  \bibnamefont {Zhang}},\ }\bibfield  {title} {\enquote {\bibinfo {title}
  {Quantum spin hall effect and topological phase transition in hgte quantum
  wells},}\ }\href {\doibase 10.1126/science.1133734} {\bibfield  {journal}
  {\bibinfo  {journal} {Science}\ }\textbf {\bibinfo {volume} {314}},\ \bibinfo
  {pages} {1757--1761} (\bibinfo {year} {2006})},\ \Eprint
  {http://arxiv.org/abs/http://science.sciencemag.org/content/314/5806/1757.full.pdf}
  {http://science.sciencemag.org/content/314/5806/1757.full.pdf} \BibitemShut
  {NoStop}%
\bibitem [{\citenamefont {Qi}\ \emph {et~al.}(2008)\citenamefont {Qi},
  \citenamefont {Hughes},\ and\ \citenamefont {Zhang}}]{Qi08}%
  \BibitemOpen
  \bibfield  {author} {\bibinfo {author} {\bibfnamefont {Xiao-Liang}\
  \bibnamefont {Qi}}, \bibinfo {author} {\bibfnamefont {Taylor~L.}\
  \bibnamefont {Hughes}}, \ and\ \bibinfo {author} {\bibfnamefont {Shou-Cheng}\
  \bibnamefont {Zhang}},\ }\bibfield  {title} {\enquote {\bibinfo {title}
  {Topological field theory of time-reversal invariant insulators},}\ }\href
  {\doibase 10.1103/PhysRevB.78.195424} {\bibfield  {journal} {\bibinfo
  {journal} {Phys. Rev. B}\ }\textbf {\bibinfo {volume} {78}},\ \bibinfo
  {pages} {195424} (\bibinfo {year} {2008})}\BibitemShut {NoStop}%
\bibitem [{\citenamefont {Moore}(2009)}]{Moore09}%
  \BibitemOpen
  \bibfield  {author} {\bibinfo {author} {\bibfnamefont {Joel}\ \bibnamefont
  {Moore}},\ }\bibfield  {title} {\enquote {\bibinfo {title} {Topological
  insulators: The next generation},}\ }\href@noop {} {\bibfield  {journal}
  {\bibinfo  {journal} {Nat. Phys.}\ }\textbf {\bibinfo {volume} {5}},\
  \bibinfo {pages} {378--380} (\bibinfo {year} {2009})}\BibitemShut {NoStop}%
\bibitem [{\citenamefont {Hasan}\ and\ \citenamefont {Kane}(2010)}]{Hasan10}%
  \BibitemOpen
  \bibfield  {author} {\bibinfo {author} {\bibfnamefont {M~Zahid}\ \bibnamefont
  {Hasan}}\ and\ \bibinfo {author} {\bibfnamefont {Charles~L}\ \bibnamefont
  {Kane}},\ }\bibfield  {title} {\enquote {\bibinfo {title} {Colloquium:
  topological insulators},}\ }\href@noop {} {\bibfield  {journal} {\bibinfo
  {journal} {\rmp}\ }\textbf {\bibinfo {volume} {82}},\ \bibinfo {pages} {3045}
  (\bibinfo {year} {2010})}\BibitemShut {NoStop}%
\bibitem [{\citenamefont {Qi}\ and\ \citenamefont {Zhang}(2011)}]{Qi11}%
  \BibitemOpen
  \bibfield  {author} {\bibinfo {author} {\bibfnamefont {Xiao-Liang}\
  \bibnamefont {Qi}}\ and\ \bibinfo {author} {\bibfnamefont {Shou-Cheng}\
  \bibnamefont {Zhang}},\ }\bibfield  {title} {\enquote {\bibinfo {title}
  {Topological insulators and superconductors},}\ }\href@noop {} {\bibfield
  {journal} {\bibinfo  {journal} {\rmp}\ }\textbf {\bibinfo {volume} {83}},\
  \bibinfo {pages} {1057} (\bibinfo {year} {2011})}\BibitemShut {NoStop}%
\bibitem [{\citenamefont {Franz}\ and\ \citenamefont
  {Molenkamp}(2013)}]{Molenkamp13}%
  \BibitemOpen
  \bibinfo {editor} {\bibfnamefont {Marcel}\ \bibnamefont {Franz}}\ and\
  \bibinfo {editor} {\bibfnamefont {Laurens}\ \bibnamefont {Molenkamp}},\
  eds.,\ \href {\doibase http://dx.doi.org/10.1016/B978-0-444-63314-9.00012-3}
  {\emph {\bibinfo {title} {Topological Insulators}}},\ \bibinfo {series}
  {Contemporary Concepts of Condensed Matter Science}, Vol.~\bibinfo {volume}
  {6}\ (\bibinfo  {publisher} {Elsevier},\ \bibinfo {year} {2013})\BibitemShut
  {NoStop}%
\bibitem [{\citenamefont {Kitaev}(2009)}]{Kitaev09}%
  \BibitemOpen
  \bibfield  {author} {\bibinfo {author} {\bibfnamefont {Alexei}\ \bibnamefont
  {Kitaev}},\ }\bibfield  {title} {\enquote {\bibinfo {title} {Periodic table
  for topological insulators and superconductors},}\ }in\ \href@noop {} {\emph
  {\bibinfo {booktitle} {AIP Conference Proceedings}}},\ Vol.\ \bibinfo
  {volume} {1134}\ (\bibinfo {organization} {AIP},\ \bibinfo {year} {2009})\
  pp.\ \bibinfo {pages} {22--30}\BibitemShut {NoStop}%
\bibitem [{\citenamefont {Schnyder}\ \emph {et~al.}(2009)\citenamefont
  {Schnyder}, \citenamefont {Ryu}, \citenamefont {Furusaki},\ and\
  \citenamefont {Ludwig}}]{Schnyder09}%
  \BibitemOpen
  \bibfield  {author} {\bibinfo {author} {\bibfnamefont {Andreas~P.}\
  \bibnamefont {Schnyder}}, \bibinfo {author} {\bibfnamefont {Shinsei}\
  \bibnamefont {Ryu}}, \bibinfo {author} {\bibfnamefont {Akira}\ \bibnamefont
  {Furusaki}}, \ and\ \bibinfo {author} {\bibfnamefont {Andreas W.~W.}\
  \bibnamefont {Ludwig}},\ }\bibfield  {title} {\enquote {\bibinfo {title}
  {Classification of topological insulators and superconductors},}\ }in\
  \href@noop {} {\emph {\bibinfo {booktitle} {AIP Conference Proceedings}}},\
  Vol.\ \bibinfo {volume} {1134}\ (\bibinfo {organization} {AIP},\ \bibinfo
  {year} {2009})\ pp.\ \bibinfo {pages} {10--21}\BibitemShut {NoStop}%
\bibitem [{\citenamefont {Ryu}\ \emph {et~al.}(2010)\citenamefont {Ryu},
  \citenamefont {Schnyder}, \citenamefont {Furusaki},\ and\ \citenamefont
  {Ludwig}}]{Ryu10}%
  \BibitemOpen
  \bibfield  {author} {\bibinfo {author} {\bibfnamefont {Shinsei}\ \bibnamefont
  {Ryu}}, \bibinfo {author} {\bibfnamefont {Andreas~P.}\ \bibnamefont
  {Schnyder}}, \bibinfo {author} {\bibfnamefont {Akira}\ \bibnamefont
  {Furusaki}}, \ and\ \bibinfo {author} {\bibfnamefont {Andreas W.~W.}\
  \bibnamefont {Ludwig}},\ }\bibfield  {title} {\enquote {\bibinfo {title}
  {Topological insulators and superconductors: tenfold way and dimensional
  hierarchy},}\ }\href@noop {} {\bibfield  {journal} {\bibinfo  {journal} {New
  J. Phys.}\ }\textbf {\bibinfo {volume} {12}},\ \bibinfo {pages} {065010}
  (\bibinfo {year} {2010})}\BibitemShut {NoStop}%
\bibitem [{\citenamefont {Alexandradinata}\ \emph {et~al.}(2016)\citenamefont
  {Alexandradinata}, \citenamefont {Wang},\ and\ \citenamefont
  {Bernevig}}]{Alexandradinata}%
  \BibitemOpen
  \bibfield  {author} {\bibinfo {author} {\bibfnamefont {A.}~\bibnamefont
  {Alexandradinata}}, \bibinfo {author} {\bibfnamefont {Zhijun}\ \bibnamefont
  {Wang}}, \ and\ \bibinfo {author} {\bibfnamefont {B.~Andrei}\ \bibnamefont
  {Bernevig}},\ }\bibfield  {title} {\enquote {\bibinfo {title} {Topological
  insulators from group cohomology},}\ }\href {\doibase
  10.1103/PhysRevX.6.021008} {\bibfield  {journal} {\bibinfo  {journal} {Phys.
  Rev. X}\ }\textbf {\bibinfo {volume} {6}},\ \bibinfo {pages} {021008}
  (\bibinfo {year} {2016})}\BibitemShut {NoStop}%
\bibitem [{\citenamefont {Trifunovic}\ and\ \citenamefont
  {Brouwer}(2019)}]{Trifunovic18}%
  \BibitemOpen
  \bibfield  {author} {\bibinfo {author} {\bibfnamefont {Luka}\ \bibnamefont
  {Trifunovic}}\ and\ \bibinfo {author} {\bibfnamefont {Piet~W.}\ \bibnamefont
  {Brouwer}},\ }\bibfield  {title} {\enquote {\bibinfo {title} {Higher-order
  bulk-boundary correspondence for topological crystalline phases},}\ }\href
  {\doibase 10.1103/PhysRevX.9.011012} {\bibfield  {journal} {\bibinfo
  {journal} {Phys. Rev. X}\ }\textbf {\bibinfo {volume} {9}},\ \bibinfo {pages}
  {011012} (\bibinfo {year} {2019})}\BibitemShut {NoStop}%
\bibitem [{\citenamefont {Dziawa}\ \emph {et~al.}(2012)\citenamefont {Dziawa},
  \citenamefont {Kowalski}, \citenamefont {Dybko}, \citenamefont {Buczko},
  \citenamefont {Szczerbakow}, \citenamefont {Szot}, \citenamefont
  {{\L}usakowska}, \citenamefont {Balasubramanian}, \citenamefont {Wojek},
  \citenamefont {Berntsen} \emph {et~al.}}]{Dziawa12}%
  \BibitemOpen
  \bibfield  {author} {\bibinfo {author} {\bibfnamefont {P}~\bibnamefont
  {Dziawa}}, \bibinfo {author} {\bibfnamefont {BJ}~\bibnamefont {Kowalski}},
  \bibinfo {author} {\bibfnamefont {K}~\bibnamefont {Dybko}}, \bibinfo {author}
  {\bibfnamefont {R}~\bibnamefont {Buczko}}, \bibinfo {author} {\bibfnamefont
  {A}~\bibnamefont {Szczerbakow}}, \bibinfo {author} {\bibfnamefont
  {M}~\bibnamefont {Szot}}, \bibinfo {author} {\bibfnamefont {E}~\bibnamefont
  {{\L}usakowska}}, \bibinfo {author} {\bibfnamefont {T}~\bibnamefont
  {Balasubramanian}}, \bibinfo {author} {\bibfnamefont {Bastian~M}\
  \bibnamefont {Wojek}}, \bibinfo {author} {\bibfnamefont {MH}~\bibnamefont
  {Berntsen}},  \emph {et~al.},\ }\bibfield  {title} {\enquote {\bibinfo
  {title} {Topological crystalline insulator states in
  {Pb\textsubscript{$1-x$}Sn\textsubscript{$x$}Se}},}\ }\href@noop {}
  {\bibfield  {journal} {\bibinfo  {journal} {Nature materials}\ }\textbf
  {\bibinfo {volume} {11}},\ \bibinfo {pages} {1023--1027} (\bibinfo {year}
  {2012})}\BibitemShut {NoStop}%
\bibitem [{\citenamefont {Tanaka}\ \emph {et~al.}(2012)\citenamefont {Tanaka},
  \citenamefont {Ren}, \citenamefont {Sato}, \citenamefont {Nakayama},
  \citenamefont {Souma}, \citenamefont {Takahashi}, \citenamefont {Segawa},\
  and\ \citenamefont {Ando}}]{Tanaka12}%
  \BibitemOpen
  \bibfield  {author} {\bibinfo {author} {\bibfnamefont {Y}~\bibnamefont
  {Tanaka}}, \bibinfo {author} {\bibfnamefont {Zhi}\ \bibnamefont {Ren}},
  \bibinfo {author} {\bibfnamefont {T}~\bibnamefont {Sato}}, \bibinfo {author}
  {\bibfnamefont {K}~\bibnamefont {Nakayama}}, \bibinfo {author} {\bibfnamefont
  {S}~\bibnamefont {Souma}}, \bibinfo {author} {\bibfnamefont {T}~\bibnamefont
  {Takahashi}}, \bibinfo {author} {\bibfnamefont {Kouji}\ \bibnamefont
  {Segawa}}, \ and\ \bibinfo {author} {\bibfnamefont {Yoichi}\ \bibnamefont
  {Ando}},\ }\bibfield  {title} {\enquote {\bibinfo {title} {Experimental
  realization of a topological crystalline insulator in {SnTe}},}\ }\href@noop
  {} {\bibfield  {journal} {\bibinfo  {journal} {Nature Physics}\ }\textbf
  {\bibinfo {volume} {8}},\ \bibinfo {pages} {800--803} (\bibinfo {year}
  {2012})}\BibitemShut {NoStop}%
\bibitem [{\citenamefont {Benalcazar}\ \emph {et~al.}(2014)\citenamefont
  {Benalcazar}, \citenamefont {Teo},\ and\ \citenamefont
  {Hughes}}]{Benalcazar14}%
  \BibitemOpen
  \bibfield  {author} {\bibinfo {author} {\bibfnamefont {Wladimir~A.}\
  \bibnamefont {Benalcazar}}, \bibinfo {author} {\bibfnamefont {Jeffrey C.~Y.}\
  \bibnamefont {Teo}}, \ and\ \bibinfo {author} {\bibfnamefont {Taylor~L.}\
  \bibnamefont {Hughes}},\ }\bibfield  {title} {\enquote {\bibinfo {title}
  {Classification of two-dimensional topological crystalline superconductors
  and majorana bound states at disclinations},}\ }\href {\doibase
  10.1103/PhysRevB.89.224503} {\bibfield  {journal} {\bibinfo  {journal} {Phys.
  Rev. B}\ }\textbf {\bibinfo {volume} {89}},\ \bibinfo {pages} {224503}
  (\bibinfo {year} {2014})}\BibitemShut {NoStop}%
\bibitem [{\citenamefont {Benalcazar}\ \emph
  {et~al.}(2017{\natexlab{a}})\citenamefont {Benalcazar}, \citenamefont
  {Bernevig},\ and\ \citenamefont {Hughes}}]{Benalcazar17}%
  \BibitemOpen
  \bibfield  {author} {\bibinfo {author} {\bibfnamefont {Wladimir~A}\
  \bibnamefont {Benalcazar}}, \bibinfo {author} {\bibfnamefont {B.~Andrei}\
  \bibnamefont {Bernevig}}, \ and\ \bibinfo {author} {\bibfnamefont {Taylor~L}\
  \bibnamefont {Hughes}},\ }\bibfield  {title} {\enquote {\bibinfo {title}
  {{Quantized electric multipole insulators}},}\ }\href {\doibase
  10.1126/science.aah6442} {\bibfield  {journal} {\bibinfo  {journal}
  {Science}\ }\textbf {\bibinfo {volume} {66}},\ \bibinfo {pages} {61--66}
  (\bibinfo {year} {2017}{\natexlab{a}})}\BibitemShut {NoStop}%
\bibitem [{\citenamefont {Benalcazar}\ \emph
  {et~al.}(2017{\natexlab{b}})\citenamefont {Benalcazar}, \citenamefont
  {Bernevig},\ and\ \citenamefont {Hughes}}]{Benalcazar17b}%
  \BibitemOpen
  \bibfield  {author} {\bibinfo {author} {\bibfnamefont {Wladimir~A}\
  \bibnamefont {Benalcazar}}, \bibinfo {author} {\bibfnamefont {B~Andrei}\
  \bibnamefont {Bernevig}}, \ and\ \bibinfo {author} {\bibfnamefont {Taylor~L}\
  \bibnamefont {Hughes}},\ }\bibfield  {title} {\enquote {\bibinfo {title}
  {{Electric multipole moments, topological multipole moment pumping, and
  chiral hinge states in crystalline insulators}},}\ }\href {\doibase
  10.1103/PhysRevB.96.245115} {\bibfield  {journal} {\bibinfo  {journal} {Phys.
  Rev. B}\ }\textbf {\bibinfo {volume} {96}},\ \bibinfo {pages} {245115}
  (\bibinfo {year} {2017}{\natexlab{b}})}\BibitemShut {NoStop}%
\bibitem [{\citenamefont {Langbehn}\ \emph {et~al.}(2017)\citenamefont
  {Langbehn}, \citenamefont {Peng}, \citenamefont {Trifunovic}, \citenamefont
  {von Oppen},\ and\ \citenamefont {Brouwer}}]{Langbehn17}%
  \BibitemOpen
  \bibfield  {author} {\bibinfo {author} {\bibfnamefont {Josias}\ \bibnamefont
  {Langbehn}}, \bibinfo {author} {\bibfnamefont {Yang}\ \bibnamefont {Peng}},
  \bibinfo {author} {\bibfnamefont {Luka}\ \bibnamefont {Trifunovic}}, \bibinfo
  {author} {\bibfnamefont {Felix}\ \bibnamefont {von Oppen}}, \ and\ \bibinfo
  {author} {\bibfnamefont {Piet~W.}\ \bibnamefont {Brouwer}},\ }\bibfield
  {title} {\enquote {\bibinfo {title} {Reflection-symmetric second-order
  topological insulators and superconductors},}\ }\href {\doibase
  10.1103/PhysRevLett.119.246401} {\bibfield  {journal} {\bibinfo  {journal}
  {Phys. Rev. Lett.}\ }\textbf {\bibinfo {volume} {119}},\ \bibinfo {pages}
  {246401} (\bibinfo {year} {2017})}\BibitemShut {NoStop}%
\bibitem [{\citenamefont {Schindler}\ \emph {et~al.}(2018)\citenamefont
  {Schindler}, \citenamefont {Cook}, \citenamefont {Vergniory}, \citenamefont
  {Wang}, \citenamefont {Parkin}, \citenamefont {Bernevig},\ and\ \citenamefont
  {Neupert}}]{Schindler17}%
  \BibitemOpen
  \bibfield  {author} {\bibinfo {author} {\bibfnamefont {Frank}\ \bibnamefont
  {Schindler}}, \bibinfo {author} {\bibfnamefont {Ashley~M.}\ \bibnamefont
  {Cook}}, \bibinfo {author} {\bibfnamefont {Maia~G.}\ \bibnamefont
  {Vergniory}}, \bibinfo {author} {\bibfnamefont {Zhijun}\ \bibnamefont
  {Wang}}, \bibinfo {author} {\bibfnamefont {Stuart S.~P.}\ \bibnamefont
  {Parkin}}, \bibinfo {author} {\bibfnamefont {B.~Andrei}\ \bibnamefont
  {Bernevig}}, \ and\ \bibinfo {author} {\bibfnamefont {Titus}\ \bibnamefont
  {Neupert}},\ }\bibfield  {title} {\enquote {\bibinfo {title} {Higher-order
  topological insulators},}\ }\href
  {http://advances.sciencemag.org/content/4/6/eaat0346} {\bibfield  {journal}
  {\bibinfo  {journal} {Science Advances}\ }\textbf {\bibinfo {volume} {4}},\
  \bibinfo {pages} {eaat0346} (\bibinfo {year} {2018})}\BibitemShut {NoStop}%
\bibitem [{\citenamefont {Song}\ \emph
  {et~al.}(2017{\natexlab{a}})\citenamefont {Song}, \citenamefont {Fang},\ and\
  \citenamefont {Fang}}]{Song17}%
  \BibitemOpen
  \bibfield  {author} {\bibinfo {author} {\bibfnamefont {Zhida}\ \bibnamefont
  {Song}}, \bibinfo {author} {\bibfnamefont {Zhong}\ \bibnamefont {Fang}}, \
  and\ \bibinfo {author} {\bibfnamefont {Chen}\ \bibnamefont {Fang}},\
  }\bibfield  {title} {\enquote {\bibinfo {title}
  {$(d\ensuremath{-}2)$-dimensional edge states of rotation symmetry protected
  topological states},}\ }\href {\doibase 10.1103/PhysRevLett.119.246402}
  {\bibfield  {journal} {\bibinfo  {journal} {Phys. Rev. Lett.}\ }\textbf
  {\bibinfo {volume} {119}},\ \bibinfo {pages} {246402} (\bibinfo {year}
  {2017}{\natexlab{a}})}\BibitemShut {NoStop}%
\bibitem [{\citenamefont {Fang}\ and\ \citenamefont {Fu}(2017)}]{Fang17}%
  \BibitemOpen
  \bibfield  {author} {\bibinfo {author} {\bibfnamefont {Chen}\ \bibnamefont
  {Fang}}\ and\ \bibinfo {author} {\bibfnamefont {Liang}\ \bibnamefont {Fu}},\
  }\bibfield  {title} {\enquote {\bibinfo {title} {Rotation anomaly and
  topological crystalline insulators},}\ }\href
  {https://arxiv.org/abs/1709.01929} {\bibfield  {journal} {\bibinfo  {journal}
  {arXiv preprint arXiv:1709.01929}\ } (\bibinfo {year} {2017})}\BibitemShut
  {NoStop}%
\bibitem [{\citenamefont {Khalaf}\ \emph {et~al.}(2018)\citenamefont {Khalaf},
  \citenamefont {Po}, \citenamefont {Vishwanath},\ and\ \citenamefont
  {Watanabe}}]{Khalaf17}%
  \BibitemOpen
  \bibfield  {author} {\bibinfo {author} {\bibfnamefont {Eslam}\ \bibnamefont
  {Khalaf}}, \bibinfo {author} {\bibfnamefont {Hoi~Chun}\ \bibnamefont {Po}},
  \bibinfo {author} {\bibfnamefont {Ashvin}\ \bibnamefont {Vishwanath}}, \ and\
  \bibinfo {author} {\bibfnamefont {Haruki}\ \bibnamefont {Watanabe}},\
  }\bibfield  {title} {\enquote {\bibinfo {title} {Symmetry indicators and
  anomalous surface states of topological crystalline insulators},}\ }\href
  {\doibase 10.1103/PhysRevX.8.031070} {\bibfield  {journal} {\bibinfo
  {journal} {Phys. Rev. X}\ }\textbf {\bibinfo {volume} {8}},\ \bibinfo {pages}
  {031070} (\bibinfo {year} {2018})}\BibitemShut {NoStop}%
\bibitem [{\citenamefont {Geier}\ \emph {et~al.}(2018)\citenamefont {Geier},
  \citenamefont {Trifunovic}, \citenamefont {Hoskam},\ and\ \citenamefont
  {Brouwer}}]{Geier18}%
  \BibitemOpen
  \bibfield  {author} {\bibinfo {author} {\bibfnamefont {Max}\ \bibnamefont
  {Geier}}, \bibinfo {author} {\bibfnamefont {Luka}\ \bibnamefont
  {Trifunovic}}, \bibinfo {author} {\bibfnamefont {Max}\ \bibnamefont
  {Hoskam}}, \ and\ \bibinfo {author} {\bibfnamefont {Piet~W.}\ \bibnamefont
  {Brouwer}},\ }\bibfield  {title} {\enquote {\bibinfo {title} {Second-order
  topological insulators and superconductors with an order-two crystalline
  symmetry},}\ }\href {\doibase 10.1103/PhysRevB.97.205135} {\bibfield
  {journal} {\bibinfo  {journal} {Phys. Rev. B}\ }\textbf {\bibinfo {volume}
  {97}},\ \bibinfo {pages} {205135} (\bibinfo {year} {2018})}\BibitemShut
  {NoStop}%
\bibitem [{\citenamefont {Khalaf}(2018)}]{Khalaf18}%
  \BibitemOpen
  \bibfield  {author} {\bibinfo {author} {\bibfnamefont {Eslam}\ \bibnamefont
  {Khalaf}},\ }\bibfield  {title} {\enquote {\bibinfo {title} {Higher-order
  topological insulators and superconductors protected by inversion
  symmetry},}\ }\href {\doibase 10.1103/PhysRevB.97.205136} {\bibfield
  {journal} {\bibinfo  {journal} {Phys. Rev. B}\ }\textbf {\bibinfo {volume}
  {97}},\ \bibinfo {pages} {205136} (\bibinfo {year} {2018})}\BibitemShut
  {NoStop}%
\bibitem [{\citenamefont {Benalcazar}\ \emph {et~al.}(2019)\citenamefont
  {Benalcazar}, \citenamefont {Li},\ and\ \citenamefont
  {Hughes}}]{Benalcazar18}%
  \BibitemOpen
  \bibfield  {author} {\bibinfo {author} {\bibfnamefont {Wladimir~A.}\
  \bibnamefont {Benalcazar}}, \bibinfo {author} {\bibfnamefont {Tianhe}\
  \bibnamefont {Li}}, \ and\ \bibinfo {author} {\bibfnamefont {Taylor~L.}\
  \bibnamefont {Hughes}},\ }\bibfield  {title} {\enquote {\bibinfo {title}
  {Quantization of fractional corner charge in ${C}_{n}$-symmetric higher-order
  topological crystalline insulators},}\ }\href {\doibase
  10.1103/PhysRevB.99.245151} {\bibfield  {journal} {\bibinfo  {journal} {Phys.
  Rev. B}\ }\textbf {\bibinfo {volume} {99}},\ \bibinfo {pages} {245151}
  (\bibinfo {year} {2019})}\BibitemShut {NoStop}%
\bibitem [{\citenamefont {Peterson}\ \emph {et~al.}(2020)\citenamefont
  {Peterson}, \citenamefont {Li}, \citenamefont {Benalcazar}, \citenamefont
  {Hughes},\ and\ \citenamefont {Bahl}}]{peterson2020}%
  \BibitemOpen
  \bibfield  {author} {\bibinfo {author} {\bibfnamefont {Christopher~W.}\
  \bibnamefont {Peterson}}, \bibinfo {author} {\bibfnamefont {Tianhe}\
  \bibnamefont {Li}}, \bibinfo {author} {\bibfnamefont {Wladimir~A.}\
  \bibnamefont {Benalcazar}}, \bibinfo {author} {\bibfnamefont {Taylor~L.}\
  \bibnamefont {Hughes}}, \ and\ \bibinfo {author} {\bibfnamefont {Gaurav}\
  \bibnamefont {Bahl}},\ }\bibfield  {title} {\enquote {\bibinfo {title} {A
  fractional corner anomaly reveals higher-order topology},}\ }\href
  {https://science.sciencemag.org/content/368/6495/1114} {\bibfield  {journal}
  {\bibinfo  {journal} {Science}\ }\textbf {\bibinfo {volume} {368}},\ \bibinfo
  {pages} {1114--1118} (\bibinfo {year} {2020})}\BibitemShut {NoStop}%
\bibitem [{\citenamefont {Teo}\ and\ \citenamefont
  {Hughes}(2017)}]{TeoHughes17}%
  \BibitemOpen
  \bibfield  {author} {\bibinfo {author} {\bibfnamefont {{Jeffrey C.Y.}}\
  \bibnamefont {Teo}}\ and\ \bibinfo {author} {\bibfnamefont {{Taylor L}}\
  \bibnamefont {Hughes}},\ }\bibfield  {title} {\enquote {\bibinfo {title}
  {Topological defects in symmetry-protected topological phases},}\ }\href
  {\doibase 10.1146/annurev-conmatphys-031016-025154} {\bibfield  {journal}
  {\bibinfo  {journal} {Annual Review of Condensed Matter Physics}\ }\textbf
  {\bibinfo {volume} {8}},\ \bibinfo {pages} {211--237} (\bibinfo {year}
  {2017})}\BibitemShut {NoStop}%
\bibitem [{\citenamefont {Liu}\ \emph {et~al.}(2019)\citenamefont {Liu},
  \citenamefont {Vishwanath},\ and\ \citenamefont {Khalaf}}]{Liu18}%
  \BibitemOpen
  \bibfield  {author} {\bibinfo {author} {\bibfnamefont {Shang}\ \bibnamefont
  {Liu}}, \bibinfo {author} {\bibfnamefont {Ashvin}\ \bibnamefont
  {Vishwanath}}, \ and\ \bibinfo {author} {\bibfnamefont {Eslam}\ \bibnamefont
  {Khalaf}},\ }\bibfield  {title} {\enquote {\bibinfo {title} {Shift
  insulators: Rotation-protected two-dimensional topological crystalline
  insulators},}\ }\href {\doibase 10.1103/PhysRevX.9.031003} {\bibfield
  {journal} {\bibinfo  {journal} {Phys. Rev. X}\ }\textbf {\bibinfo {volume}
  {9}},\ \bibinfo {pages} {031003} (\bibinfo {year} {2019})}\BibitemShut
  {NoStop}%
\bibitem [{\citenamefont {Li}\ \emph {et~al.}(2019)\citenamefont {Li},
  \citenamefont {Zhu}, \citenamefont {Benalcazar},\ and\ \citenamefont
  {Hughes}}]{Li19}%
  \BibitemOpen
  \bibfield  {author} {\bibinfo {author} {\bibfnamefont {Tianhe}\ \bibnamefont
  {Li}}, \bibinfo {author} {\bibfnamefont {Penghao}\ \bibnamefont {Zhu}},
  \bibinfo {author} {\bibfnamefont {Wladimir~A.}\ \bibnamefont {Benalcazar}}, \
  and\ \bibinfo {author} {\bibfnamefont {Taylor~L.}\ \bibnamefont {Hughes}},\
  }\bibfield  {title} {\enquote {\bibinfo {title} {Fractional disclination
  charge in two-dimensional $c_n$-symmetric topological crystalline
  insulators},}\ }\href {https://arxiv.org/abs/1906.02752} {\bibfield
  {journal} {\bibinfo  {journal} {arxiv:1906.02752}\ } (\bibinfo {year}
  {2019})}\BibitemShut {NoStop}%
\bibitem [{\citenamefont {Po}\ \emph {et~al.}(2017)\citenamefont {Po},
  \citenamefont {Vishwanath},\ and\ \citenamefont {Watanabe}}]{Po17}%
  \BibitemOpen
  \bibfield  {author} {\bibinfo {author} {\bibfnamefont {Hoi~Chun}\
  \bibnamefont {Po}}, \bibinfo {author} {\bibfnamefont {Ashvin}\ \bibnamefont
  {Vishwanath}}, \ and\ \bibinfo {author} {\bibfnamefont {Haruki}\ \bibnamefont
  {Watanabe}},\ }\bibfield  {title} {\enquote {\bibinfo {title} {Symmetry-based
  indicators of band topology in the 230 space groups},}\ }\href {\doibase
  10.1038/s41467-017-00133-2} {\bibfield  {journal} {\bibinfo  {journal}
  {Nature Communications}\ }\textbf {\bibinfo {volume} {8}},\ \bibinfo {pages}
  {50} (\bibinfo {year} {2017})}\BibitemShut {NoStop}%
\bibitem [{\citenamefont {Bradlyn}\ \emph
  {et~al.}(2017{\natexlab{a}})\citenamefont {Bradlyn}, \citenamefont {Elcoro},
  \citenamefont {Cano}, \citenamefont {Vergniory}, \citenamefont {Wang},
  \citenamefont {Felser}, \citenamefont {Aroyo},\ and\ \citenamefont
  {Bernevig}}]{bradlyn2017topological}%
  \BibitemOpen
  \bibfield  {author} {\bibinfo {author} {\bibfnamefont {Barry}\ \bibnamefont
  {Bradlyn}}, \bibinfo {author} {\bibfnamefont {L}~\bibnamefont {Elcoro}},
  \bibinfo {author} {\bibfnamefont {Jennifer}\ \bibnamefont {Cano}}, \bibinfo
  {author} {\bibfnamefont {MG}~\bibnamefont {Vergniory}}, \bibinfo {author}
  {\bibfnamefont {Zhijun}\ \bibnamefont {Wang}}, \bibinfo {author}
  {\bibfnamefont {C}~\bibnamefont {Felser}}, \bibinfo {author} {\bibfnamefont
  {MI}~\bibnamefont {Aroyo}}, \ and\ \bibinfo {author} {\bibfnamefont
  {B~Andrei}\ \bibnamefont {Bernevig}},\ }\bibfield  {title} {\enquote
  {\bibinfo {title} {Topological quantum chemistry},}\ }\href@noop {}
  {\bibfield  {journal} {\bibinfo  {journal} {Nature}\ }\textbf {\bibinfo
  {volume} {547}},\ \bibinfo {pages} {298} (\bibinfo {year}
  {2017}{\natexlab{a}})}\BibitemShut {NoStop}%
\bibitem [{\citenamefont {Hwang}\ \emph {et~al.}(2019)\citenamefont {Hwang},
  \citenamefont {Ahn},\ and\ \citenamefont {Yang}}]{AhnJung}%
  \BibitemOpen
  \bibfield  {author} {\bibinfo {author} {\bibfnamefont {Yoonseok}\
  \bibnamefont {Hwang}}, \bibinfo {author} {\bibfnamefont {Junyeong}\
  \bibnamefont {Ahn}}, \ and\ \bibinfo {author} {\bibfnamefont {Bohm-Jung}\
  \bibnamefont {Yang}},\ }\bibfield  {title} {\enquote {\bibinfo {title}
  {Fragile topology protected by inversion symmetry: Diagnosis, bulk-boundary
  correspondence, and wilson loop},}\ }\href {\doibase
  10.1103/PhysRevB.100.205126} {\bibfield  {journal} {\bibinfo  {journal}
  {Phys. Rev. B}\ }\textbf {\bibinfo {volume} {100}},\ \bibinfo {pages}
  {205126} (\bibinfo {year} {2019})}\BibitemShut {NoStop}%
\bibitem [{\citenamefont {Dubinkin}\ and\ \citenamefont
  {Hughes}(2020)}]{dubinkin2020entanglement}%
  \BibitemOpen
  \bibfield  {author} {\bibinfo {author} {\bibfnamefont {Oleg}\ \bibnamefont
  {Dubinkin}}\ and\ \bibinfo {author} {\bibfnamefont {Taylor~L}\ \bibnamefont
  {Hughes}},\ }\bibfield  {title} {\enquote {\bibinfo {title} {Entanglement
  signatures of multipolar higher order topological phases},}\ }\href@noop {}
  {\bibfield  {journal} {\bibinfo  {journal} {arXiv preprint arXiv:2002.08385}\
  } (\bibinfo {year} {2020})}\BibitemShut {NoStop}%
\bibitem [{\citenamefont {Wieder}\ \emph {et~al.}(2020)\citenamefont {Wieder},
  \citenamefont {Wang}, \citenamefont {Cano}, \citenamefont {Dai},
  \citenamefont {Schoop}, \citenamefont {Bradlyn},\ and\ \citenamefont
  {Bernevig}}]{Wieder2020}%
  \BibitemOpen
  \bibfield  {author} {\bibinfo {author} {\bibfnamefont {Benjamin~J}\
  \bibnamefont {Wieder}}, \bibinfo {author} {\bibfnamefont {Zhijun}\
  \bibnamefont {Wang}}, \bibinfo {author} {\bibfnamefont {Jennifer}\
  \bibnamefont {Cano}}, \bibinfo {author} {\bibfnamefont {Xi}~\bibnamefont
  {Dai}}, \bibinfo {author} {\bibfnamefont {Leslie~M}\ \bibnamefont {Schoop}},
  \bibinfo {author} {\bibfnamefont {Barry}\ \bibnamefont {Bradlyn}}, \ and\
  \bibinfo {author} {\bibfnamefont {B~Andrei}\ \bibnamefont {Bernevig}},\
  }\bibfield  {title} {\enquote {\bibinfo {title} {{Strong and fragile
  topological Dirac semimetals with higher-order Fermi arcs}},}\ }\href
  {\doibase 10.1038/s41467-020-14443-5} {\bibfield  {journal} {\bibinfo
  {journal} {Nature Communications}\ }\textbf {\bibinfo {volume} {11}},\
  \bibinfo {pages} {627} (\bibinfo {year} {2020})}\BibitemShut {NoStop}%
\bibitem [{\citenamefont {Marzari}\ and\ \citenamefont
  {Vanderbilt}(1997)}]{Marzari97}%
  \BibitemOpen
  \bibfield  {author} {\bibinfo {author} {\bibfnamefont {Nicola}\ \bibnamefont
  {Marzari}}\ and\ \bibinfo {author} {\bibfnamefont {David}\ \bibnamefont
  {Vanderbilt}},\ }\bibfield  {title} {\enquote {\bibinfo {title} {Maximally
  localized generalized wannier functions for composite energy bands},}\ }\href
  {\doibase 10.1103/PhysRevB.56.12847} {\bibfield  {journal} {\bibinfo
  {journal} {Phys. Rev. B}\ }\textbf {\bibinfo {volume} {56}},\ \bibinfo
  {pages} {12847--12865} (\bibinfo {year} {1997})}\BibitemShut {NoStop}%
\bibitem [{\citenamefont {Marzari}\ \emph {et~al.}(2012)\citenamefont
  {Marzari}, \citenamefont {Mostofi}, \citenamefont {Yates}, \citenamefont
  {Souza},\ and\ \citenamefont {Vanderbilt}}]{Marzari12}%
  \BibitemOpen
  \bibfield  {author} {\bibinfo {author} {\bibfnamefont {Nicola}\ \bibnamefont
  {Marzari}}, \bibinfo {author} {\bibfnamefont {Arash~A.}\ \bibnamefont
  {Mostofi}}, \bibinfo {author} {\bibfnamefont {Jonathan~R.}\ \bibnamefont
  {Yates}}, \bibinfo {author} {\bibfnamefont {Ivo}\ \bibnamefont {Souza}}, \
  and\ \bibinfo {author} {\bibfnamefont {David}\ \bibnamefont {Vanderbilt}},\
  }\bibfield  {title} {\enquote {\bibinfo {title} {Maximally localized wannier
  functions: Theory and applications},}\ }\href {\doibase
  10.1103/RevModPhys.84.1419} {\bibfield  {journal} {\bibinfo  {journal} {Rev.
  Mod. Phys.}\ }\textbf {\bibinfo {volume} {84}},\ \bibinfo {pages}
  {1419--1475} (\bibinfo {year} {2012})}\BibitemShut {NoStop}%
\bibitem [{\citenamefont {Yu}\ \emph {et~al.}(2011)\citenamefont {Yu},
  \citenamefont {Qi}, \citenamefont {Bernevig}, \citenamefont {Fang},\ and\
  \citenamefont {Dai}}]{qibernevigwilsonloop}%
  \BibitemOpen
  \bibfield  {author} {\bibinfo {author} {\bibfnamefont {Rui}\ \bibnamefont
  {Yu}}, \bibinfo {author} {\bibfnamefont {Xiao~Liang}\ \bibnamefont {Qi}},
  \bibinfo {author} {\bibfnamefont {Andrei}\ \bibnamefont {Bernevig}}, \bibinfo
  {author} {\bibfnamefont {Zhong}\ \bibnamefont {Fang}}, \ and\ \bibinfo
  {author} {\bibfnamefont {Xi}~\bibnamefont {Dai}},\ }\bibfield  {title}
  {\enquote {\bibinfo {title} {Equivalent expression of ${\mathbb{z}}_{2}$
  topological invariant for band insulators using the non-abelian berry
  connection},}\ }\href {\doibase 10.1103/PhysRevB.84.075119} {\bibfield
  {journal} {\bibinfo  {journal} {Phys. Rev. B}\ }\textbf {\bibinfo {volume}
  {84}},\ \bibinfo {pages} {075119} (\bibinfo {year} {2011})}\BibitemShut
  {NoStop}%
\bibitem [{\citenamefont {You}\ \emph {et~al.}(2019)\citenamefont {You},
  \citenamefont {Burnell},\ and\ \citenamefont {Hughes}}]{YouBurnell}%
  \BibitemOpen
  \bibfield  {author} {\bibinfo {author} {\bibfnamefont {Yizhi}\ \bibnamefont
  {You}}, \bibinfo {author} {\bibfnamefont {FJ}~\bibnamefont {Burnell}}, \ and\
  \bibinfo {author} {\bibfnamefont {Taylor~L}\ \bibnamefont {Hughes}},\
  }\bibfield  {title} {\enquote {\bibinfo {title} {Multipolar topological field
  theories: Bridging higher order topological insulators and fractons},}\
  }\href@noop {} {\bibfield  {journal} {\bibinfo  {journal} {arXiv preprint
  arXiv:1909.05868}\ } (\bibinfo {year} {2019})}\BibitemShut {NoStop}%
\bibitem [{\citenamefont {Zak}(1980)}]{zak80}%
  \BibitemOpen
  \bibfield  {author} {\bibinfo {author} {\bibfnamefont {J.}~\bibnamefont
  {Zak}},\ }\bibfield  {title} {\enquote {\bibinfo {title} {Symmetry
  specification of bands in solids},}\ }\href {\doibase
  10.1103/PhysRevLett.45.1025} {\bibfield  {journal} {\bibinfo  {journal}
  {Phys. Rev. Lett.}\ }\textbf {\bibinfo {volume} {45}},\ \bibinfo {pages}
  {1025--1028} (\bibinfo {year} {1980})}\BibitemShut {NoStop}%
\bibitem [{\citenamefont {Zak}(1982)}]{zak1982band}%
  \BibitemOpen
  \bibfield  {author} {\bibinfo {author} {\bibfnamefont {J}~\bibnamefont
  {Zak}},\ }\bibfield  {title} {\enquote {\bibinfo {title} {Band
  representations of space groups},}\ }\href@noop {} {\bibfield  {journal}
  {\bibinfo  {journal} {Physical Review B}\ }\textbf {\bibinfo {volume} {26}},\
  \bibinfo {pages} {3010} (\bibinfo {year} {1982})}\BibitemShut {NoStop}%
\bibitem [{\citenamefont {Michel}\ and\ \citenamefont
  {Zak}(2000)}]{michel2000elementary}%
  \BibitemOpen
  \bibfield  {author} {\bibinfo {author} {\bibfnamefont {L}~\bibnamefont
  {Michel}}\ and\ \bibinfo {author} {\bibfnamefont {J}~\bibnamefont {Zak}},\
  }\bibfield  {title} {\enquote {\bibinfo {title} {Elementary energy bands in
  crystalline solids},}\ }\href@noop {} {\bibfield  {journal} {\bibinfo
  {journal} {EPL (Europhysics Letters)}\ }\textbf {\bibinfo {volume} {50}},\
  \bibinfo {pages} {519} (\bibinfo {year} {2000})}\BibitemShut {NoStop}%
\bibitem [{\citenamefont {Vergniory}\ \emph {et~al.}(2017)\citenamefont
  {Vergniory}, \citenamefont {Elcoro}, \citenamefont {Wang}, \citenamefont
  {Cano}, \citenamefont {Felser}, \citenamefont {Aroyo}, \citenamefont
  {Bernevig},\ and\ \citenamefont {Bradlyn}}]{vergniory2017graph}%
  \BibitemOpen
  \bibfield  {author} {\bibinfo {author} {\bibfnamefont {MG}~\bibnamefont
  {Vergniory}}, \bibinfo {author} {\bibfnamefont {L}~\bibnamefont {Elcoro}},
  \bibinfo {author} {\bibfnamefont {Zhijun}\ \bibnamefont {Wang}}, \bibinfo
  {author} {\bibfnamefont {Jennifer}\ \bibnamefont {Cano}}, \bibinfo {author}
  {\bibfnamefont {C}~\bibnamefont {Felser}}, \bibinfo {author} {\bibfnamefont
  {MI}~\bibnamefont {Aroyo}}, \bibinfo {author} {\bibfnamefont {B~Andrei}\
  \bibnamefont {Bernevig}}, \ and\ \bibinfo {author} {\bibfnamefont {Barry}\
  \bibnamefont {Bradlyn}},\ }\bibfield  {title} {\enquote {\bibinfo {title}
  {Graph theory data for topological quantum chemistry},}\ }\href@noop {}
  {\bibfield  {journal} {\bibinfo  {journal} {Physical Review E}\ }\textbf
  {\bibinfo {volume} {96}},\ \bibinfo {pages} {023310} (\bibinfo {year}
  {2017})}\BibitemShut {NoStop}%
\bibitem [{Wik()}]{Wiki}%
  \BibitemOpen
  \href@noop {} {\enquote {\bibinfo {title} {Wikipedia entry for wallpaper
  group},}\ }\bibinfo {howpublished}
  {\url{https://en.wikipedia.org/wiki/Wallpaper_group}}\BibitemShut {NoStop}%
\bibitem [{\citenamefont {Cano}\ \emph
  {et~al.}(2018{\natexlab{a}})\citenamefont {Cano}, \citenamefont {Bradlyn},
  \citenamefont {Wang}, \citenamefont {Elcoro}, \citenamefont {Vergniory},
  \citenamefont {Felser}, \citenamefont {Aroyo},\ and\ \citenamefont
  {Bernevig}}]{cano2018building}%
  \BibitemOpen
  \bibfield  {author} {\bibinfo {author} {\bibfnamefont {Jennifer}\
  \bibnamefont {Cano}}, \bibinfo {author} {\bibfnamefont {Barry}\ \bibnamefont
  {Bradlyn}}, \bibinfo {author} {\bibfnamefont {Zhijun}\ \bibnamefont {Wang}},
  \bibinfo {author} {\bibfnamefont {L}~\bibnamefont {Elcoro}}, \bibinfo
  {author} {\bibfnamefont {MG}~\bibnamefont {Vergniory}}, \bibinfo {author}
  {\bibfnamefont {C}~\bibnamefont {Felser}}, \bibinfo {author} {\bibfnamefont
  {MI}~\bibnamefont {Aroyo}}, \ and\ \bibinfo {author} {\bibfnamefont
  {B~Andrei}\ \bibnamefont {Bernevig}},\ }\bibfield  {title} {\enquote
  {\bibinfo {title} {Building blocks of topological quantum chemistry:
  Elementary band representations},}\ }\href@noop {} {\bibfield  {journal}
  {\bibinfo  {journal} {Physical Review B}\ }\textbf {\bibinfo {volume} {97}},\
  \bibinfo {pages} {035139} (\bibinfo {year} {2018}{\natexlab{a}})}\BibitemShut
  {NoStop}%
\bibitem [{\citenamefont {Fidkowski}\ \emph {et~al.}(2011)\citenamefont
  {Fidkowski}, \citenamefont {Jackson},\ and\ \citenamefont
  {Klich}}]{Fidkowski11}%
  \BibitemOpen
  \bibfield  {author} {\bibinfo {author} {\bibfnamefont {Lukasz}\ \bibnamefont
  {Fidkowski}}, \bibinfo {author} {\bibfnamefont {T.~S.}\ \bibnamefont
  {Jackson}}, \ and\ \bibinfo {author} {\bibfnamefont {Israel}\ \bibnamefont
  {Klich}},\ }\bibfield  {title} {\enquote {\bibinfo {title} {{Model
  characterization of gapless edge modes of topological insulators using
  intermediate Brillouin-zone functions}},}\ }\href {\doibase
  10.1103/PhysRevLett.107.036601} {\bibfield  {journal} {\bibinfo  {journal}
  {Physical Review Letters}\ }\textbf {\bibinfo {volume} {107}},\ \bibinfo
  {pages} {1--4} (\bibinfo {year} {2011})}\BibitemShut {NoStop}%
\bibitem [{\citenamefont {Yang}\ \emph {et~al.}(2020)\citenamefont {Yang},
  \citenamefont {Li}, \citenamefont {Duan},\ and\ \citenamefont
  {Xu}}]{TypeIIQI}%
  \BibitemOpen
  \bibfield  {author} {\bibinfo {author} {\bibfnamefont {Yan-Bin}\ \bibnamefont
  {Yang}}, \bibinfo {author} {\bibfnamefont {Kai}\ \bibnamefont {Li}}, \bibinfo
  {author} {\bibfnamefont {L.-M.}\ \bibnamefont {Duan}}, \ and\ \bibinfo
  {author} {\bibfnamefont {Yong}\ \bibnamefont {Xu}},\ }\bibfield  {title}
  {\enquote {\bibinfo {title} {Type-ii quadrupole topological insulators},}\
  }\href {\doibase 10.1103/PhysRevResearch.2.033029} {\bibfield  {journal}
  {\bibinfo  {journal} {Phys. Rev. Research}\ }\textbf {\bibinfo {volume}
  {2}},\ \bibinfo {pages} {033029} (\bibinfo {year} {2020})}\BibitemShut
  {NoStop}%
\bibitem [{\citenamefont {Goldstone}\ and\ \citenamefont
  {Wilczek}(1981)}]{Goldstone1981}%
  \BibitemOpen
  \bibfield  {author} {\bibinfo {author} {\bibfnamefont {Jeffrey}\ \bibnamefont
  {Goldstone}}\ and\ \bibinfo {author} {\bibfnamefont {Frank}\ \bibnamefont
  {Wilczek}},\ }\bibfield  {title} {\enquote {\bibinfo {title} {Fractional
  quantum numbers on solitons},}\ }\href {\doibase 10.1103/PhysRevLett.47.986}
  {\bibfield  {journal} {\bibinfo  {journal} {Phys. Rev. Lett.}\ }\textbf
  {\bibinfo {volume} {47}},\ \bibinfo {pages} {986--989} (\bibinfo {year}
  {1981})}\BibitemShut {NoStop}%
\bibitem [{\citenamefont {Wieder}\ and\ \citenamefont
  {Bernevig}(2018)}]{Wieder18}%
  \BibitemOpen
  \bibfield  {author} {\bibinfo {author} {\bibfnamefont {Benjamin~J}\
  \bibnamefont {Wieder}}\ and\ \bibinfo {author} {\bibfnamefont {B~Andrei}\
  \bibnamefont {Bernevig}},\ }\bibfield  {title} {\enquote {\bibinfo {title}
  {The axion insulator as a pump of fragile topology},}\ }\href@noop {}
  {\bibfield  {journal} {\bibinfo  {journal} {arXiv preprint arXiv:1810.02373}\
  } (\bibinfo {year} {2018})}\BibitemShut {NoStop}%
\bibitem [{\citenamefont {van Miert}\ and\ \citenamefont
  {Ortix}(2018)}]{MiertOrtix}%
  \BibitemOpen
  \bibfield  {author} {\bibinfo {author} {\bibfnamefont {Guido}\ \bibnamefont
  {van Miert}}\ and\ \bibinfo {author} {\bibfnamefont {Carmine}\ \bibnamefont
  {Ortix}},\ }\bibfield  {title} {\enquote {\bibinfo {title} {Higher-order
  topological insulators protected by inversion and rotoinversion
  symmetries},}\ }\href {\doibase 10.1103/PhysRevB.98.081110} {\bibfield
  {journal} {\bibinfo  {journal} {Phys. Rev. B}\ }\textbf {\bibinfo {volume}
  {98}},\ \bibinfo {pages} {081110} (\bibinfo {year} {2018})}\BibitemShut
  {NoStop}%
\bibitem [{\citenamefont {Rhim}\ \emph {et~al.}(2017)\citenamefont {Rhim},
  \citenamefont {Behrends},\ and\ \citenamefont {Bardarson}}]{rhim2017}%
  \BibitemOpen
  \bibfield  {author} {\bibinfo {author} {\bibfnamefont {Jun-Won}\ \bibnamefont
  {Rhim}}, \bibinfo {author} {\bibfnamefont {Jan}\ \bibnamefont {Behrends}}, \
  and\ \bibinfo {author} {\bibfnamefont {Jens~H.}\ \bibnamefont {Bardarson}},\
  }\bibfield  {title} {\enquote {\bibinfo {title} {Bulk-boundary correspondence
  from the intercellular zak phase},}\ }\href {\doibase
  10.1103/PhysRevB.95.035421} {\bibfield  {journal} {\bibinfo  {journal} {Phys.
  Rev. B}\ }\textbf {\bibinfo {volume} {95}},\ \bibinfo {pages} {035421}
  (\bibinfo {year} {2017})}\BibitemShut {NoStop}%
\bibitem [{\citenamefont {Song}\ \emph
  {et~al.}(2017{\natexlab{b}})\citenamefont {Song}, \citenamefont {Huang},
  \citenamefont {Fu},\ and\ \citenamefont {Hermele}}]{song2017topological}%
  \BibitemOpen
  \bibfield  {author} {\bibinfo {author} {\bibfnamefont {Hao}\ \bibnamefont
  {Song}}, \bibinfo {author} {\bibfnamefont {Sheng-Jie}\ \bibnamefont {Huang}},
  \bibinfo {author} {\bibfnamefont {Liang}\ \bibnamefont {Fu}}, \ and\ \bibinfo
  {author} {\bibfnamefont {Michael}\ \bibnamefont {Hermele}},\ }\bibfield
  {title} {\enquote {\bibinfo {title} {Topological phases protected by point
  group symmetry},}\ }\href@noop {} {\bibfield  {journal} {\bibinfo  {journal}
  {Physical Review X}\ }\textbf {\bibinfo {volume} {7}},\ \bibinfo {pages}
  {011020} (\bibinfo {year} {2017}{\natexlab{b}})}\BibitemShut {NoStop}%
\bibitem [{\citenamefont {Huang}\ \emph {et~al.}(2017)\citenamefont {Huang},
  \citenamefont {Song}, \citenamefont {Huang},\ and\ \citenamefont
  {Hermele}}]{Huang17}%
  \BibitemOpen
  \bibfield  {author} {\bibinfo {author} {\bibfnamefont {Sheng-Jie}\
  \bibnamefont {Huang}}, \bibinfo {author} {\bibfnamefont {Hao}\ \bibnamefont
  {Song}}, \bibinfo {author} {\bibfnamefont {Yi-Ping}\ \bibnamefont {Huang}}, \
  and\ \bibinfo {author} {\bibfnamefont {Michael}\ \bibnamefont {Hermele}},\
  }\bibfield  {title} {\enquote {\bibinfo {title} {Building crystalline
  topological phases from lower-dimensional states},}\ }\href {\doibase
  10.1103/PhysRevB.96.205106} {\bibfield  {journal} {\bibinfo  {journal} {Phys.
  Rev. B}\ }\textbf {\bibinfo {volume} {96}},\ \bibinfo {pages} {205106}
  (\bibinfo {year} {2017})}\BibitemShut {NoStop}%
\bibitem [{\citenamefont {Song}\ \emph
  {et~al.}(2018{\natexlab{a}})\citenamefont {Song}, \citenamefont {Zhang},
  \citenamefont {Fang},\ and\ \citenamefont {Fang}}]{Song2018}%
  \BibitemOpen
  \bibfield  {author} {\bibinfo {author} {\bibfnamefont {Zhida}\ \bibnamefont
  {Song}}, \bibinfo {author} {\bibfnamefont {Tiantian}\ \bibnamefont {Zhang}},
  \bibinfo {author} {\bibfnamefont {Zhong}\ \bibnamefont {Fang}}, \ and\
  \bibinfo {author} {\bibfnamefont {Chen}\ \bibnamefont {Fang}},\ }\bibfield
  {title} {\enquote {\bibinfo {title} {{Quantitative mappings between symmetry
  and topology in solids}},}\ }\href {\doibase 10.1038/s41467-018-06010-w}
  {\bibfield  {journal} {\bibinfo  {journal} {Nature Communications}\ }\textbf
  {\bibinfo {volume} {9}},\ \bibinfo {pages} {3530} (\bibinfo {year}
  {2018}{\natexlab{a}})}\BibitemShut {NoStop}%
\bibitem [{\citenamefont {Bradlyn}\ \emph
  {et~al.}(2017{\natexlab{b}})\citenamefont {Bradlyn}, \citenamefont {Elcoro},
  \citenamefont {Cano}, \citenamefont {Vergniory}, \citenamefont {Wang},
  \citenamefont {Felser}, \citenamefont {Aroyo},\ and\ \citenamefont
  {Bernevig}}]{Bradlyn17}%
  \BibitemOpen
  \bibfield  {author} {\bibinfo {author} {\bibfnamefont {Barry}\ \bibnamefont
  {Bradlyn}}, \bibinfo {author} {\bibfnamefont {L.}~\bibnamefont {Elcoro}},
  \bibinfo {author} {\bibfnamefont {Jennifer}\ \bibnamefont {Cano}}, \bibinfo
  {author} {\bibfnamefont {M.~G.}\ \bibnamefont {Vergniory}}, \bibinfo {author}
  {\bibfnamefont {Zhijun}\ \bibnamefont {Wang}}, \bibinfo {author}
  {\bibfnamefont {C.}~\bibnamefont {Felser}}, \bibinfo {author} {\bibfnamefont
  {M.~I.}\ \bibnamefont {Aroyo}}, \ and\ \bibinfo {author} {\bibfnamefont
  {B.~Andrei}\ \bibnamefont {Bernevig}},\ }\bibfield  {title} {\enquote
  {\bibinfo {title} {Topological quantum chemistry},}\ }\href
  {https://doi.org/10.1038/nature23268} {\bibfield  {journal} {\bibinfo
  {journal} {Nature}\ }\textbf {\bibinfo {volume} {547}},\ \bibinfo {pages}
  {298 EP --} (\bibinfo {year} {2017}{\natexlab{b}})},\ \bibinfo {note}
  {article}\BibitemShut {NoStop}%
\bibitem [{\citenamefont {Gross}\ and\ \citenamefont
  {Bloch}(2017)}]{ColdAtoms}%
  \BibitemOpen
  \bibfield  {author} {\bibinfo {author} {\bibfnamefont {Christian}\
  \bibnamefont {Gross}}\ and\ \bibinfo {author} {\bibfnamefont {Immanuel}\
  \bibnamefont {Bloch}},\ }\bibfield  {title} {\enquote {\bibinfo {title}
  {Quantum simulations with ultracold atoms in optical lattices},}\ }\href
  {\doibase 10.1126/science.aal3837} {\bibfield  {journal} {\bibinfo  {journal}
  {Science}\ }\textbf {\bibinfo {volume} {357}},\ \bibinfo {pages} {995--1001}
  (\bibinfo {year} {2017})},\ \Eprint
  {http://arxiv.org/abs/https://science.sciencemag.org/content/357/6355/995.full.pdf}
  {https://science.sciencemag.org/content/357/6355/995.full.pdf} \BibitemShut
  {NoStop}%
\bibitem [{\citenamefont {Teo}\ and\ \citenamefont {Kane}(2010)}]{Teo10}%
  \BibitemOpen
  \bibfield  {author} {\bibinfo {author} {\bibfnamefont {Jeffrey C.~Y.}\
  \bibnamefont {Teo}}\ and\ \bibinfo {author} {\bibfnamefont {C.~L.}\
  \bibnamefont {Kane}},\ }\bibfield  {title} {\enquote {\bibinfo {title}
  {Topological defects and gapless modes in insulators and superconductors},}\
  }\href {\doibase 10.1103/PhysRevB.82.115120} {\bibfield  {journal} {\bibinfo
  {journal} {Phys. Rev. B}\ }\textbf {\bibinfo {volume} {82}},\ \bibinfo
  {pages} {115120} (\bibinfo {year} {2010})}\BibitemShut {NoStop}%
\bibitem [{\citenamefont {Po}\ \emph {et~al.}(2018)\citenamefont {Po},
  \citenamefont {Watanabe},\ and\ \citenamefont {Vishwanath}}]{PoFragile}%
  \BibitemOpen
  \bibfield  {author} {\bibinfo {author} {\bibfnamefont {Hoi~Chun}\
  \bibnamefont {Po}}, \bibinfo {author} {\bibfnamefont {Haruki}\ \bibnamefont
  {Watanabe}}, \ and\ \bibinfo {author} {\bibfnamefont {Ashvin}\ \bibnamefont
  {Vishwanath}},\ }\bibfield  {title} {\enquote {\bibinfo {title} {Fragile
  topology and wannier obstructions},}\ }\href {\doibase
  10.1103/PhysRevLett.121.126402} {\bibfield  {journal} {\bibinfo  {journal}
  {Phys. Rev. Lett.}\ }\textbf {\bibinfo {volume} {121}},\ \bibinfo {pages}
  {126402} (\bibinfo {year} {2018})}\BibitemShut {NoStop}%
\bibitem [{\citenamefont {Fukui}(2019)}]{Fukui18}%
  \BibitemOpen
  \bibfield  {author} {\bibinfo {author} {\bibfnamefont {Takahiro}\
  \bibnamefont {Fukui}},\ }\bibfield  {title} {\enquote {\bibinfo {title}
  {Dirac fermion model associated with a second-order topological insulator},}\
  }\href@noop {} {\bibfield  {journal} {\bibinfo  {journal} {Physical Review
  B}\ }\textbf {\bibinfo {volume} {99}},\ \bibinfo {pages} {165129} (\bibinfo
  {year} {2019})}\BibitemShut {NoStop}%
\bibitem [{\citenamefont {Song}\ \emph
  {et~al.}(2018{\natexlab{b}})\citenamefont {Song}, \citenamefont {Zhang},\
  and\ \citenamefont {Fang}}]{SongSemimetal}%
  \BibitemOpen
  \bibfield  {author} {\bibinfo {author} {\bibfnamefont {Zhida}\ \bibnamefont
  {Song}}, \bibinfo {author} {\bibfnamefont {Tiantian}\ \bibnamefont {Zhang}},
  \ and\ \bibinfo {author} {\bibfnamefont {Chen}\ \bibnamefont {Fang}},\
  }\bibfield  {title} {\enquote {\bibinfo {title} {Diagnosis for nonmagnetic
  topological semimetals in the absence of spin-orbital coupling},}\ }\href
  {\doibase 10.1103/PhysRevX.8.031069} {\bibfield  {journal} {\bibinfo
  {journal} {Phys. Rev. X}\ }\textbf {\bibinfo {volume} {8}},\ \bibinfo {pages}
  {031069} (\bibinfo {year} {2018}{\natexlab{b}})}\BibitemShut {NoStop}%
\bibitem [{\citenamefont {Hughes}\ \emph {et~al.}(2011)\citenamefont {Hughes},
  \citenamefont {Prodan},\ and\ \citenamefont {Bernevig}}]{hughes2011}%
  \BibitemOpen
  \bibfield  {author} {\bibinfo {author} {\bibfnamefont {Taylor~L}\
  \bibnamefont {Hughes}}, \bibinfo {author} {\bibfnamefont {Emil}\ \bibnamefont
  {Prodan}}, \ and\ \bibinfo {author} {\bibfnamefont {B~Andrei}\ \bibnamefont
  {Bernevig}},\ }\bibfield  {title} {\enquote {\bibinfo {title}
  {Inversion-symmetric topological insulators},}\ }\href@noop {} {\bibfield
  {journal} {\bibinfo  {journal} {Phys. Rev. B}\ }\textbf {\bibinfo {volume}
  {83}},\ \bibinfo {pages} {245132} (\bibinfo {year} {2011})}\BibitemShut
  {NoStop}%
\bibitem [{\citenamefont {Turner}\ \emph {et~al.}(2012)\citenamefont {Turner},
  \citenamefont {Zhang}, \citenamefont {Mong},\ and\ \citenamefont
  {Vishwanath}}]{turner2012}%
  \BibitemOpen
  \bibfield  {author} {\bibinfo {author} {\bibfnamefont {Ari~M}\ \bibnamefont
  {Turner}}, \bibinfo {author} {\bibfnamefont {Yi}~\bibnamefont {Zhang}},
  \bibinfo {author} {\bibfnamefont {Roger~SK}\ \bibnamefont {Mong}}, \ and\
  \bibinfo {author} {\bibfnamefont {Ashvin}\ \bibnamefont {Vishwanath}},\
  }\bibfield  {title} {\enquote {\bibinfo {title} {Quantized response and
  topology of magnetic insulators with inversion symmetry},}\ }\href@noop {}
  {\bibfield  {journal} {\bibinfo  {journal} {Phys. Rev. B}\ }\textbf {\bibinfo
  {volume} {85}},\ \bibinfo {pages} {165120} (\bibinfo {year}
  {2012})}\BibitemShut {NoStop}%
\bibitem [{\citenamefont {Baez}(2009)}]{Torsor}%
  \BibitemOpen
  \bibfield  {author} {\bibinfo {author} {\bibfnamefont {John}\ \bibnamefont
  {Baez}},\ }\href {http://math.ucr.edu/home/baez/torsors.html} {\enquote
  {\bibinfo {title} {Torsors made easy},}\ } (\bibinfo {year}
  {2009})\BibitemShut {NoStop}%
\bibitem [{\citenamefont {Bradlyn}\ \emph
  {et~al.}(2017{\natexlab{c}})\citenamefont {Bradlyn}, \citenamefont {Elcoro},
  \citenamefont {Cano}, \citenamefont {Vergniory}, \citenamefont {Wang},
  \citenamefont {Felser}, \citenamefont {Aroyo},\ and\ \citenamefont
  {Bernevig}}]{Bradlyn2017c}%
  \BibitemOpen
  \bibfield  {author} {\bibinfo {author} {\bibfnamefont {Barry}\ \bibnamefont
  {Bradlyn}}, \bibinfo {author} {\bibfnamefont {L.}~\bibnamefont {Elcoro}},
  \bibinfo {author} {\bibfnamefont {Jennifer}\ \bibnamefont {Cano}}, \bibinfo
  {author} {\bibfnamefont {M.~G.}\ \bibnamefont {Vergniory}}, \bibinfo {author}
  {\bibfnamefont {Zhijun}\ \bibnamefont {Wang}}, \bibinfo {author}
  {\bibfnamefont {C.}~\bibnamefont {Felser}}, \bibinfo {author} {\bibfnamefont
  {Mois~I.}\ \bibnamefont {Aroyo}}, \ and\ \bibinfo {author} {\bibfnamefont
  {B.~Andrei}\ \bibnamefont {Bernevig}},\ }\bibfield  {title} {\enquote
  {\bibinfo {title} {{Topological quantum chemistry}},}\ }\href {\doibase
  10.1038/nature23268} {\bibfield  {journal} {\bibinfo  {journal} {Nature}\
  }\textbf {\bibinfo {volume} {547}},\ \bibinfo {pages} {298--305} (\bibinfo
  {year} {2017}{\natexlab{c}})},\ \Eprint {http://arxiv.org/abs/NIHMS150003}
  {arXiv:NIHMS150003} \BibitemShut {NoStop}%
\bibitem [{\citenamefont {Cano}\ \emph
  {et~al.}(2018{\natexlab{b}})\citenamefont {Cano}, \citenamefont {Bradlyn},
  \citenamefont {Wang}, \citenamefont {Elcoro}, \citenamefont {Vergniory},
  \citenamefont {Felser}, \citenamefont {Aroyo},\ and\ \citenamefont
  {Bernevig}}]{Cano2018a}%
  \BibitemOpen
  \bibfield  {author} {\bibinfo {author} {\bibfnamefont {Jennifer}\
  \bibnamefont {Cano}}, \bibinfo {author} {\bibfnamefont {Barry}\ \bibnamefont
  {Bradlyn}}, \bibinfo {author} {\bibfnamefont {Zhijun}\ \bibnamefont {Wang}},
  \bibinfo {author} {\bibfnamefont {L.}~\bibnamefont {Elcoro}}, \bibinfo
  {author} {\bibfnamefont {M.~G.}\ \bibnamefont {Vergniory}}, \bibinfo {author}
  {\bibfnamefont {C.}~\bibnamefont {Felser}}, \bibinfo {author} {\bibfnamefont
  {Mois~I.}\ \bibnamefont {Aroyo}}, \ and\ \bibinfo {author} {\bibfnamefont
  {B.~Andrei}\ \bibnamefont {Bernevig}},\ }\bibfield  {title} {\enquote
  {\bibinfo {title} {{Building blocks of topological quantum chemistry:
  Elementary band representations}},}\ }\href {\doibase
  10.1103/PhysRevB.97.035139} {\bibfield  {journal} {\bibinfo  {journal}
  {Physical Review B}\ }\textbf {\bibinfo {volume} {97}},\ \bibinfo {pages}
  {035139} (\bibinfo {year} {2018}{\natexlab{b}})},\ \Eprint
  {http://arxiv.org/abs/1709.01935} {arXiv:1709.01935} \BibitemShut {NoStop}%
\bibitem [{\citenamefont {Evarestov}(2007)}]{evarestov2007quantum}%
  \BibitemOpen
  \bibfield  {author} {\bibinfo {author} {\bibfnamefont {Robert~A}\
  \bibnamefont {Evarestov}},\ }\href@noop {} {\emph {\bibinfo {title} {Quantum
  chemistry of solids: the LCAO first principles treatment of crystals}}},\
  Vol.\ \bibinfo {volume} {153}\ (\bibinfo  {publisher} {Springer Science \&
  Business Media},\ \bibinfo {year} {2007})\BibitemShut {NoStop}%
\bibitem [{\citenamefont {Wieder}\ \emph {et~al.}(2018)\citenamefont {Wieder},
  \citenamefont {Bradlyn}, \citenamefont {Wang}, \citenamefont {Cano},
  \citenamefont {Kim}, \citenamefont {Kim}, \citenamefont {Rappe},
  \citenamefont {Kane},\ and\ \citenamefont {Bernevig}}]{WiederScience}%
  \BibitemOpen
  \bibfield  {author} {\bibinfo {author} {\bibfnamefont {Benjamin~J.}\
  \bibnamefont {Wieder}}, \bibinfo {author} {\bibfnamefont {Barry}\
  \bibnamefont {Bradlyn}}, \bibinfo {author} {\bibfnamefont {Zhijun}\
  \bibnamefont {Wang}}, \bibinfo {author} {\bibfnamefont {Jennifer}\
  \bibnamefont {Cano}}, \bibinfo {author} {\bibfnamefont {Youngkuk}\
  \bibnamefont {Kim}}, \bibinfo {author} {\bibfnamefont {Hyeong-Seok~D.}\
  \bibnamefont {Kim}}, \bibinfo {author} {\bibfnamefont {Andrew~M.}\
  \bibnamefont {Rappe}}, \bibinfo {author} {\bibfnamefont {C.~L.}\ \bibnamefont
  {Kane}}, \ and\ \bibinfo {author} {\bibfnamefont {B.~Andrei}\ \bibnamefont
  {Bernevig}},\ }\bibfield  {title} {\enquote {\bibinfo {title} {Wallpaper
  fermions and the nonsymmorphic dirac insulator},}\ }\href {\doibase
  10.1126/science.aan2802} {\bibfield  {journal} {\bibinfo  {journal}
  {Science}\ }\textbf {\bibinfo {volume} {361}},\ \bibinfo {pages} {246--251}
  (\bibinfo {year} {2018})},\ \Eprint
  {http://arxiv.org/abs/https://science.sciencemag.org/content/361/6399/246.full.pdf}
  {https://science.sciencemag.org/content/361/6399/246.full.pdf} \BibitemShut
  {NoStop}%
\bibitem [{\citenamefont {Kim}\ \emph {et~al.}(2019)\citenamefont {Kim},
  \citenamefont {Shiozaki},\ and\ \citenamefont {Murakami}}]{Kim2019}%
  \BibitemOpen
  \bibfield  {author} {\bibinfo {author} {\bibfnamefont {Heejae}\ \bibnamefont
  {Kim}}, \bibinfo {author} {\bibfnamefont {Ken}\ \bibnamefont {Shiozaki}}, \
  and\ \bibinfo {author} {\bibfnamefont {Shuichi}\ \bibnamefont {Murakami}},\
  }\bibfield  {title} {\enquote {\bibinfo {title} {Glide-symmetric magnetic
  topological crystalline insulators with inversion symmetry},}\ }\href
  {\doibase 10.1103/PhysRevB.100.165202} {\bibfield  {journal} {\bibinfo
  {journal} {Phys. Rev. B}\ }\textbf {\bibinfo {volume} {100}},\ \bibinfo
  {pages} {165202} (\bibinfo {year} {2019})}\BibitemShut {NoStop}%
\bibitem [{\citenamefont {Wieder}\ and\ \citenamefont
  {Kane}(2016)}]{WiederKane}%
  \BibitemOpen
  \bibfield  {author} {\bibinfo {author} {\bibfnamefont {Benjamin~J.}\
  \bibnamefont {Wieder}}\ and\ \bibinfo {author} {\bibfnamefont {C.~L.}\
  \bibnamefont {Kane}},\ }\bibfield  {title} {\enquote {\bibinfo {title}
  {Spin-orbit semimetals in the layer groups},}\ }\href {\doibase
  10.1103/PhysRevB.94.155108} {\bibfield  {journal} {\bibinfo  {journal} {Phys.
  Rev. B}\ }\textbf {\bibinfo {volume} {94}},\ \bibinfo {pages} {155108}
  (\bibinfo {year} {2016})}\BibitemShut {NoStop}%
\bibitem [{\citenamefont {Aroyo}\ \emph {et~al.}(2006)\citenamefont {Aroyo},
  \citenamefont {Perez-Mato}, \citenamefont {Capillas}, \citenamefont
  {Kroumova}, \citenamefont {Ivantchev}, \citenamefont {Madariaga},
  \citenamefont {Kirov},\ and\ \citenamefont {Wondratschek}}]{aroyo2006bilbao}%
  \BibitemOpen
  \bibfield  {author} {\bibinfo {author} {\bibfnamefont {Mois~Ilia}\
  \bibnamefont {Aroyo}}, \bibinfo {author} {\bibfnamefont {Juan~Manuel}\
  \bibnamefont {Perez-Mato}}, \bibinfo {author} {\bibfnamefont {Cesar}\
  \bibnamefont {Capillas}}, \bibinfo {author} {\bibfnamefont {Eli}\
  \bibnamefont {Kroumova}}, \bibinfo {author} {\bibfnamefont {Svetoslav}\
  \bibnamefont {Ivantchev}}, \bibinfo {author} {\bibfnamefont {Gotzon}\
  \bibnamefont {Madariaga}}, \bibinfo {author} {\bibfnamefont {Asen}\
  \bibnamefont {Kirov}}, \ and\ \bibinfo {author} {\bibfnamefont {Hans}\
  \bibnamefont {Wondratschek}},\ }\bibfield  {title} {\enquote {\bibinfo
  {title} {Bilbao crystallographic server: I. databases and crystallographic
  computing programs},}\ }\href@noop {} {\bibfield  {journal} {\bibinfo
  {journal} {Zeitschrift f{\"u}r Kristallographie-Crystalline Materials}\
  }\textbf {\bibinfo {volume} {221}},\ \bibinfo {pages} {15--27} (\bibinfo
  {year} {2006})}\BibitemShut {NoStop}%
\bibitem [{\citenamefont {{Crackwell}}\ and\ \citenamefont
  {{Wong}}(1967)}]{doublemagneticrep}%
  \BibitemOpen
  \bibfield  {author} {\bibinfo {author} {\bibfnamefont {A.~P.}\ \bibnamefont
  {{Crackwell}}}\ and\ \bibinfo {author} {\bibfnamefont {K.~C.}\ \bibnamefont
  {{Wong}}},\ }\bibfield  {title} {\enquote {\bibinfo {title} {{Double-valued
  corepresentations of magnetic point groups}},}\ }\href {\doibase
  10.1071/PH670173} {\bibfield  {journal} {\bibinfo  {journal} {Australian
  Journal of Physics}\ }\textbf {\bibinfo {volume} {20}},\ \bibinfo {pages}
  {173} (\bibinfo {year} {1967})}\BibitemShut {NoStop}%
\bibitem [{\citenamefont {Varnava}\ and\ \citenamefont
  {Vanderbilt}(2018)}]{Venderbilt}%
  \BibitemOpen
  \bibfield  {author} {\bibinfo {author} {\bibfnamefont {Nicodemos}\
  \bibnamefont {Varnava}}\ and\ \bibinfo {author} {\bibfnamefont {David}\
  \bibnamefont {Vanderbilt}},\ }\bibfield  {title} {\enquote {\bibinfo {title}
  {{Surfaces of axion insulators}},}\ }\href {\doibase
  10.1103/PhysRevB.98.245117} {\bibfield  {journal} {\bibinfo  {journal}
  {Physical Review B}\ }\textbf {\bibinfo {volume} {98}},\ \bibinfo {pages}
  {245117} (\bibinfo {year} {2018})}\BibitemShut {NoStop}%
\bibitem [{\citenamefont {Fu}\ and\ \citenamefont {Kane}()}]{FuKane}%
  \BibitemOpen
  \bibfield  {author} {\bibinfo {author} {\bibfnamefont {Liang}\ \bibnamefont
  {Fu}}\ and\ \bibinfo {author} {\bibfnamefont {C~L}\ \bibnamefont {Kane}},\
  }\bibfield  {title} {\enquote {\bibinfo {title} {{Topological insulators with
  inversion symmetry}},}\ }\href {\doibase 10.1103/PhysRevB.76.045302} {\
  10.1103/PhysRevB.76.045302}\BibitemShut {NoStop}%
\bibitem [{\citenamefont {Fu}\ and\ \citenamefont {Kane}(2006)}]{fu2006time}%
  \BibitemOpen
  \bibfield  {author} {\bibinfo {author} {\bibfnamefont {Liang}\ \bibnamefont
  {Fu}}\ and\ \bibinfo {author} {\bibfnamefont {Charles~L}\ \bibnamefont
  {Kane}},\ }\bibfield  {title} {\enquote {\bibinfo {title} {Time reversal
  polarization and a z 2 adiabatic spin pump},}\ }\href@noop {} {\bibfield
  {journal} {\bibinfo  {journal} {Physical Review B}\ }\textbf {\bibinfo
  {volume} {74}},\ \bibinfo {pages} {195312} (\bibinfo {year}
  {2006})}\BibitemShut {NoStop}%
\bibitem [{\citenamefont {Noh}\ \emph {et~al.}(2018)\citenamefont {Noh},
  \citenamefont {Benalcazar}, \citenamefont {Huang}, \citenamefont {Collins},
  \citenamefont {Chen}, \citenamefont {Hughes},\ and\ \citenamefont
  {Rechtsman}}]{noh2018}%
  \BibitemOpen
  \bibfield  {author} {\bibinfo {author} {\bibfnamefont {Jiho}\ \bibnamefont
  {Noh}}, \bibinfo {author} {\bibfnamefont {Wladimir~A.}\ \bibnamefont
  {Benalcazar}}, \bibinfo {author} {\bibfnamefont {Sheng}\ \bibnamefont
  {Huang}}, \bibinfo {author} {\bibfnamefont {Matthew~J.}\ \bibnamefont
  {Collins}}, \bibinfo {author} {\bibfnamefont {Kevin~P.}\ \bibnamefont
  {Chen}}, \bibinfo {author} {\bibfnamefont {Taylor~L.}\ \bibnamefont
  {Hughes}}, \ and\ \bibinfo {author} {\bibfnamefont {Mikael~C.}\ \bibnamefont
  {Rechtsman}},\ }\bibfield  {title} {\enquote {\bibinfo {title} {Topological
  protection of photonic mid-gap defect modes},}\ }\href {\doibase
  10.1038/s41566-018-0179-3} {\bibfield  {journal} {\bibinfo  {journal} {Nature
  Photonics}\ } (\bibinfo {year} {2018}),\
  10.1038/s41566-018-0179-3}\BibitemShut {NoStop}%
\bibitem [{\citenamefont {Peterson}\ \emph {et~al.}(2018)\citenamefont
  {Peterson}, \citenamefont {Benalcazar}, \citenamefont {Hughes},\ and\
  \citenamefont {Bahl}}]{peterson2018}%
  \BibitemOpen
  \bibfield  {author} {\bibinfo {author} {\bibfnamefont {Christopher~W.}\
  \bibnamefont {Peterson}}, \bibinfo {author} {\bibfnamefont {Wladimir~A.}\
  \bibnamefont {Benalcazar}}, \bibinfo {author} {\bibfnamefont {Taylor~L.}\
  \bibnamefont {Hughes}}, \ and\ \bibinfo {author} {\bibfnamefont {Gaurav}\
  \bibnamefont {Bahl}},\ }\bibfield  {title} {\enquote {\bibinfo {title} {A
  quantized microwave quadrupole insulator with topologically protected corner
  states},}\ }\href {http://dx.doi.org/10.1038/nature25777} {\bibfield
  {journal} {\bibinfo  {journal} {Nature}\ }\textbf {\bibinfo {volume} {555}},\
  \bibinfo {pages} {346 EP --} (\bibinfo {year} {2018})}\BibitemShut {NoStop}%
\bibitem [{\citenamefont {Serra-Garcia}\ \emph {et~al.}(2019)\citenamefont
  {Serra-Garcia}, \citenamefont {S\"usstrunk},\ and\ \citenamefont
  {Huber}}]{serragarcia2019}%
  \BibitemOpen
  \bibfield  {author} {\bibinfo {author} {\bibfnamefont {Marc}\ \bibnamefont
  {Serra-Garcia}}, \bibinfo {author} {\bibfnamefont {Roman}\ \bibnamefont
  {S\"usstrunk}}, \ and\ \bibinfo {author} {\bibfnamefont {Sebastian~D.}\
  \bibnamefont {Huber}},\ }\bibfield  {title} {\enquote {\bibinfo {title}
  {Observation of quadrupole transitions and edge mode topology in an lc
  circuit network},}\ }\href {\doibase 10.1103/PhysRevB.99.020304} {\bibfield
  {journal} {\bibinfo  {journal} {Phys. Rev. B}\ }\textbf {\bibinfo {volume}
  {99}},\ \bibinfo {pages} {020304} (\bibinfo {year} {2019})}\BibitemShut
  {NoStop}%
\bibitem [{\citenamefont {Imhof}\ \emph {et~al.}(2018)\citenamefont {Imhof},
  \citenamefont {Berger}, \citenamefont {Bayer}, \citenamefont {Brehm},
  \citenamefont {Molenkamp}, \citenamefont {Kiessling}, \citenamefont
  {Schindler}, \citenamefont {Lee}, \citenamefont {Greiter}, \citenamefont
  {Neupert},\ and\ \citenamefont {Thomale}}]{imhof2018}%
  \BibitemOpen
  \bibfield  {author} {\bibinfo {author} {\bibfnamefont {Stefan}\ \bibnamefont
  {Imhof}}, \bibinfo {author} {\bibfnamefont {Christian}\ \bibnamefont
  {Berger}}, \bibinfo {author} {\bibfnamefont {Florian}\ \bibnamefont {Bayer}},
  \bibinfo {author} {\bibfnamefont {Johannes}\ \bibnamefont {Brehm}}, \bibinfo
  {author} {\bibfnamefont {Laurens~W.}\ \bibnamefont {Molenkamp}}, \bibinfo
  {author} {\bibfnamefont {Tobias}\ \bibnamefont {Kiessling}}, \bibinfo
  {author} {\bibfnamefont {Frank}\ \bibnamefont {Schindler}}, \bibinfo {author}
  {\bibfnamefont {Ching~Hua}\ \bibnamefont {Lee}}, \bibinfo {author}
  {\bibfnamefont {Martin}\ \bibnamefont {Greiter}}, \bibinfo {author}
  {\bibfnamefont {Titus}\ \bibnamefont {Neupert}}, \ and\ \bibinfo {author}
  {\bibfnamefont {Ronny}\ \bibnamefont {Thomale}},\ }\bibfield  {title}
  {\enquote {\bibinfo {title} {Topolectrical-circuit realization of topological
  corner modes},}\ }\href {\doibase 10.1038/s41567-018-0246-1} {\bibfield
  {journal} {\bibinfo  {journal} {Nature Physics}\ }\textbf {\bibinfo {volume}
  {14}},\ \bibinfo {pages} {925--929} (\bibinfo {year} {2018})}\BibitemShut
  {NoStop}%
\bibitem [{\citenamefont {Xie}\ \emph {et~al.}(2018)\citenamefont {Xie},
  \citenamefont {Wang}, \citenamefont {Wang}, \citenamefont {Zhu},
  \citenamefont {Jiang}, \citenamefont {Lu},\ and\ \citenamefont
  {Chen}}]{xie2018}%
  \BibitemOpen
  \bibfield  {author} {\bibinfo {author} {\bibfnamefont {Bi~Ye}\ \bibnamefont
  {Xie}}, \bibinfo {author} {\bibfnamefont {Hong~Fei}\ \bibnamefont {Wang}},
  \bibinfo {author} {\bibfnamefont {Hai-Xiao}\ \bibnamefont {Wang}}, \bibinfo
  {author} {\bibfnamefont {Xue~Yi}\ \bibnamefont {Zhu}}, \bibinfo {author}
  {\bibfnamefont {Jian-Hua}\ \bibnamefont {Jiang}}, \bibinfo {author}
  {\bibfnamefont {Ming~Hui}\ \bibnamefont {Lu}}, \ and\ \bibinfo {author}
  {\bibfnamefont {Yan~Feng}\ \bibnamefont {Chen}},\ }\bibfield  {title}
  {\enquote {\bibinfo {title} {Second-order photonic topological insulator with
  corner states},}\ }\href@noop {} {\bibfield  {journal} {\bibinfo  {journal}
  {arxiv:1805.07555}\ } (\bibinfo {year} {2018})}\BibitemShut {NoStop}%
\bibitem [{\citenamefont {Ni}\ \emph {et~al.}(2018)\citenamefont {Ni},
  \citenamefont {Weiner}, \citenamefont {Alu},\ and\ \citenamefont
  {Khanikaev}}]{khanikaev2018}%
  \BibitemOpen
  \bibfield  {author} {\bibinfo {author} {\bibfnamefont {Xiang}\ \bibnamefont
  {Ni}}, \bibinfo {author} {\bibfnamefont {Matthew}\ \bibnamefont {Weiner}},
  \bibinfo {author} {\bibfnamefont {Andrea}\ \bibnamefont {Alu}}, \ and\
  \bibinfo {author} {\bibfnamefont {Alexander~B.}\ \bibnamefont {Khanikaev}},\
  }\bibfield  {title} {\enquote {\bibinfo {title} {Observation of bulk
  polarization transitions and higher-order embedded topological eigenstates
  for sound},}\ }\href@noop {} {\bibfield  {journal} {\bibinfo  {journal}
  {arxiv:1807.00896}\ } (\bibinfo {year} {2018})}\BibitemShut {NoStop}%
\bibitem [{\citenamefont {Xue}\ \emph {et~al.}(2018)\citenamefont {Xue},
  \citenamefont {Yang}, \citenamefont {Gao}, \citenamefont {Chong},\ and\
  \citenamefont {Zhang}}]{xue2018}%
  \BibitemOpen
  \bibfield  {author} {\bibinfo {author} {\bibfnamefont {Haoran}\ \bibnamefont
  {Xue}}, \bibinfo {author} {\bibfnamefont {Yahui}\ \bibnamefont {Yang}},
  \bibinfo {author} {\bibfnamefont {Fei}\ \bibnamefont {Gao}}, \bibinfo
  {author} {\bibfnamefont {Yidong}\ \bibnamefont {Chong}}, \ and\ \bibinfo
  {author} {\bibfnamefont {Baile}\ \bibnamefont {Zhang}},\ }\bibfield  {title}
  {\enquote {\bibinfo {title} {Acoustic higher-order topological insulator on a
  kagome lattice},}\ }\href@noop {} {\bibfield  {journal} {\bibinfo  {journal}
  {arXiv:1806.09418}\ } (\bibinfo {year} {2018})}\BibitemShut {NoStop}%
\bibitem [{\citenamefont {Kempkes}\ \emph {et~al.}(2019)\citenamefont
  {Kempkes}, \citenamefont {Slot}, \citenamefont {van~den Broeke},
  \citenamefont {Capiod}, \citenamefont {Benalcazar}, \citenamefont
  {Vanmaekelbergh}, \citenamefont {Bercioux}, \citenamefont {Swart},\ and\
  \citenamefont {Morais~Smith}}]{kempkes2019}%
  \BibitemOpen
  \bibfield  {author} {\bibinfo {author} {\bibfnamefont {S.~N.}\ \bibnamefont
  {Kempkes}}, \bibinfo {author} {\bibfnamefont {M.~R.}\ \bibnamefont {Slot}},
  \bibinfo {author} {\bibfnamefont {J.~J.}\ \bibnamefont {van~den Broeke}},
  \bibinfo {author} {\bibfnamefont {P.}~\bibnamefont {Capiod}}, \bibinfo
  {author} {\bibfnamefont {W.~A.}\ \bibnamefont {Benalcazar}}, \bibinfo
  {author} {\bibfnamefont {D.}~\bibnamefont {Vanmaekelbergh}}, \bibinfo
  {author} {\bibfnamefont {D.}~\bibnamefont {Bercioux}}, \bibinfo {author}
  {\bibfnamefont {I.}~\bibnamefont {Swart}}, \ and\ \bibinfo {author}
  {\bibfnamefont {C.}~\bibnamefont {Morais~Smith}},\ }\bibfield  {title}
  {\enquote {\bibinfo {title} {Robust zero-energy modes in an electronic
  higher-order topological insulator},}\ }\href {\doibase
  10.1038/s41563-019-0483-4} {\bibfield  {journal} {\bibinfo  {journal} {Nature
  Materials}\ }\textbf {\bibinfo {volume} {18}},\ \bibinfo {pages} {1292--1297}
  (\bibinfo {year} {2019})}\BibitemShut {NoStop}%
\bibitem [{\citenamefont {Xue}\ \emph {et~al.}(2019{\natexlab{a}})\citenamefont
  {Xue}, \citenamefont {Yang}, \citenamefont {Liu}, \citenamefont {Gao},
  \citenamefont {Chong},\ and\ \citenamefont {Zhang}}]{xue2019b}%
  \BibitemOpen
  \bibfield  {author} {\bibinfo {author} {\bibfnamefont {Haoran}\ \bibnamefont
  {Xue}}, \bibinfo {author} {\bibfnamefont {Yahui}\ \bibnamefont {Yang}},
  \bibinfo {author} {\bibfnamefont {Guigeng}\ \bibnamefont {Liu}}, \bibinfo
  {author} {\bibfnamefont {Fei}\ \bibnamefont {Gao}}, \bibinfo {author}
  {\bibfnamefont {Yidong}\ \bibnamefont {Chong}}, \ and\ \bibinfo {author}
  {\bibfnamefont {Baile}\ \bibnamefont {Zhang}},\ }\bibfield  {title} {\enquote
  {\bibinfo {title} {Realization of an acoustic third-order topological
  insulator},}\ }\href {\doibase 10.1103/PhysRevLett.122.244301} {\bibfield
  {journal} {\bibinfo  {journal} {Phys. Rev. Lett.}\ }\textbf {\bibinfo
  {volume} {122}},\ \bibinfo {pages} {244301} (\bibinfo {year}
  {2019}{\natexlab{a}})}\BibitemShut {NoStop}%
\bibitem [{\citenamefont {Mittal}\ \emph {et~al.}(2019)\citenamefont {Mittal},
  \citenamefont {Orre}, \citenamefont {Zhu}, \citenamefont {Gorlach},
  \citenamefont {Poddubny},\ and\ \citenamefont {Hafezi}}]{mittal2019}%
  \BibitemOpen
  \bibfield  {author} {\bibinfo {author} {\bibfnamefont {Sunil}\ \bibnamefont
  {Mittal}}, \bibinfo {author} {\bibfnamefont {Venkata~Vikram}\ \bibnamefont
  {Orre}}, \bibinfo {author} {\bibfnamefont {Guanyu}\ \bibnamefont {Zhu}},
  \bibinfo {author} {\bibfnamefont {Maxim~A.}\ \bibnamefont {Gorlach}},
  \bibinfo {author} {\bibfnamefont {Alexander}\ \bibnamefont {Poddubny}}, \
  and\ \bibinfo {author} {\bibfnamefont {Mohammad}\ \bibnamefont {Hafezi}},\
  }\bibfield  {title} {\enquote {\bibinfo {title} {Photonic quadrupole
  topological phases},}\ }\href {\doibase 10.1038/s41566-019-0452-0} {\bibfield
   {journal} {\bibinfo  {journal} {Nature Photonics}\ }\textbf {\bibinfo
  {volume} {13}},\ \bibinfo {pages} {692--696} (\bibinfo {year}
  {2019})}\BibitemShut {NoStop}%
\bibitem [{\citenamefont {Bao}\ \emph {et~al.}(2019)\citenamefont {Bao},
  \citenamefont {Zou}, \citenamefont {Zhang}, \citenamefont {He}, \citenamefont
  {Sun},\ and\ \citenamefont {Zhang}}]{bao2019}%
  \BibitemOpen
  \bibfield  {author} {\bibinfo {author} {\bibfnamefont {Jiacheng}\
  \bibnamefont {Bao}}, \bibinfo {author} {\bibfnamefont {Deyuan}\ \bibnamefont
  {Zou}}, \bibinfo {author} {\bibfnamefont {Weixuan}\ \bibnamefont {Zhang}},
  \bibinfo {author} {\bibfnamefont {Wenjing}\ \bibnamefont {He}}, \bibinfo
  {author} {\bibfnamefont {Houjun}\ \bibnamefont {Sun}}, \ and\ \bibinfo
  {author} {\bibfnamefont {Xiangdong}\ \bibnamefont {Zhang}},\ }\bibfield
  {title} {\enquote {\bibinfo {title} {Topoelectrical circuit octupole
  insulator with topologically protected corner states},}\ }\href {\doibase
  10.1103/PhysRevB.100.201406} {\bibfield  {journal} {\bibinfo  {journal}
  {Phys. Rev. B}\ }\textbf {\bibinfo {volume} {100}},\ \bibinfo {pages}
  {201406} (\bibinfo {year} {2019})}\BibitemShut {NoStop}%
\bibitem [{\citenamefont {Xue}\ \emph {et~al.}(2019{\natexlab{b}})\citenamefont
  {Xue}, \citenamefont {Ge}, \citenamefont {Sun}, \citenamefont {Wang},
  \citenamefont {Jia}, \citenamefont {Guan}, \citenamefont {Yuan},
  \citenamefont {Chong},\ and\ \citenamefont {Zhang}}]{xue2019}%
  \BibitemOpen
  \bibfield  {author} {\bibinfo {author} {\bibfnamefont {Haoran}\ \bibnamefont
  {Xue}}, \bibinfo {author} {\bibfnamefont {Yong}\ \bibnamefont {Ge}}, \bibinfo
  {author} {\bibfnamefont {Hong-Xiang}\ \bibnamefont {Sun}}, \bibinfo {author}
  {\bibfnamefont {Qiang}\ \bibnamefont {Wang}}, \bibinfo {author}
  {\bibfnamefont {Ding}\ \bibnamefont {Jia}}, \bibinfo {author} {\bibfnamefont
  {Yi-Jun}\ \bibnamefont {Guan}}, \bibinfo {author} {\bibfnamefont {Shou-Qi}\
  \bibnamefont {Yuan}}, \bibinfo {author} {\bibfnamefont {Yidong}\ \bibnamefont
  {Chong}}, \ and\ \bibinfo {author} {\bibfnamefont {Baile}\ \bibnamefont
  {Zhang}},\ }\href@noop {} {\enquote {\bibinfo {title} {Quantized octupole
  acoustic topological insulator},}\ } (\bibinfo {year} {2019}{\natexlab{b}}),\
  \Eprint {http://arxiv.org/abs/1911.06068} {arXiv:1911.06068
  [cond-mat.mes-hall]} \BibitemShut {NoStop}%
\bibitem [{\citenamefont {Ni}\ \emph {et~al.}(2019)\citenamefont {Ni},
  \citenamefont {Li}, \citenamefont {Weiner}, \citenamefont {Al{\`u}},\ and\
  \citenamefont {Khanikaev}}]{khanikaev2019}%
  \BibitemOpen
  \bibfield  {author} {\bibinfo {author} {\bibfnamefont {Xiang}\ \bibnamefont
  {Ni}}, \bibinfo {author} {\bibfnamefont {Mengyao}\ \bibnamefont {Li}},
  \bibinfo {author} {\bibfnamefont {Matthew}\ \bibnamefont {Weiner}}, \bibinfo
  {author} {\bibfnamefont {Andrea}\ \bibnamefont {Al{\`u}}}, \ and\ \bibinfo
  {author} {\bibfnamefont {Alexander~B.}\ \bibnamefont {Khanikaev}},\
  }\href@noop {} {\enquote {\bibinfo {title} {Demonstration of a quantized
  acoustic octupole topological insulator},}\ } (\bibinfo {year} {2019}),\
  \Eprint {http://arxiv.org/abs/1911.06469} {arXiv:1911.06469
  [cond-mat.mes-hall]} \BibitemShut {NoStop}%
\bibitem [{\citenamefont {Aidelsburger}\ \emph {et~al.}(2011)\citenamefont
  {Aidelsburger}, \citenamefont {Atala}, \citenamefont {Nascimb\`ene},
  \citenamefont {Trotzky}, \citenamefont {Chen},\ and\ \citenamefont
  {Bloch}}]{aidelsburger2011}%
  \BibitemOpen
  \bibfield  {author} {\bibinfo {author} {\bibfnamefont {M.}~\bibnamefont
  {Aidelsburger}}, \bibinfo {author} {\bibfnamefont {M.}~\bibnamefont {Atala}},
  \bibinfo {author} {\bibfnamefont {S.}~\bibnamefont {Nascimb\`ene}}, \bibinfo
  {author} {\bibfnamefont {S.}~\bibnamefont {Trotzky}}, \bibinfo {author}
  {\bibfnamefont {Y.-A.}\ \bibnamefont {Chen}}, \ and\ \bibinfo {author}
  {\bibfnamefont {I.}~\bibnamefont {Bloch}},\ }\bibfield  {title} {\enquote
  {\bibinfo {title} {Experimental realization of strong effective magnetic
  fields in an optical lattice},}\ }\href {\doibase
  10.1103/PhysRevLett.107.255301} {\bibfield  {journal} {\bibinfo  {journal}
  {Phys. Rev. Lett.}\ }\textbf {\bibinfo {volume} {107}},\ \bibinfo {pages}
  {255301} (\bibinfo {year} {2011})}\BibitemShut {NoStop}%
\bibitem [{\citenamefont {Li}\ \emph {et~al.}(2013)\citenamefont {Li},
  \citenamefont {Zhao},\ and\ \citenamefont {Vincent~Liu}}]{li2013}%
  \BibitemOpen
  \bibfield  {author} {\bibinfo {author} {\bibfnamefont {Xiaopeng}\
  \bibnamefont {Li}}, \bibinfo {author} {\bibfnamefont {Erhai}\ \bibnamefont
  {Zhao}}, \ and\ \bibinfo {author} {\bibfnamefont {W.}~\bibnamefont
  {Vincent~Liu}},\ }\bibfield  {title} {\enquote {\bibinfo {title} {Topological
  states in a ladder-like optical lattice containing ultracold atoms in higher
  orbital bands},}\ }\href {\doibase 10.1038/ncomms2523} {\bibfield  {journal}
  {\bibinfo  {journal} {Nature Communications}\ }\textbf {\bibinfo {volume}
  {4}},\ \bibinfo {pages} {1523} (\bibinfo {year} {2013})}\BibitemShut
  {NoStop}%
\bibitem [{\citenamefont {Atala}\ \emph {et~al.}(2013)\citenamefont {Atala},
  \citenamefont {Aidelsburger}, \citenamefont {Barreiro}, \citenamefont
  {Abanin}, \citenamefont {Kitagawa}, \citenamefont {Demler},\ and\
  \citenamefont {Bloch}}]{atala2013}%
  \BibitemOpen
  \bibfield  {author} {\bibinfo {author} {\bibfnamefont {Marcos}\ \bibnamefont
  {Atala}}, \bibinfo {author} {\bibfnamefont {Monika}\ \bibnamefont
  {Aidelsburger}}, \bibinfo {author} {\bibfnamefont {Julio~T.}\ \bibnamefont
  {Barreiro}}, \bibinfo {author} {\bibfnamefont {Dmitry}\ \bibnamefont
  {Abanin}}, \bibinfo {author} {\bibfnamefont {Takuya}\ \bibnamefont
  {Kitagawa}}, \bibinfo {author} {\bibfnamefont {Eugene}\ \bibnamefont
  {Demler}}, \ and\ \bibinfo {author} {\bibfnamefont {Immanuel}\ \bibnamefont
  {Bloch}},\ }\bibfield  {title} {\enquote {\bibinfo {title} {Direct
  measurement of the zak phase in topological bloch bands},}\ }\href {\doibase
  10.1038/nphys2790} {\bibfield  {journal} {\bibinfo  {journal} {Nature
  Physics}\ }\textbf {\bibinfo {volume} {9}},\ \bibinfo {pages} {795--800}
  (\bibinfo {year} {2013})}\BibitemShut {NoStop}%
\bibitem [{\citenamefont {Atala}\ \emph {et~al.}(2014)\citenamefont {Atala},
  \citenamefont {Aidelsburger}, \citenamefont {Lohse}, \citenamefont
  {Barreiro}, \citenamefont {Paredes},\ and\ \citenamefont
  {Bloch}}]{atala2014}%
  \BibitemOpen
  \bibfield  {author} {\bibinfo {author} {\bibfnamefont {Marcos}\ \bibnamefont
  {Atala}}, \bibinfo {author} {\bibfnamefont {Monika}\ \bibnamefont
  {Aidelsburger}}, \bibinfo {author} {\bibfnamefont {Michael}\ \bibnamefont
  {Lohse}}, \bibinfo {author} {\bibfnamefont {Julio~T.}\ \bibnamefont
  {Barreiro}}, \bibinfo {author} {\bibfnamefont {Bel{\'e}n}\ \bibnamefont
  {Paredes}}, \ and\ \bibinfo {author} {\bibfnamefont {Immanuel}\ \bibnamefont
  {Bloch}},\ }\bibfield  {title} {\enquote {\bibinfo {title} {Observation of
  chiral currents with ultracold atoms in bosonic ladders},}\ }\href {\doibase
  10.1038/nphys2998} {\bibfield  {journal} {\bibinfo  {journal} {Nature
  Physics}\ }\textbf {\bibinfo {volume} {10}},\ \bibinfo {pages} {588--593}
  (\bibinfo {year} {2014})}\BibitemShut {NoStop}%
\bibitem [{\citenamefont {Aidelsburger}\ \emph {et~al.}(2015)\citenamefont
  {Aidelsburger}, \citenamefont {Lohse}, \citenamefont {Schweizer},
  \citenamefont {Atala}, \citenamefont {Barreiro}, \citenamefont
  {Nascimb{\`e}ne}, \citenamefont {Cooper}, \citenamefont {Bloch},\ and\
  \citenamefont {Goldman}}]{aidelsburger2015}%
  \BibitemOpen
  \bibfield  {author} {\bibinfo {author} {\bibfnamefont {M.}~\bibnamefont
  {Aidelsburger}}, \bibinfo {author} {\bibfnamefont {M.}~\bibnamefont {Lohse}},
  \bibinfo {author} {\bibfnamefont {C.}~\bibnamefont {Schweizer}}, \bibinfo
  {author} {\bibfnamefont {M.}~\bibnamefont {Atala}}, \bibinfo {author}
  {\bibfnamefont {J.~T.}\ \bibnamefont {Barreiro}}, \bibinfo {author}
  {\bibfnamefont {S.}~\bibnamefont {Nascimb{\`e}ne}}, \bibinfo {author}
  {\bibfnamefont {N.~R.}\ \bibnamefont {Cooper}}, \bibinfo {author}
  {\bibfnamefont {I.}~\bibnamefont {Bloch}}, \ and\ \bibinfo {author}
  {\bibfnamefont {N.}~\bibnamefont {Goldman}},\ }\bibfield  {title} {\enquote
  {\bibinfo {title} {Measuring the chern number of hofstadter bands with
  ultracold bosonic atoms},}\ }\href {\doibase 10.1038/nphys3171} {\bibfield
  {journal} {\bibinfo  {journal} {Nature Physics}\ }\textbf {\bibinfo {volume}
  {11}},\ \bibinfo {pages} {162--166} (\bibinfo {year} {2015})}\BibitemShut
  {NoStop}%
\bibitem [{\citenamefont {Mancini}\ \emph {et~al.}(2015)\citenamefont
  {Mancini}, \citenamefont {Pagano}, \citenamefont {Cappellini}, \citenamefont
  {Livi}, \citenamefont {Rider}, \citenamefont {Catani}, \citenamefont {Sias},
  \citenamefont {Zoller}, \citenamefont {Inguscio}, \citenamefont {Dalmonte},\
  and\ \citenamefont {Fallani}}]{mancini2015}%
  \BibitemOpen
  \bibfield  {author} {\bibinfo {author} {\bibfnamefont {M.}~\bibnamefont
  {Mancini}}, \bibinfo {author} {\bibfnamefont {G.}~\bibnamefont {Pagano}},
  \bibinfo {author} {\bibfnamefont {G.}~\bibnamefont {Cappellini}}, \bibinfo
  {author} {\bibfnamefont {L.}~\bibnamefont {Livi}}, \bibinfo {author}
  {\bibfnamefont {M.}~\bibnamefont {Rider}}, \bibinfo {author} {\bibfnamefont
  {J.}~\bibnamefont {Catani}}, \bibinfo {author} {\bibfnamefont
  {C.}~\bibnamefont {Sias}}, \bibinfo {author} {\bibfnamefont {P.}~\bibnamefont
  {Zoller}}, \bibinfo {author} {\bibfnamefont {M.}~\bibnamefont {Inguscio}},
  \bibinfo {author} {\bibfnamefont {M.}~\bibnamefont {Dalmonte}}, \ and\
  \bibinfo {author} {\bibfnamefont {L.}~\bibnamefont {Fallani}},\ }\bibfield
  {title} {\enquote {\bibinfo {title} {Observation of chiral edge states with
  neutral fermions in synthetic hall ribbons},}\ }\href {\doibase
  10.1126/science.aaa8736} {\bibfield  {journal} {\bibinfo  {journal}
  {Science}\ }\textbf {\bibinfo {volume} {349}},\ \bibinfo {pages} {1510--1513}
  (\bibinfo {year} {2015})},\ \Eprint
  {http://arxiv.org/abs/https://science.sciencemag.org/content/349/6255/1510.full.pdf}
  {https://science.sciencemag.org/content/349/6255/1510.full.pdf} \BibitemShut
  {NoStop}%
\bibitem [{\citenamefont {Li}\ \emph {et~al.}(2016)\citenamefont {Li},
  \citenamefont {Duca}, \citenamefont {Reitter}, \citenamefont {Grusdt},
  \citenamefont {Demler}, \citenamefont {Endres}, \citenamefont
  {Schleier-Smith}, \citenamefont {Bloch},\ and\ \citenamefont
  {Schneider}}]{li2016}%
  \BibitemOpen
  \bibfield  {author} {\bibinfo {author} {\bibfnamefont {Tracy}\ \bibnamefont
  {Li}}, \bibinfo {author} {\bibfnamefont {Lucia}\ \bibnamefont {Duca}},
  \bibinfo {author} {\bibfnamefont {Martin}\ \bibnamefont {Reitter}}, \bibinfo
  {author} {\bibfnamefont {Fabian}\ \bibnamefont {Grusdt}}, \bibinfo {author}
  {\bibfnamefont {Eugene}\ \bibnamefont {Demler}}, \bibinfo {author}
  {\bibfnamefont {Manuel}\ \bibnamefont {Endres}}, \bibinfo {author}
  {\bibfnamefont {Monika}\ \bibnamefont {Schleier-Smith}}, \bibinfo {author}
  {\bibfnamefont {Immanuel}\ \bibnamefont {Bloch}}, \ and\ \bibinfo {author}
  {\bibfnamefont {Ulrich}\ \bibnamefont {Schneider}},\ }\bibfield  {title}
  {\enquote {\bibinfo {title} {Bloch state tomography using wilson lines},}\
  }\href {\doibase 10.1126/science.aad5812} {\bibfield  {journal} {\bibinfo
  {journal} {Science}\ }\textbf {\bibinfo {volume} {352}},\ \bibinfo {pages}
  {1094--1097} (\bibinfo {year} {2016})},\ \Eprint
  {http://arxiv.org/abs/https://science.sciencemag.org/content/352/6289/1094.full.pdf}
  {https://science.sciencemag.org/content/352/6289/1094.full.pdf} \BibitemShut
  {NoStop}%
\bibitem [{\citenamefont {Leder}\ \emph {et~al.}(2016)\citenamefont {Leder},
  \citenamefont {Grossert}, \citenamefont {Sitta}, \citenamefont {Genske},
  \citenamefont {Rosch},\ and\ \citenamefont {Weitz}}]{leder2016}%
  \BibitemOpen
  \bibfield  {author} {\bibinfo {author} {\bibfnamefont {Martin}\ \bibnamefont
  {Leder}}, \bibinfo {author} {\bibfnamefont {Christopher}\ \bibnamefont
  {Grossert}}, \bibinfo {author} {\bibfnamefont {Lukas}\ \bibnamefont {Sitta}},
  \bibinfo {author} {\bibfnamefont {Maximilian}\ \bibnamefont {Genske}},
  \bibinfo {author} {\bibfnamefont {Achim}\ \bibnamefont {Rosch}}, \ and\
  \bibinfo {author} {\bibfnamefont {Martin}\ \bibnamefont {Weitz}},\ }\bibfield
   {title} {\enquote {\bibinfo {title} {Real-space imaging of a topologically
  protected edge state with ultracold atoms in an amplitude-chirped optical
  lattice},}\ }\href {\doibase 10.1038/ncomms13112} {\bibfield  {journal}
  {\bibinfo  {journal} {Nature Communications}\ }\textbf {\bibinfo {volume}
  {7}},\ \bibinfo {pages} {13112} (\bibinfo {year} {2016})}\BibitemShut
  {NoStop}%
\bibitem [{\citenamefont {Trifunovic}\ and\ \citenamefont
  {Brouwer}()}]{HOTIReview}%
  \BibitemOpen
  \bibfield  {author} {\bibinfo {author} {\bibfnamefont {Luka}\ \bibnamefont
  {Trifunovic}}\ and\ \bibinfo {author} {\bibfnamefont {Piet~W}\ \bibnamefont
  {Brouwer}},\ }\href@noop {} {\emph {\bibinfo {title} {{Higher-order
  topological band structures}}}},\ \bibinfo {type} {Tech. Rep.},\ \Eprint
  {http://arxiv.org/abs/2003.01144v1} {arXiv:2003.01144v1} \BibitemShut
  {NoStop}%
\end{thebibliography}%
\end{document}